\documentclass[fleqn,usenatbib]{mnras}

\usepackage{newtxtext,newtxmath}

\usepackage[T1]{fontenc}

\DeclareRobustCommand{\VAN}[3]{#2}
\let\VANthebibliography\thebibliography
\def\thebibliography{\DeclareRobustCommand{\VAN}[3]{##3}\VANthebibliography}

\usepackage{graphicx}	
\usepackage{amsmath}	

\defcitealias{2019A&A...622A.132M}{M19}

\title[Global dust distribution in edge-on galaxies]{The distribution of dust in edge-on galaxies: I. The global structure}

\author[A. V. Mosenkov et al.]{
Aleksandr V. Mosenkov,$^{1,2}$\thanks{E-mail: aleksandr\_mosenkov@byu.edu}
Pavel A. Usachev,$^{2,3,4}$
Zacory Shakespear,$^{1}$
Jacob Guerrette,$^{1}$
\newauthor
Maarten~Baes,$^{5}$
Simone~Bianchi,$^{6}$
Emmanuel M. Xilouris,$^{7}$
George~A.~Gontcharov,$^{2}$
Vladimir~B.~Il'in,$^{2,3,8}$
\newauthor
Alexander~A.~Marchuk,$^{2,3}$
Sergey~S.~Savchenko,$^{2,3,4}$
Anton~A.~Smirnov$^{2,3}$
\\
$^{1}$Department of Physics and Astronomy, N283 ESC, Brigham Young University, Provo, UT 84602, USA\\
$^{2}$Central (Pulkovo) Astronomical Observatory, Russian Academy of Sciences, Pulkovskoye Chaussee 65/1, 196140 St. Petersburg, Russia\\
$^{3}$St.Petersburg State University, Universitetskii pr. 28, St.Petersburg, 198504 Russia\\
$^{4}$Special Astrophysical Observatory, Russian Academy of Sciences, Nizhnii Arkhyz, 369167 Russia\\
$^{5}$Sterrenkundig Observatorium, Department of Physics and Astronomy, Universiteit Gent Krijgslaan 281 S9, B-9000 Gent, Belgium\\
$^{6}$INAF-Osservatorio Astrofisico di Arcetri, Largo E. Fermi 5, I-50125, Florence, Italy\\
$^{7}$National Observatory of Athens, Institute for Astronomy, Astrophysics, Space Applications and Remote Sensing, 
Ioannou Metaxa\\ and Vasileos Pavlou GR-15236, Athens, Greece\\
$^{8}$Saint Petersburg University of Aerospace Instrumentation, Bol. Morskaya ul. 67A, St. Petersburg 190000, Russia\\
}

\date{Accepted XXX. Received YYY; in original form ZZZ}

\pubyear{2021}

\begin{document}
\label{firstpage}
\pagerange{\pageref{firstpage}--\pageref{lastpage}}
\maketitle

\begin{abstract}
In this first paper in a series we present a study of the global dust emission distribution in nearby edge-on spiral galaxies. Our sample consists of 16 angularly large and 13 less spatially resolved galaxies selected from the DustPedia sample. To explore the dust emission distribution, we exploit the {\it Herschel} photometry in the range 100--500~$\mu$m. We employ S\'ersic and three-dimensional disc models to fit the observed two-dimensional profiles of the galaxies. Both approaches give similar results. Our analysis unequivocally states the case for the presence of extraplanar dust in between 6 to 10 large galaxies. The results reveal that both the disc scale length and height increase as a function of wavelength between 100 and 500~$\mu$m. The dust disc scale height positively correlates with the dust disc scale length, similar to what is observed for the stellar discs. We also find correlations between the scale lengths and scale heights in the near- and far-infrared which suggest that the stellar discs and their dust counterparts are tightly connected. Furthermore, the intrinsic flattening of the dust disc is inversely proportional to the maximum rotation velocity and the dust mass of the galaxy: more massive spiral galaxies host, on average, relatively thinner dust discs. Also, there is a tendency for the dust-to-stellar scale height ratio to decrease with the dust mass and rotation velocity. We conclude that low-mass spiral galaxies host a diffuse, puffed-up dust disc with a thickness similar to that of the stellar disc.
\end{abstract}




\section{Introduction}
\label{sec:intro}

Galaxies represent complex gravitationally bound systems, the main constituents of which are enigmatic dark matter, stars, gas, and dust. The subject of interstellar dust, which manifests itself by the absorption of ultraviolet (UV) and optical photons, light scattering, and thermal emission at far-infrared/submillimetre (FIR/submm) wavelengths due to the absorbed radiation, has been extensively studied for decades \citep[see e.g., reviews by][]{1979ARA&A..17...73S,1983ARA&A..21..177S, 2003ARA&A..41..241D, 2011piim.book.....D,2018ARA&A..56..673G}. The optical manifestation of dust in the form of a sharp dust lane, observed in many edge-on galaxies, serves as clear evidence that, while only contributing minorly to the total mass of the interstellar medium (ISM), it plays a crucial role in the radiative transfer (RT) in galaxies, and represents a tremendous obstacle for studying the interior of galaxies in the UV and optical domains. Moreover, the physical role of dust in the ISM is of great importance. Dust serves as a catalyst for the transformation of atomic to molecular hydrogen \citep{1963ApJ...138..393G,2007ApJ...654..273G} and participates in the cooling and heating of the ISM \citep{1978ApJS...36..595D,1994ApJ...427..822B,2005pcim.book.....T}.
Observations and physical laboratories help us infer the characteristics of dust which are then linked to physical processes in the ISM at work. This builds up our understanding of how dust forms, evolves, and is destroyed in the context of the formation and evolution of galaxies.

The overall spatial distribution of dust in galaxies can be inferred in several ways. One way is by fitting optical images of edge-on galaxies with a prominent dust lane using RT simulations which allows one to retrieve the parameters of the dust disc \citep{1997A&A...325..135X,1998A&A...331..894X,1999A&A...344..868X,2007A&A...471..765B,2013A&A...550A..74D,2014MNRAS.441..869D,2016A&A...592A..71M,2018A&A...616A.120M}. In RT modelling, it is generally assumed that dust, similar to stars, follows an exponential distribution both in the radial and vertical directions. The fact that the global radial distribution of dust in galaxies is essentially exponential has been confirmed in a number of studies where one-dimensional, azimuthally-averaged profiles for individual galaxies \citep{1998A&A...335..807A,2009ApJ...701.1965M,2015A&A...576A..33H,2017A&A...605A..18C} and averaged profiles for a sample of galaxies \citep{2016MNRAS.462..331S} have been analysed. However, recently \citet[][hereafter \citetalias{2019A&A...622A.132M}]{2019A&A...622A.132M} found that for 320 disc galaxies observed with the {\it Herschel} Space Observatory \citep{2010A&A...518L...1P} and fitted with a two-dimensional S\'ersic \citep{1963BAAA....6...41S,1968adga.book.....S} profile, the pure exponential decline of the radial dust profile is not entirely correct --- the radial distribution of cold dust emission has, on average, a S\'ersic index (which characterizes the shape of the model surface brightness profile) of $0.7\pm0.4$ at 250~$\mu$m. They concluded that a substantial fraction of dust discs can be well-fitted with a Gaussian-like profile.  

As to the vertical distribution of dust in galaxies, an exponential law is also commonly used \citep[see e.g.,][]{2013A&A...556A..54V}, but observational evidence for employing this law instead of, for example, a sech$^2$-law is missing, mainly due to the poor resolution of contemporary FIR observations. Moreover, there is not even a consensus on which law to use for describing the {\it stellar} vertical distribution in galaxies (see \citealt{2011ApJ...741...28C}, a review by \citealt{2011ARA&A..49..301V}, and a recent study on the Milky Way by \citealt{2021MNRAS.507.5246M}). 

Apart from our Galaxy, which is viewed exactly edge-on from the Earth, dozens of nearby edge-on galaxies offer the best opportunity for exploring both the radial and vertical distributions of stars, gas, and dust in great detail. The natural and most common approach for studying the dust distribution in edge-on galaxies is by exploiting and analysing FIR/submm observations of the thermal emission of dust \citep{1998ApJ...507L.125A,2000A&A...356..795A,2013A&A...556A..54V,2014A&A...565A...4H,2016A&A...586A...8B,2021MNRAS.502..969Y}. However, other wavelength ranges  are also widely used for studying the dust structure in galaxies: UV \citep[mainly through absorption and photon scattering off dust grains,][]{2014ApJ...785L..18S,2014ApJ...789..131H,2015ApJ...815..133S,2016ApJ...833...58H,2016A&A...587A..86B,2018ApJ...862...25J,2019MNRAS.489.4690S}, optical \citep[through absorbing dust structures viewed against the background stellar light of the galaxy,][]{1997AJ....114.2463H,1999AJ....117.2077H,2000AJ....119..644H,2004AJ....128..662T,2004AJ....128..674R,2004ApJ...608..189D,2005ASPC..331..287H,2010MNRAS.404..792M,2012ApJ...753...25H,2019AJ....158..103H,2020MNRAS.495.3705N}, mid-infrared (MIR, \citealt{2006A&A...445..123I,2007A&A...474..461I,2007ApJ...668..918B}), as well as near-infrared \citep[NIR,][]{2014ApJ...786...41M} and optical polarimetry \citep{2018ApJ...862...87S}. According to these studies, there is now ample evidence for the presence of extraplanar (i.e. at high vertical distances) clouds and filaments of dust in disc galaxies.

The high-latitude clouds of dust in galaxies can be produced by several possible mechanisms. First, since thick stellar discs and haloes are ubiquitous in disc galaxies \citep[see e.g.,][]{2006AJ....131..226Y,2018A&A...610A...5C}, it is reasonable to suggest that thick dust discs or haloes are related to their stellar counterparts through the population of AGB stars present in these thick components \citep{howk2012}.  Second, the dust in the thin disc could be ejected by hydrodynamic or magnetohydrodynamic flows, hit by interstellar shocks or galactic fountains or chimneys \citep{1976ApJ...205..762S,1980ApJ...236..577B,1990ApJ...352..506H,1998LNP...506..495D} and entrained together with gas until it is finally expelled out above the midplane, forming a thick dust disc or halo, or even expelled further out in the intergalactic medium \citep{1994AJ....108.1619Z,2004AJ....128..662T,2010MNRAS.405.1025M}. This scenario suggests that small grains are partly destroyed, which is supported both observationally and theoretically \citep{2006ApJ...642L.141B,2006ApJ...652L..33W,2010A&A...510A..36M,2014A&A...570A..32B,2016A&A...586A...8B}. Other mechanisms include flows driven by magnetic field instabilities \citep{1966ApJ...145..811P,1992ApJ...401..137P}, slow global winds (e.g., cosmic ray driven winds, \citealt{2000A&A...354..480P}), and radiation pressure \citep{1991ApJ...366..443F,1991ApJ...381..137F,1993ApJ...407..157F,1998MNRAS.300.1006D}. 

Motivated by the study of \citetalias{2019A&A...622A.132M}, who considered two-dimensional  {\it Herschel} profiles for a large sample of DustPedia \citep{2017PASP..129d4102D} galaxies, oriented at {\it arbitrary} angles, we present a detailed analysis of the global structure of the cold dust emission with a focus on the vertical dust structure in nearby, well-resolved {\it edge-on} galaxies.
As such, this is the first systematic study for probing the 3D distribution of dust emission and exploring the general scaling relations between the stellar and dust structural parameters. Given the superior angular resolution and sensitivity of the {\it Herschel} observations, we are able to carry out a spatially resolved analysis for the off-plane dust structure. \citet{2016A&A...586A...8B} has explored the vertical extent of dust emission around NGC\,0891 and revealed the presence of a second, thick dust disc component. In the present paper one of our goals is to expand their study and demonstrate the existence of extraplanar FIR emission in other nearby, well-resolved edge-on galaxies. 

For this purpose we use two alternate models for describing the dust emission profiles in galaxies: S\'ersic and 3D exponential disc.
We also study the dependence of the retrieved parameters on wavelength in the FIR domain, compare the dust structure with the stellar distribution, and examine several general scaling relations of galaxies which involve the parameters of the dust structure. Finally, we link the obtained results with the above proposed scenarios for the presence of extraplanar dust in galaxies.

In this article we study the distribution of the FIR/submm emission, which is not the same as the mass density distribution of dust. The monochromatic luminosity density distribution can be approximated by a modified blackbody (MBB, see e.g. sect.~3 in \citealt{2019ApJ...874..141U}) model written as

\begin{equation}
\label{eq:mbb}
I_\nu (\lambda) = \kappa ({\lambda_0}) \ \Sigma_{\rm d} \left(\frac{\lambda_0}{\lambda}\right)^{\beta} B_\nu(T_{\rm d}, \lambda)\,,
\end{equation}
where $\kappa ({\lambda_0})$ is the dust absorption cross-section per unit mass (or opacity) at a reference wavelength $\lambda_0$, $\Sigma_{\rm d}$ is the dust mass surface density, $\beta$ is the dust emissivity index (which characterises the composition of dust grains and is typically $\lesssim 2.0$, see \citealt{2018ARA&A..56..673G} and references therein), and $B_\nu(T_{\rm d},\lambda)$ is the Planck function for the temperature $T_{\rm d}$ at wavelength $\lambda$. As one can see, apart from the dust mass surface density, the luminosity density at the given wavelength also depends on the dust temperature, which, in its turn, relates to the interstellar radiation field (ISRF), and optical properties of dust --- each of them can be a function of the coordinates inside the galaxy. \citet{2016MNRAS.462..331S} fitted MBB spectral-energy distributions (SEDs) to combined radial profiles of 117 galaxies at 100--500~$\mu$m and showed that the dust temperature gradually decreases from $\sim25$~K in the galaxy centre to $\sim15$~K at the optical radius (see their fig.~4). The behaviour of the emissivity index $\beta$ is uncertain:  in some galaxies it increases with radius, whereas in others it decreases \citep{2015A&A...576A..33H}. The dust mass absorption coefficient is poorly constrained, and it is highly uncertain how it varies with radius, albeit \citet{2019MNRAS.489.5256C} found a significant variation in $\kappa (500)$ (measured at 500~$\mu$m) for M\,74 (by a factor of 2.3) and M\,83 (by a factor of 5.3). If we assume that $\beta$ and $\kappa ({\lambda_0})$ do not vary significantly inside the galaxy (usually they are fixed in SED modelling), the constructed mass density profiles of cold dust in galaxies generally follow the FIR/submm luminosity density distribution (although with a different scale length depending on wavelength, see e.g. \citealt{2016MNRAS.462..331S,2017A&A...605A..18C}). In this paper by the dust distribution we mean the FIR surface brightness distribution, although, in fact, these terms should not be used interchangeably. 

The outline of this paper is as follows. In Section~\ref{sec:sample}, we present the sample of nearby edge-on galaxies with available {\it Herschel} observations. In Section~\ref{sec:preparation}, we describe how we prepare the data for a subsequent analysis presented in Section~\ref{sec:method}. The results of our structural decomposition are provided in Section~\ref{sec:results}. In Section~\ref{sec:relations}, we present several new galaxy scaling relations based on the obtained results and summarise our findings in Section~\ref{sec:conclusions}.


\section{The data and sample}
\label{sec:sample}

In our study, we exploit the DustPedia imagery (including the error maps) in the five {\it Herschel} bands (where applicable, see below): PACS\,100, PACS\,160, SPIRE\,250, SPIRE\,350, and SPIRE\,500~$\mu$m. In these bands, the pixel scales are as follows: $3\arcsec$, $4\arcsec$, $6\arcsec$, $8\arcsec$, and $12\arcsec$, respectively. The point spread function (PSF) FWHMs for these wavebands are $8\arcsec$ (PACS\,100), $12\arcsec$ (PACS\,160), $18\arcsec$ (SPIRE\,250), $24\arcsec$ (SPIRE\,350), and $36\arcsec$ (SPIRE\,500). The PACS\,100 waveband has thus the best resolution among the {\it Herschel} bands that we consider in this study. The dedicated reduction of the {\it Herschel} maps is presented in \citet{2018A&A...609A..37C}.

To select edge-on galaxies with available {\it Herschel} observations, we use the DustPedia \citep{2017PASP..129d4102D} galaxy sample and database\footnote{\url{http://dustpedia.astro.noa.gr/}} \citep{2018A&A...609A..37C}. From the whole DustPedia sample, we pre-selected 50 edge-on galaxies based on the galaxy inclination angle $i\geq85\degr$ estimated in \citetalias{2019A&A...622A.132M}. From this preliminary sample we rejected galaxies in which the disc is i) obviously less inclined so that the spirals are well-seen (e.g., NGC\,4359, NGC\,7090), and/or ii) the galaxy has a deformed, irregular shape (e.g., NGC\,4747). Also, we should take into account that the poor resolution of the {\it Herschel} telescope constrains our ability to recover the structural parameters of the galaxy. Therefore, we removed galaxies with a radial extent, determined at S/N=3, of less than 2\,arcmin in the PACS\,100 band and/or with a vertical extent of a few pixels only (e.g., NGC\,4289, NGC\,4710, NGC\,4758). 

After this careful selection, we created a final sample of 29 genuine edge-on galaxies with an inclination angle $i\gtrsim 85\degr$, which is either taken from the literature (mostly based on RT modelling or kinematical analysis) or roughly estimated by positioning the dust lane with respect to the galaxy mid-plane (see \citealt{2015MNRAS.451.2376M}). In the latter case or when the dust lane is barely visible or warped (e.g. in NGC\,3628), the uncertainty on $i$ is set to a conservative value of $1.5\degr$.
For NGC\,4244, the value of its inclination angle in the literature varies from $82\degr$ \citep{2011ApJ...729...18C} to $88\degr$ \citep{2012A&A...541L...5H}. We decided to use the latter estimate as it is retrieved from RT modelling, and, due to the large difference in the estimated inclination angles, assign to it a large error of $2\degr$. The uncertainty on the galaxy inclination is taken into account when we compute the uncertainties of the model parameters of our photometric fitting described in Sect.~\ref{sec:method}.

We split by hand our final sample into two sub-samples roughly based on their angular size D25: the Main sample ($\mathrm{D25}\gtrsim4$~arcmin) and the Additional sample ($\mathrm{D25}\lesssim4$~arcmin). The Main sample comprises 16 large galaxies, most of which have RT models in the literature (see Table~\ref{tab:sample}). The high spatial resolution of these nearby galaxies makes them particularly well-suited for studying the structure of their dust component in several PACS and SPIRE wavebands. The Additional sample has 13 galaxies smaller in angular size and none of them have RT models. As these galaxies have a lower spatial resolution, we only consider the PACS\,100 band for them. This supplementary sample supports the results of our study, especially for low-mass galaxies (see below). In Sect.~\ref{sec:results}, we examine if our selected galaxies are resolved in the vertical direction based on the retrieved parameters of our modelling. 

\begin{figure*}
\label{fig:Images}
\centering
\includegraphics[width=8.8cm]{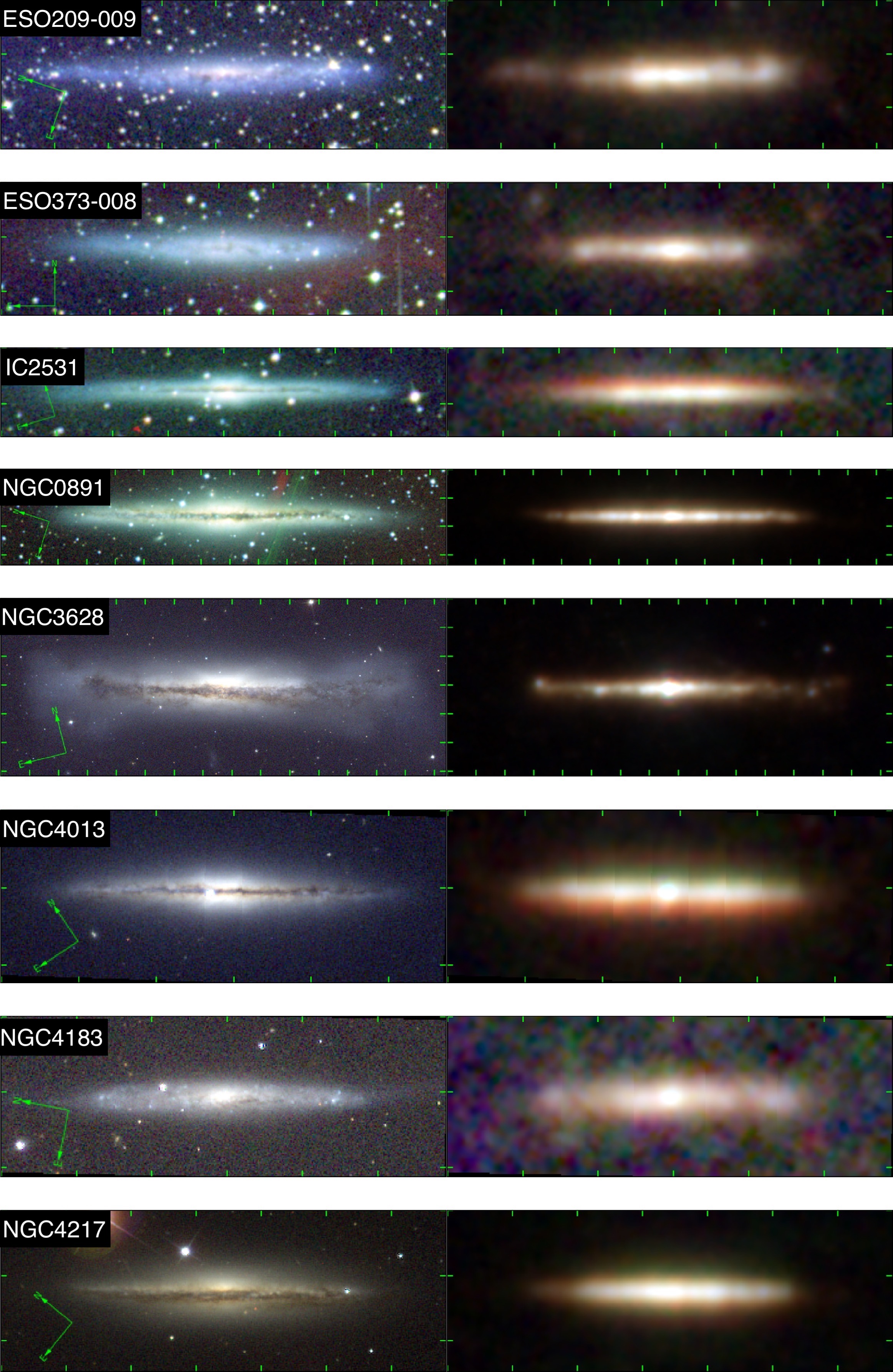}
\includegraphics[width=8.8cm]{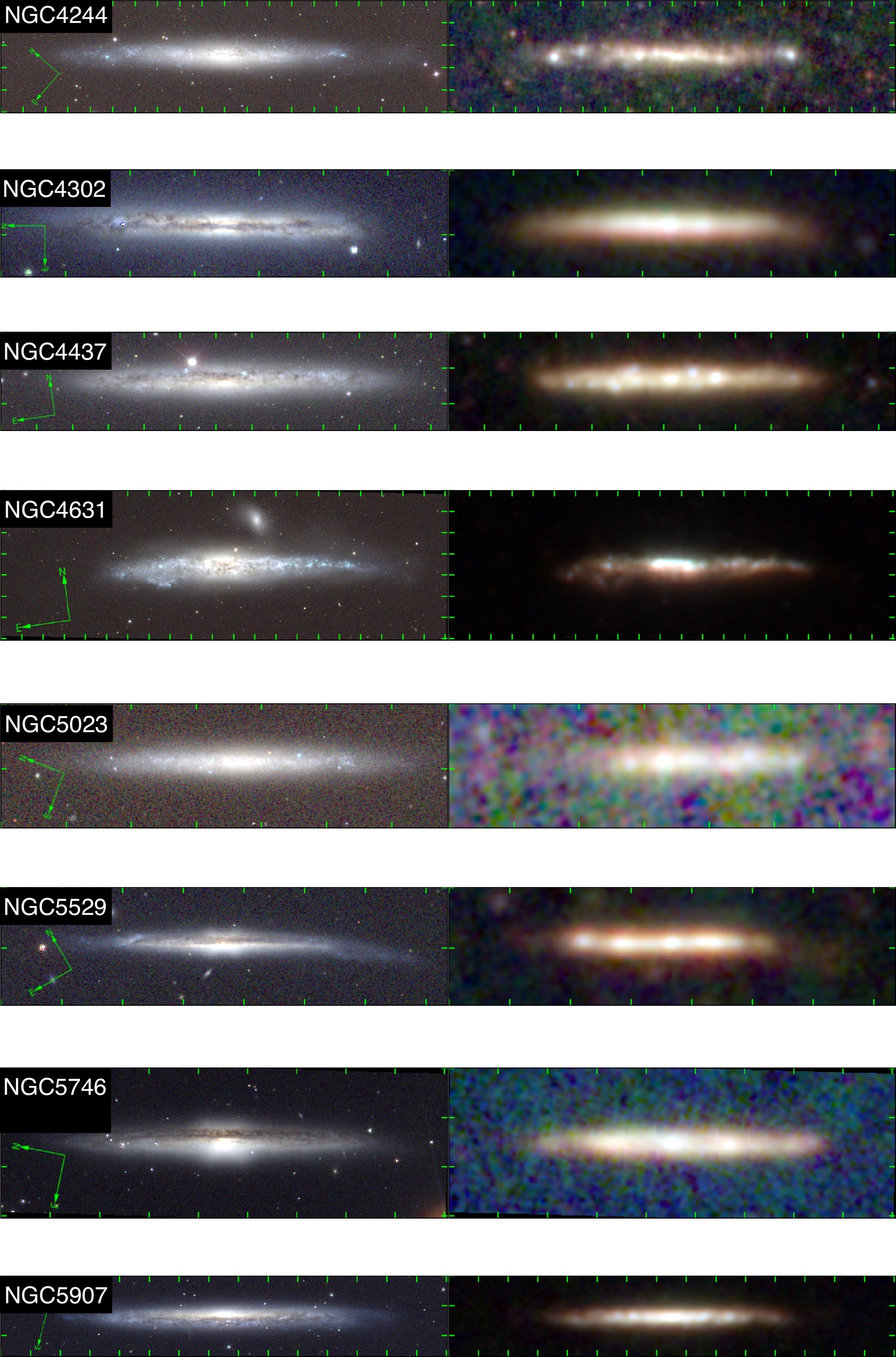}
\caption{In each panel, {\it left images:} composite RGB-snapshots for the Main sample galaxies of the $g$-, $r$- and $i$-passband frames from the SDSS or Digitized Sky Survey (POSS2/UKSTU Red, Blue and IR); {\it right images:} composite RGB-snapshots based on the {\it Herschel} 100, 160 and 250~$\mu$m images. The green ticks on the axes showcase a 1~arcmin size.}
\end{figure*}

We list the general properties of the galaxies under study in Table~\ref{tab:sample} and display their composite optical Sloan Digital Sky Survey (SDSS, \citealt{2000AJ....120.1579Y,2020ApJS..249....3A}) $gri$ (or Digitized Sky Survey, \citealt{1991PASP..103..661R,1996ASPC..101...88L}) and composite {\it Herschel} 100, 160, 250~$\mu$m images in Fig.~\ref{fig:Images}. As one can see from the table, the galaxies in our sample are nearby, have a large angular extent, and are mostly of type Sb--Sc, but span fairly wide ranges of stellar masses $9 \lesssim \log\,M_{*}/M_{\sun}  \lesssim 11$, dust masses ($6 \lesssim \log\,M_\mathrm{dust}/M_{\sun}  \lesssim 8$), and maximum rotation velocities ($70\,\mathrm{km\,s}^{-1} \lesssim v_\mathrm{rot} \lesssim 300$~km\,s$^{-1}$). Most galaxies demonstrate well-defined optical dust lanes, although in low-mass galaxies (ESO\,373-008, NGC\,4183, NGC\,4244, NGC\,5023) the dust distribution looks more diffuse and less structured. This is compatible with the results of \citet{2004ApJ...608..189D} and \citet{2019AJ....158..103H} who found that dust lanes start to appear in galaxies with $M_{*} \gtrsim 10^9\,M_{\sun}$. In NGC\,3628 and NGC\,4631, the stellar (and dust) discs are strongly disturbed due to interaction with close neighbours, which is manifested in various conspicuous tidal features in deep optical images (see \citealt{2020MNRAS.494.1751M} and references therein). 

To trace the old stellar population in the galaxies (which makes up the bulk of the stellar mass), we use the 3.4~$\mu$m (W1) imaging available thanks to the WISE space observatory \citep{2010AJ....140.1868W}. The pixel size is $1.375\arcsec$ and FWHM$=6.1\arcsec$. The reason why we use WISE data instead of, for example, the {\it Spitzer} Space Telescope \citep{2004ApJS..154....1W} with a much better resolution is that i) all galaxies in our sample are imaged by WISE and ii) the poor WISE resolution is more compatible with the {\it Herschel} resolution as fine structural details in the large nearby galaxies are smeared out and their global stellar structure (decomposed into a single disc and a bulge) can be easily compared to the dust structure traced by {\it Herschel}. We note that the emission in the WISE W1 band can be biased by hot dust from massive star-forming regions or dust around active nuclei and emission from the 3.3~$\mu$m polycyclic aromatic hydrocarbon (PAH) feature \citep{2012ApJ...744...17M}. As shown by \citet{2015ApJS..219....5Q} using {\it Spitzer} 3.6~$\mu$m and 4.5~$\mu$m data, the contamination by dust can be as much as 20--30\%  for Sa--Sc galaxies which constitute our sample. The distribution of this contaminating emission in edge-on galaxies is concentrated along a very thin disc the vertical scale of which is typically smaller than the WISE resolution and much smaller than the thickness of the general stellar disc --- see stellar and non-stellar maps provided by the Spitzer Survey of Stellar Structure in Galaxies (S$^4$G, \citealt{2010PASP..122.1397S,2013ApJ...771...59M}) collaboration in \citet{2015ApJS..219....5Q}\footnote{\url{https://irsa.ipac.caltech.edu/data/SPITZER/S4G/galaxies/}} for edge-on galaxies including some of the galaxies from our sample. In principle, this can introduce a bias in tracing the stellar distribution and make its vertical profile thinner than it is in reality. However, given that the contamination of the dust emission does not usually exceed 30\% of the total galaxy flux at 3.6~$\mu$m, the influence of this contamination should not significantly change the results of our profile fitting of the stellar distribution described in Sect.~\ref{sec:method}.

\begin{table*}
	\centering
	\caption{The general properties of edge-on galaxies from the Main (upper part) and Additional (bottom part) samples.}
	\label{tab:sample}
	\begin{tabular}{ccccccccccc}
		\hline
		Galaxy & RA     & Dec.   & D   & D25 & $\log M_{*}$ & Type & $i$   & Ref. & $v_\mathrm{rot}$ & $\log M_\mathrm{dust}$\\
                & (J2000)&(J2000) &(Mpc)&(arcmin)& (M$_\odot$)   &     & (deg) & &  (km\,s$^{-1}$) & (M$_\odot$)  \\
                &        &        & (1) & (2)    &  (3)       & (4) &  (5)  & (6) & (7) & (8)  \\     
		\hline
ESO\,209-009 &07:58:15&-49:51:15&12.59&6.49&10.12& SBc &$87.1\pm1.5$& this study$^{*}$ & $141.4\pm3.6$ &$7.21\pm0.05$\\
ESO\,373-008 &09:33:22&-33:02:01&9.68&3.7&9.56& Sc &$89.5\pm1.5$& this study$^{*}$ & $93.7\pm2.8$ &$6.7\pm0.06$\\
IC\,2531 &09:59:56&-29:37:01&36.65&6.53&10.83& Sc &$89.5\pm0.2$& 1,2,3 & $228.3\pm3.2$ &$7.75\pm0.06$\\
NGC\,0891 &02:22:33&+42:20:57&9.86&13.03&10.82& Sb &$89.7\pm0.3$& 1,4,5 & $212.1\pm5.0$ &$7.7\pm0.05$\\
NGC\,3628 &11:20:17&+13:35:23&10.81&11.04&10.82& SBb &$88.4\pm1.5$&5& $212.1\pm3.4$ &$7.61\pm0.05$\\
NGC\,4013 &11:58:31&+43:56:48&18.79&4.88&10.73& SABb &$89.9\pm0.1$& 1,6,7  & $181.7\pm4.3$ &$7.36\pm0.04$\\
NGC\,4183 &12:13:17&+43:41:55&16.67&4.29&9.68& Sc &$85.5\pm0.1$&8& $103.7\pm2.0$ &$6.88\pm0.06$\\
NGC\,4217 &12:15:51&+47:05:30&18.54&5.53&10.77& Sb &$87.9\pm0.1$& 6,7  & $187.6\pm4.1$ &$7.51\pm0.04$\\
NGC\,4244 &12:17:30&+37:48:26&4.35&16.22&9.2& Sc &$88\pm2.0$& 9$^{*}$,10$^{*}$,11,12  & $89.0\pm2.0$ &$6.38\pm0.07$\\
NGC\,4302 &12:21:42&+14:35:54&14.32&6&10.37& Sc &$89.6\pm0.1$&6& $167.8\pm3.1$ &$7.2\pm0.05$\\
NGC\,4437 &12:32:46&+00:06:54&8.39&9.02&9.97& Sc &$86\pm1.5$& 13$^{*}$  & $139.5\pm2.6$ &$7.11\pm0.05$\\
NGC\,4631 &12:42:08&+32:32:29&7.35&14.45&10.42& SBcd &$86\pm1.5$& 14$^{*}$  & $138.5\pm3.3$ &$7.28\pm0.06$\\
NGC\,5023 &13:12:13&+44:02:28&6.58&6.22&8.81& Sc &$87\pm1.5$&15& $80.3\pm1.8$ &$6.02\pm0.08$\\
NGC\,5529 &14:15:34&+36:13:36&44.46&5.79&11.06& SABc &$87.1\pm0.1$& 1,6,7  & $282.4\pm5.2$ &$7.97\pm0.07$\\
NGC\,5746 &14:44:56&+01:57:18&27.04&7.24&11.1& Sb &$86.8\pm0.1$&6& $311.1\pm7.8$ &$7.85\pm0.06$\\
NGC\,5907 &15:15:54&+56:19:44&17.22&11.3&10.87& SABc &$86.5\pm0.3$& 16$^{*}$,1,17,7  & $226.6\pm4.3$ &$7.85\pm0.05$\\
		\hline
IC\,0610 &10:26:28&+20:13:41&29.24&1.89&10.26& Sbc &$90\pm1.5$& this study$^{*}$ & $138.1\pm4.1$ &$6.83\pm0.06$\\
IC\,2233 &08:13:59&+45:44:32&14&2.88&9.21& SBc &$88.5\pm1.5$& 18$^{*}$ & $77.1\pm1.8$ &$5.91\pm0.16$\\
NGC\,3454 &10:54:30&+17:20:39&20.61&2.41&9.7& SBc &$85\pm1.5$& this study$^{*}$ & $92.7\pm2.5$ &$6.69\pm0.08$\\
NGC\,3501 &11:02:47&+17:59:22&22.59&4.31&10.2& Sc &$87\pm1.5$& this study$^{*}$ & $138.8\pm3.0$ &$7.05\pm0.06$\\
NGC\,3592 &11:14:27&+17:15:36&20.51&2.13&9.53& Sc &$86\pm1.5$& this study$^{*}$ & $78.9\pm1.9$ &$6.45\pm0.18$\\
NGC\,4222 &12:16:23&+13:18:25&19.77&2.83&9.8& Scd &$90\pm1.5$& this study$^{*}$ & $99.8\pm2.2$ &$6.89\pm0.08$\\
NGC\,4330 &12:23:17&+11:22:05&18.71&2.29&9.91& Sc &$84\pm1.5$& 19$^{*}$ & $112.3\pm3.3$ &$6.75\pm0.06$\\
NGC\,4607 &12:41:12&+11:53:12&19.95&2.86&10.03& SBbc &$90\pm1.5$& this study$^{*}$ & $99.1\pm3.4$ &$6.84\pm0.06$\\
UGC\,07321 &12:17:34&+22:32:25&23.33&4.82&9.61& Scd &$90\pm1.5$& this study$^{*}$ & $93.9\pm2.3$ &$6.97\pm0.12$\\
UGC\,07387 &12:20:17&+04:12:06&38.19&1.99&9.98& Scd &$87\pm1.5$& this study$^{*}$ & $116.8\pm3.3$ &$6.91\pm0.1$\\
UGC\,07522 &12:25:58&+03:25:49&35.48&2.62&10.36& Sc &$90\pm1.5$& this study$^{*}$ & $143.4\pm3.7$ &$7.3\pm0.07$\\
UGC\,07982 &12:49:50&+02:51:05&19.23&2.91&9.74& Sc &$87\pm1.5$& this study$^{*}$ & $98.4\pm2.8$ &$6.63\pm0.1$\\
UGC\,09242 &14:25:21&+39:32:22&24.32&4.21&9.21& Scd &$90\pm1.5$& this study$^{*}$ & $78.7\pm2.5$ &$6.15\pm0.19$\\
\hline		
	\end{tabular}
   \parbox[t]{180mm}{ {\bf Columns:}
   (1) Distance taken from \citet{2018A&A...609A..37C} as ``best'' (preferred) distance estimate,
   (2) optical diameter taken from HyperLeda\footnote{\url{http://leda.univ-lyon1.fr/}} \citep{2014A&A...570A..13M} based on their  \texttt{logd25} parameter,
   (3) stellar mass from \citetalias{2019A&A...622A.132M},
   (4) morphological type from HyperLeda,
   (5) weighted average inclination angle and its weighted average uncertainty based on the references listed in column Ref.,
   (6) references for inclination angle mainly estimated from RT modelling or using the approach proposed in \citet{2015MNRAS.451.2376M} (* denotes non-RT methods), 
   (7) apparent maximum rotation velocity of gas and its error taken from HyperLeda,
   (8) total dust mass from \citet{2019A&A...624A..80N} derived from SED modeling using a Modified Black-Body (MBB) model.\\   
{\bf References:}  
(1) \citet{1999A&A...344..868X}.
(2) \citet{2016A&A...592A..71M}.
(3) \citet{2017MNRAS.464...48P}.
(4) \citet{1998A&A...331..894X}.
(5) \citet{2015ApJ...815..133S}.
(6) \citet{2007A&A...471..765B}.
(7) \citet{2018A&A...616A.120M}.
(8) \citet{2018ApJS..239...21S}.
(9) \citet{1996AJ....112..457O}.
(10) \citet{2011ApJ...729...18C}.
(11) \citet{2011ApJ...741....6M}.
(12) \citet{2012A&A...541L...5H}.
(13) \citet{2020MNRAS.494.1751M}.
(14) \citet{2004A&A...414..475D}.
(15) \citet{2000A&AS..145...83A}.
(16) \citet{1982A&A...110...61V}.
(17) \citet{2006A&A...459..703J}.
(18) \citet{2008AJ....135..291M}.
(19) \citet{2018A&A...614A..57F}.}
\end{table*}

\section{Data preparation}
\label{sec:preparation}

Below we describe our preparation of the {\it Herschel} data. For WISE 3.4~$\mu$m, we use already prepared data from \citetalias{2019A&A...622A.132M}. As the present study is focused on the global structure of dust in edge-on spiral galaxies, especially on detecting extraplanar dust emission, we should pay special attention to the preparation of the {\it Herschel} data. The WISE data is only used for tracing the global stellar structure of the galaxies as compared to their dust counterpart, with no emphasis on the thick stellar disc. Thus, we are satisfied with the preparation of the WISE data which has been done in \citetalias{2019A&A...622A.132M}.

For each galaxy image, we carefully examined the background emission to remove extrinsic sources. To mask out all objects, including the target galaxy, we employed {\tt{SExtractor}} \citep{1996A&AS..117..393B} with the parameters \texttt{DETECT\_THRESH=2.0} and \texttt{DETECT\_MINAREA=3}. The size of the created mask was enlarged by a factor of 1.5, as determined empirically, to totally cover all scattered light. Careful masking of the galaxy image was crucial at this point to completely exclude all emission from any sources except for the background. Therefore, we manually revisited the created masks to ensure that all detected objects had been completely masked and no scattered light from them was present. Then, we iteratively fitted the background using polynomials of varying degrees starting from zero (a flat background). We examined the background in the resultant background-subtracted image to make a decision about which polynomial of the background to use (typically, the  second degree was sufficient to fit the background around the galaxy). Some images were contaminated by Galactic cirrus, which structure cannot be fitted using a single polynomial. In such cases, we partitioned the frame into several large-scale areas, each fitted with a separate polynomial. The typical size of such structures in our images was much larger than the area occupied by the detected sources, and, thus, the fitting for these parts using a sufficient number of pixels was possible. This procedure allowed us to clean the frames from the undesired large-scale background in our images.     

All the {\it Herschel} galaxy images were rebinned to the PACS\,100 image, so that they all would have the same pixel scale (for comparison purposes) and a unified mask (to take into account the same number of pixels while fitting, see below). All galaxy images were rotated by an average angle estimated by \citetalias{2019A&A...622A.132M}, who fitted the galaxy profiles with a single S\'ersic model with a free position angle (see Sect.~\ref{sec:method}). Thus, in our final images the galactic discs are aligned with the horizontal direction. This was done to easily extract the cumulative (integrated) galaxy profiles along the major (parallel to the mid-plane) and minor (perpendicular to the mid-plane) axes in our subsequent analysis. After the rotation, we equally cropped the images for each galaxy, so that the final frames in all five bands would have the same size.  

The rebinning, rotation and cropping procedures were also applied to the error maps which contain information about the weight for each pixel as a measure of the photometric quality.

Finally, for each galaxy we created a unified mask image consisting of a sum of the individual masks in each band which we initially created for the background subtraction (excluding, of course, the target galaxy itself from this mask). Despite the presence of some contaminating objects which appear in one filter and not in others, the unified mask allows us to create cumulative radial and vertical profiles which take into account the same pixels from the galaxy in all bands used. This unified mask is exploited in our further photometric fitting of the galaxy images so that we exclude the possibility of the dependence of our models on the masks used in different bands.

Knowing the extended point spread function (PSF) is of crucial importance for accurate fitting of galaxy profiles, especially when exploring the vertical surface brightness distribution in edge-on galaxies \citep[see e.g., a comprehensive analysis in][]{2014A&A...567A..97S}. As shown by \citet{2014A&A...567A..97S} for the Hubble Space Telescope, the size of the PSF model should be ``at least 1.5 times as extended as the vertical distance of the edge-on galaxy''. In our study, we exploit deep PSFs for the PACS instrument from \citet{2016A&A...591A.117B}, combined from Vesta and Mars observations, and for the SPIRE instrument based on Neptune observations which were used for the flux calibration of this photometer \citep{2013MNRAS.433.3062B}. These PSF images have a large dynamical range ($\sim10^6$) and radius ($\sim8$~arcmin), which are sufficient for the purposes of this study --- for all galaxies in our sample, the size of the PSF is larger than the maximal vertical/radial distance in our final galaxy frames, meaning the whole range of vertical (and radial) distances is covered by the PSF. The PSF images have been rotated (to align the direction of the spacecraft's Z-axis with its direction on the galaxy images) and rebinned to 3~$\arcsec$/pix to match the respecting final galaxy images. The PSF image for the WISE~3.4~$\mu$m data was taken from the Explanatory Supplement to the WISE All-Sky Data Release Products\footnote{\url{https://wise2.ipac.caltech.edu/docs/release/allsky/expsup/}}. Along with the PSF, we use the term ``line'' spread function (LSF) as a one-dimensional beam produced by averaging the two dimensional PSF along one (vertical or radial) direction (see \citealt{2013A&A...556A..54V,2016A&A...586A...8B}).

All the steps in data preparation described above were carried out using the {\tt{IMAN}} Python package\footnote{\url{https://bitbucket.org/mosenkov/iman_new/src/master/}}.

\section{The fitting method}
\label{sec:method}

Using a single S\'ersic model, \citetalias{2019A&A...622A.132M} performed 2D fitting of cold dust emission profiles for 320 DustPedia galaxies, including all objects from our sample. Here, we modify the modelling approach which was applied in \citetalias{2019A&A...622A.132M} in several aspects.

First, in contrast to \citetalias{2019A&A...622A.132M}, who used automatic image preparation for galaxy fitting, the background subtraction and masking in this study were revised to accurately account for all possible sources that could potentially contaminate the galaxy emission. This is of high importance for studying the off-plane emission in edge-on galaxies which can be easily confused (and misinterpreted) with an under-subtracted sky background or scattered light from other objects in the image. 

Second, for galaxies with bright FIR emission clumps or other unresolved components located in the plane of the galaxy, these features are fitted using a Gaussian or S\'ersic function. This is done to minimise the influence of such compact sources on the overall galaxy profile instead of masking them out.

Third, we use two alternative models in our study: a simple S\'ersic model and a 3D broken exponential disc (see the description of these profiles in Sect.~\ref{sec:sersic} and \ref{sec:3Ddisc}, respectively). Using such a complex, non-single exponential profile to describe the FIR emission in spiral galaxies is based on the \citetalias{2019A&A...622A.132M} modelling results: the dust emission profiles often exhibit a deficit of emission in the centre (a plateau, or a core) with respect to the general exponential fitting function (also noted by \citealt{2009ApJ...701.1965M}). As shown by \citetalias{2019A&A...622A.132M}, the average S\'ersic index for 250~$\mu$m galaxy profiles is $0.7\pm0.4$, and half of the sample galaxies they studied have $n_{250}<0.7$. This indicates that the dust discs are not purely exponential and are often better described by a Gaussian function. They also concluded that the dust emission deficiency often occurs within 0.3--0.4 optical radii. In Sect.~\ref{sec:results}, we will demonstrate that this is also the case for many edge-on galaxies in our sample. 

Fourth, in this study we employ the {\tt{IMFIT}} code \citep{2015ApJ...799..226E} instead of {\tt{GALFIT}} \citep{2002AJ....124..266P,2010AJ....139.2097P} and {\tt{GALFITM}} \citep{2011ASPC..442..479B,2013MNRAS.430..330H}, which were used in \citetalias{2019A&A...622A.132M}. This is done because the latter two codes do not include 3D fitting models. In Sect.~\ref{sec:results}, we compare the results of our single S\'ersic modelling with those from \citetalias{2019A&A...622A.132M} to assess the consistency of the two fitted models with each other.   

Below, we briefly describe the two alternate functions (S\'ersic and 3D disc models) for our photometric analysis. In our fitting, we use the Levenberg-Marquardt algorithm to retrieve the best-fit parameters by minimising the $\chi^2$ value  and convolve the models using the PSF images described in Sect.~\ref{sec:preparation}. Weight maps for each galaxy image are taken into account while modelling. For each galaxy image in each band, we do not keep fixed the center or the position angle of the component used, as the astrometric calibration in the DustPedia images is not ideal and, for example, the bright center of the galaxy can be slightly shifted (by 1--2 pixels at maximum) in different bands. For galaxies with a complex FIR profile with bright clumps and/or a bright central source (a bulge, a bar, or an AGN), we add other components to our model which can be described by S\'ersic, Gaussian, or PSF profiles.

We do not repeat fitting of the WISE 3.4~$\mu$m images with a single S\'ersic model since it has been carried out in \citetalias{2019A&A...622A.132M} for the whole DustPedia sample. However, we perform a bulge-disc decomposition for all edge-on galaxies in our sample to retrieve the parameters of their stellar discs and compare them with those of the dust emission profiles.

\subsection{S\'ersic model}
\label{sec:sersic}

The S\'ersic function \citep{1963BAAA....6...41S,1968adga.book.....S} is characterised by the following major-axis intensity profile (see also \citealt{2019A&A...630A.113B}):

\begin{equation}
\label{sersic}
I(r) \; = \; I_\mathrm{e} \: \exp \left\{ -b_\mathrm{n} \left[ \left( \frac{r}{r_\mathrm{e}} \right)^{1/n} \! - \: 1 \right] \right\}\,,
\end{equation}
where $r$ is the radius, $I_\mathrm{e}$ is the surface brightness at the effective (half-light) radius $r_\mathrm{e}$
and the S\'ersic index $n$ controls the shape of the intensity profile. The
function $b_\mathrm{n}$ is calculated via the polynomial approximation found by
\citet{1999A&A...352..447C} (when $n > 0.36$) and \citet{2003ApJ...582..689M} (when $n \leq 0.36$).

For galaxies from the Main sample, we start fitting with a $250~\mu$m galaxy image. We chose this band as it has an average resolution among the five {\it Herschel} bands considered and negligible contamination of the warm dust emission as compared to the PACS bands with higher resolution. The derived $n$ parameter is then kept fixed during the fitting of the galaxy images in the remaining bands. This approach is justified  because, according to \citetalias{2019A&A...622A.132M} (see their fig.~6), the S\'ersic index of the dust emission in the $100$ -- $500\mu$m wavebands varies within 10--15\% of the average value without any systematics with wavelength, and, thus, can be taken independent (constant) of wavelength. 

For galaxies from the Additional sample, the fitting, of course, is done in only one PACS\,100 waveband.

Since for some sample galaxies the inclination angle is slightly different from $90\degr$, we correct the apparent flattening $q$ (another free parameter in our model) for the inclination to compute the intrinsic flattening $q_0$ using the Hubble formula \citep{1926ApJ....64..321H}:
\begin{equation}
\label{hubble}
q_0^\mathrm{sersic} = \frac{\sqrt{q^2-\cos^2 i}}{\sin i}\,.
\end{equation}

The uncertainty of this parameter includes the error on $q$ from {\tt{IMFIT}} and the inclination uncertainty given in Table~\ref{tab:sample}.

\subsection{3D disc model}
\label{sec:3Ddisc}

The function we describe below represents a 3D luminosity density model for an axisymmetric disc, the vertical $Z$-axis of which is inclined by the angle $i$ to the line of sight. The radial profile in this model is a broken exponential function \citep{1970ApJ...160..811F} and the vertical profile has a simple exponential decline. The 3D luminosity density model $j(R,z)$ at radius $R$ from the central axis and at height $z$ from the galaxy midplane is given by

\begin{equation}
\label{3d_disc:z}
j(R,z) \; = \; j_{\rm rad}(R) \:  \mathrm{e}^{-|z|/h_{z}}\,,
\end{equation}
where $h_{z}$ is the vertical scale height of the disc.

The radial distribution in the disc plane is described by
\begin{equation}
\label{3d_disc:r}
j_{\rm rad}(R) \; = \; S \, j_{0} \, \mathrm{e}^{-\frac{R}{h_\mathrm{in}}} [1 + \mathrm{e}^{\alpha(R \, - \,
R_\mathrm{b})}]^{\frac{1}{\alpha} (\frac{1}{h_\mathrm{in}} \, - \, \frac{1}{h_\mathrm{out}})}\,,
\end{equation}
where $j_{0}$ is the central luminosity density, $h_\mathrm{in}$ is the exponential scale length of the inner disc region before the break at the radius $R_\mathrm{b}$, and $h_\mathrm{out}$ is the exponential scale length of the outer disc region beyond the break radius (see \citealt{2008AJ....135...20E}). The parameter $\alpha$ controls the abruptness of the transition between the inner and outer regions of the disc. In our fitting, we constrain this parameter between 0.1 (steady change of the profile) and 50 (abrupt break). The scaling factor $S$ is given by

\begin{equation}
\label{3d_disc:S}
S \; = \; (1 + {\rm e}^{-\alpha \, R_\mathrm{b}}) ^{-\frac{1}{\alpha} (\frac{1}{h_\mathrm{in}} \, - \, \frac{1}{h_\mathrm{out}})}\,.
\end{equation}

If the 3D luminosity density distribution has no break along the radial profile, then it can be given by

\begin{equation}
\label{3d_disc:simple}
j(R,z) \; = \;  J_{0} \, \mathrm{e}^{-R/h_R - |z|/h_z}\,.
\end{equation}

As noted above, the inner galaxy region often demonstrates a deficit of FIR emission which means that $h_\mathrm{in}$ should have a very large value. For all such profiles, we fix $h_\mathrm{in}=1000$~pix, for simplicity. Also, for all galaxies with a broken exponential profile, we do not fit the break radius $R_\mathrm{b}$, but manually set it (and keep it fixed) by analysing cumulative radial profiles of the galaxy in all wavebands with the sufficient resolution in the vertical direction. Further on in the paper, we define $h_R = h_\mathrm{in}$ if $h_\mathrm{in}$ is of the same order as $h_\mathrm{out}$ (i.e. if the intensity within the break radius is not constant) and $h_R = h_\mathrm{out}$ otherwise.   

In our fitting, the model inclination $i$ is kept fixed at the galaxy inclination listed in Table~\ref{tab:sample}. This is done to reduce the number of free parameters and to avoid possible degeneracies in our modelling. For calculating the uncertainties of the fit parameters and to take into account the effect of inclination (for very thin discs, even a very little change of inclination can be dramatic for the observed profile, see e.g., \citealt{2013A&A...556A..54V}), we run 50 sets of models with different (but fixed during fitting) inclination angles. The values of the inclination are  drawn from a normal distribution with the mean and standard deviation taken from Table~\ref{tab:sample}.

For WISE 3.4~$\mu$m, we only consider a simple exponential 3D disc without a break as almost all galaxies in our sample do not show strong breaks in the WISE radial profiles. This is likely due to the poor resolution of the WISE data coupled with the fact that the stellar populations do not show a flat profile in the inner part of the galaxy which is observed in many dust emission profiles, as shown in  \citetalias{2019A&A...622A.132M}.

As in the case of the S\'ersic modelling, we start our fitting with the SPIRE\,250 band and then use the output parameters for this band as the first guess parameters for the other {\it Herschel} bands. For the Additional sample, we only fit the images at 100~$\mu$m.

To find the intrinsic flattening $q_0^\mathrm{disc}$ for both the broken exponential and simple 3D disc models, we determine the apparent flattening for an outer isophote of 0.003 Jy/pix of the non-convolved model rotated to $90\degr$. This isophote level was estimated empirically for all galaxies in our sample. It has a sufficient S/N ratio $>3$ in all wavebands used and is outside of the break radius if it exists for all the galaxies in the sample.

\section{Results}
\label{sec:results}

In this section, we present the results of our fitting and compare them with the literature. We consider dependencies of the retrieved parameters on wavelength and explore different scaling relations including correlations with the general galaxy observables. 

The results of our fitting for the S\'ersic and 3D disc models are provided in Table~A1, with the respective errors if available. Cumulative radial and vertical profiles of the sample galaxies with the superimposed models in the PACS\,100 waveband are shown in Fig.~\ref{fig:profiles}\footnote{Profiles in all wavebands used and for the S\'ersic model are provided in the online material.}. These profiles were created by averaging the intensities in the four quadrants created if we dissect the galaxy along the major and minor axes. Then, we find an average for all the pixels perpendicular to the mid-plane for the given radius (cumulative radial profile) or parallel to the mid-plane for the given height (cumulative vertical profile). This approach is used to significantly increase the S/N ratio in the resultant profile and to smooth out bright non-axisymmetric features. In our plots, we also add the LSF to demonstrate how the resolution can affect the observed and model profiles.

\subsection{Notes on the fit results}
\label{sec:notes}

For the whole sample (if we exclude outliers with extraplanar dust emission, see Sect.~\ref{sec:profiles}), we find a reduced $\chi^2=1.04\pm0.45$ for the 3D disc models and $0.97\pm0.44$ for the S\'ersic models in all {\it Herschel} bands under study. These similar values of the reduced $\chi^2$ suggest that both models are almost equally good in describing the 2D galaxy profiles. Therefore, in Sect.~\ref{sec:relations} we will use the results for both models to strengthen our conclusions on galaxy scaling relations.

According to the results of our modelling of the {\it Herschel} images, not all galaxies in our Main and Additional samples have reliable values of the apparent flattening and, consequently, the disc scale height in the wavebands we selected for our analysis. The apparent width of the emission profile in the vertical direction serves as a good criterion for examining the reliability of the fitting: it should be larger than the PSF HWHM=FWHM/2 in this specific waveband, $r_\mathrm{e,\lambda}\cdot q_{\lambda}>0.8\,\mathrm{HWHM}_{\lambda}$ \citep[see fig.~7 and corresponding text in][]{2009MNRAS.393.1531G}. 
Indeed, for some galaxies in Fig.~\ref{fig:profiles}, the FWHM of the vertical cumulative profile is closely or almost completely aligned with the PSF profile so that the shape of the inner part of the vertical galaxy profile is virtually described by the PSF shape. We mark such galaxies in Table~A1 by *. In our Main and Additional samples, 13 and 8 galaxies, respectively, are vertically resolved in PACS\,100. 7 galaxies from our Main sample appeared to be vertically resolved in the three PACS\,100--SPIRE\,250 bands, and only four galaxies are vertically resolved in all five PACS\,100--SPIRE\,500 wavebands. 

We note that the galaxy can have a reliable estimate of the exponential scale length and a poorly constrained estimate of the scale height due to the insufficient {\it Herschel} resolution in this waveband. We do not consider such galaxies in our statistics (marked by * in Table~A1) and scaling relations in Sect.~\ref{sec:relations} where the disc scale height $h_{z,\lambda}$, the intrinsic disc flattening $q_{0,\lambda}^\mathrm{disc}$, and the S\'ersic apparent flattening $q_{\lambda}$ or intrinsic flattening $q_{0,\lambda}^\mathrm{sersic}$ are present.

\subsection{Validity of the results}
\label{sec:comparison}

\begin{figure*}
\label{fig:M19_compar}
\centering
\includegraphics[height=4.3cm]{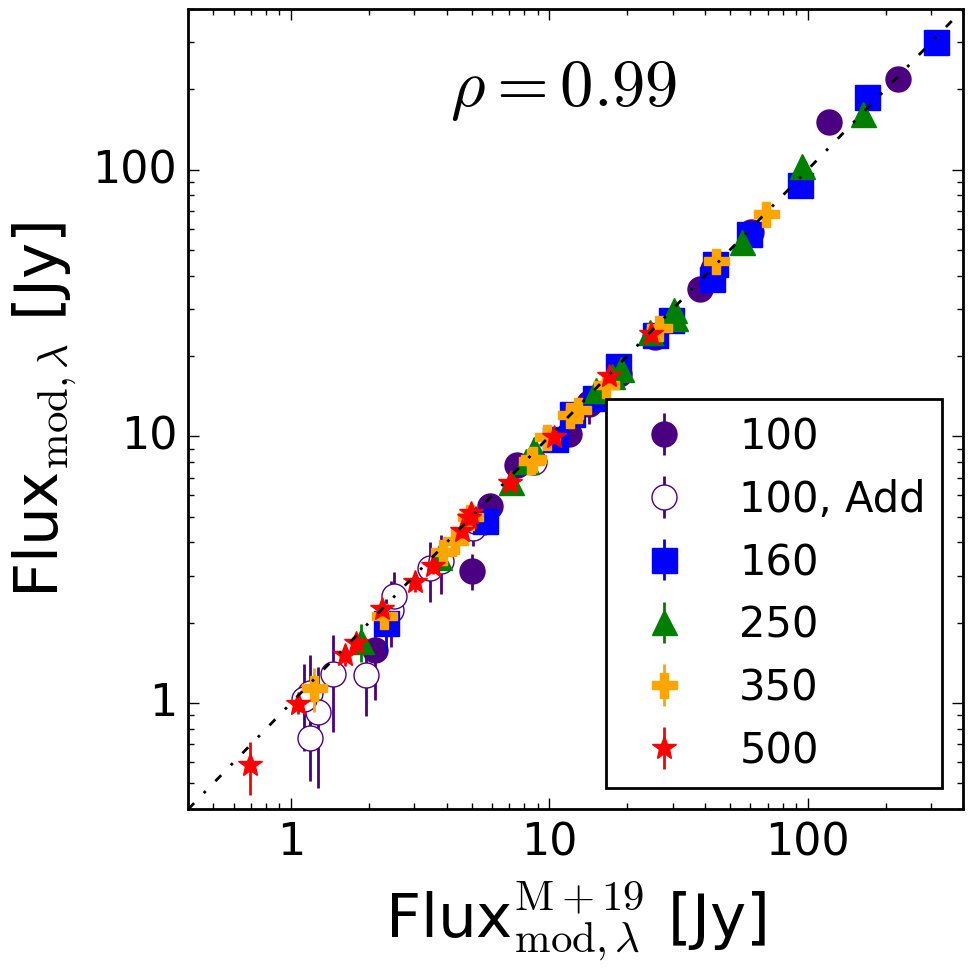}
\includegraphics[height=4.3cm]{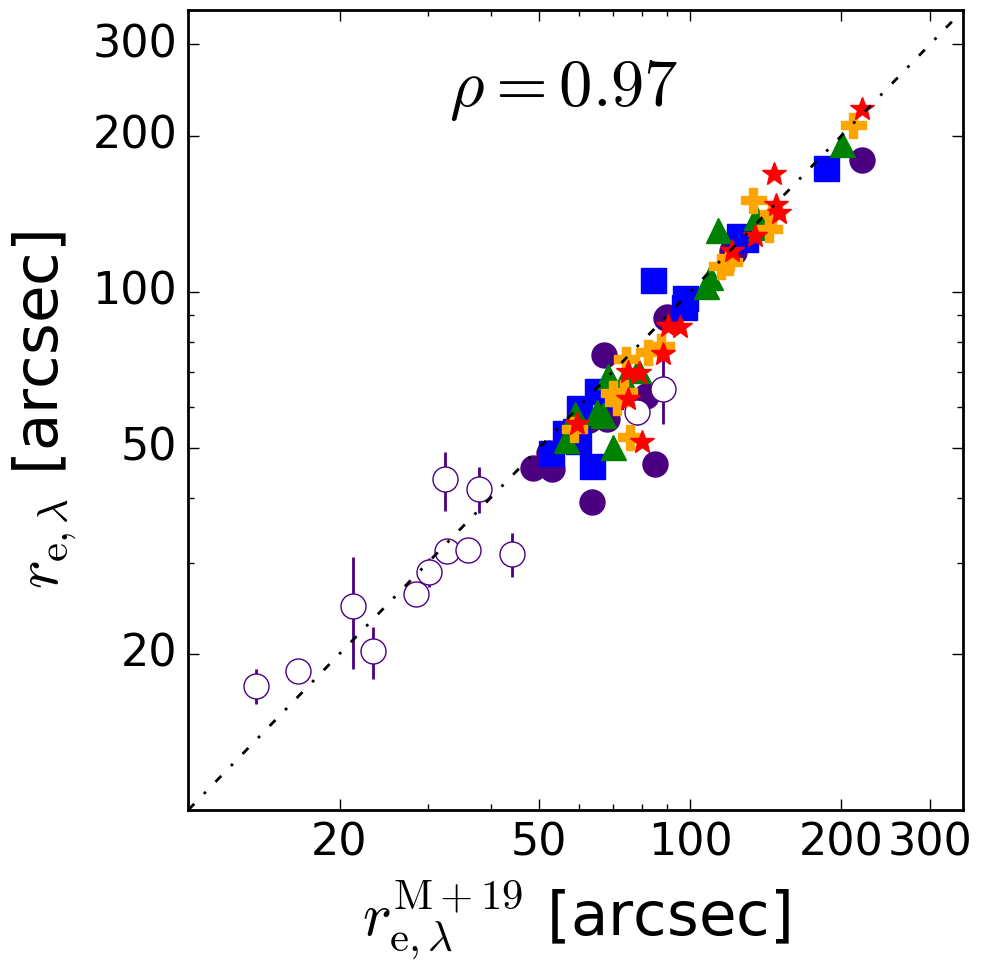}
\includegraphics[height=4.3cm]{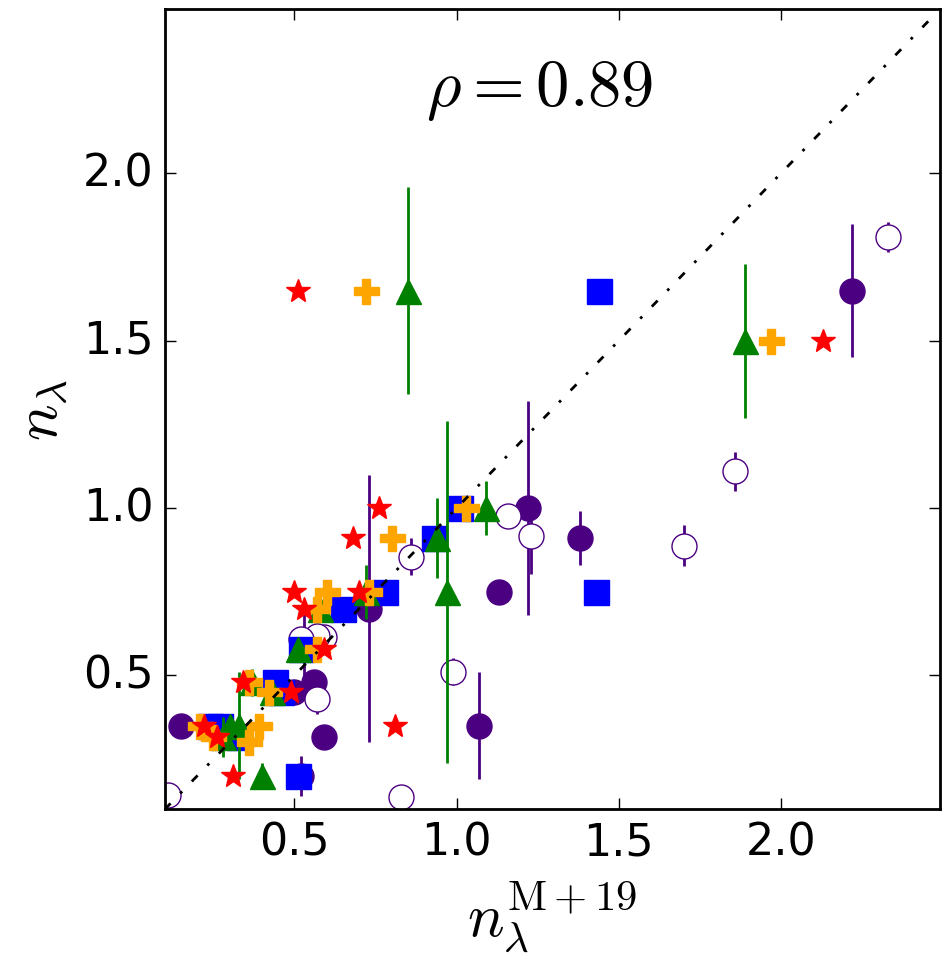}
\includegraphics[height=4.3cm]{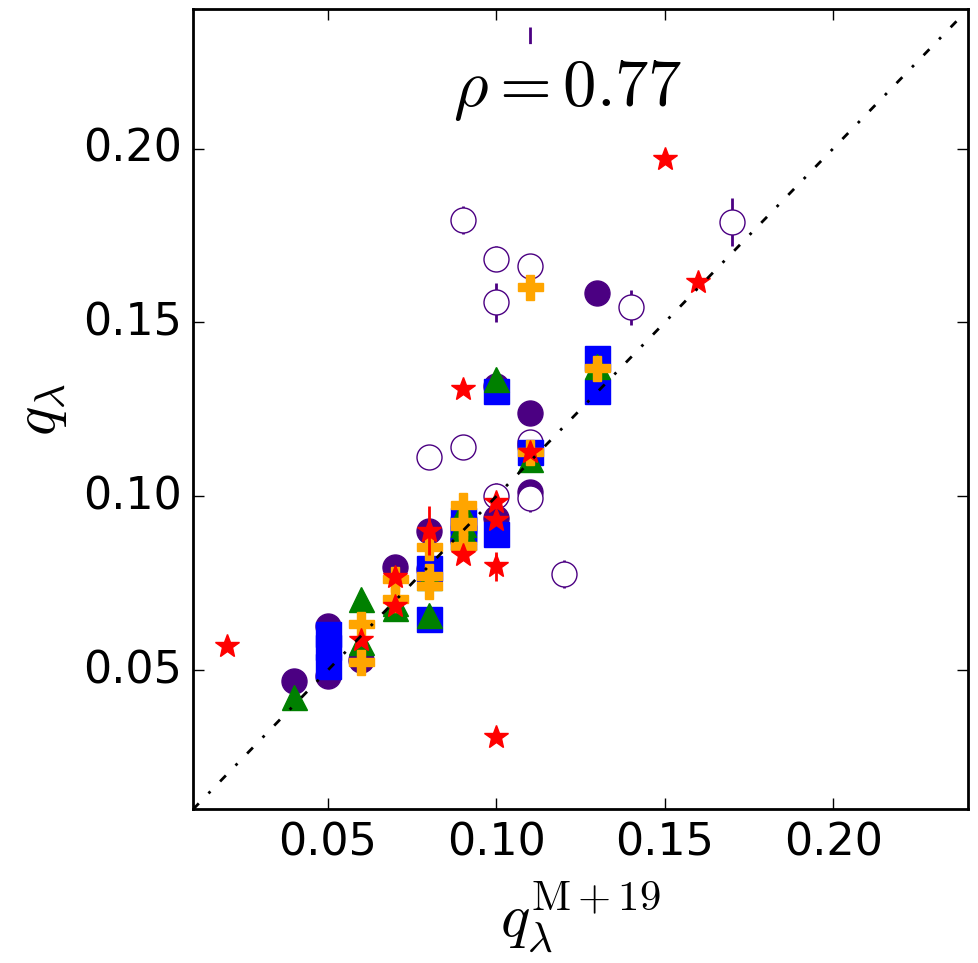}
\caption{Comparison of the best-fit parameters for the models obtained in \citetalias{2019A&A...622A.132M} (X-axis) and those obtained in this work (Y-axis) for the S\'ersic model. The results for the different bands are depicted by different symbols and colours. The dot-dashed lines show one-to-one relations.}
\end{figure*}

In Fig.~\ref{fig:M19_compar}, we compare the results of our S\'ersic fitting with those presented in \citetalias{2019A&A...622A.132M} for the same edge-on galaxies. As one can see, the results agree fairly well. The fluxes are in almost perfect agreement (the Pearson correlation coefficient $\rho=0.99$), with slightly lower values for smaller galaxies as compared to the \citetalias{2019A&A...622A.132M} results. For the effective radius, the agreement is excellent ($\rho=0.97$). The S\'ersic index is the most sensitive parameter in our fitting, but even for it, a strong correlation ($\rho=0.89$) with a few outliers is seen. These outliers have a larger S\'ersic index $n_{\lambda}^\mathrm{M+19}$ in \citetalias{2019A&A...622A.132M} because in the present fitting, the models of these galaxies consist of several components (S\'ersic dust disc + dust clumps, a bright point-like source at the centre, a bulge, or a bar etc. --- see Sect.~\ref{sec:profiles}) instead of a single S\'ersic profile in \citetalias{2019A&A...622A.132M}. As to the apparent flattening, most galaxies have consistent values in both sets of S\'ersic models, with several outliers being preferentially among galaxies from the Additional sample. This discrepancy is likely related to the poorer spatial resolution of these galaxies: their vertical profiles are essentially described by the PSF (see Sect.~\ref{sec:notes}). Overall, we conclude that our fits generally agree with those in \citetalias{2019A&A...622A.132M}.

\begin{figure*}
\label{fig:dust_disk_compar}
\centering
\includegraphics[height=3.8cm]{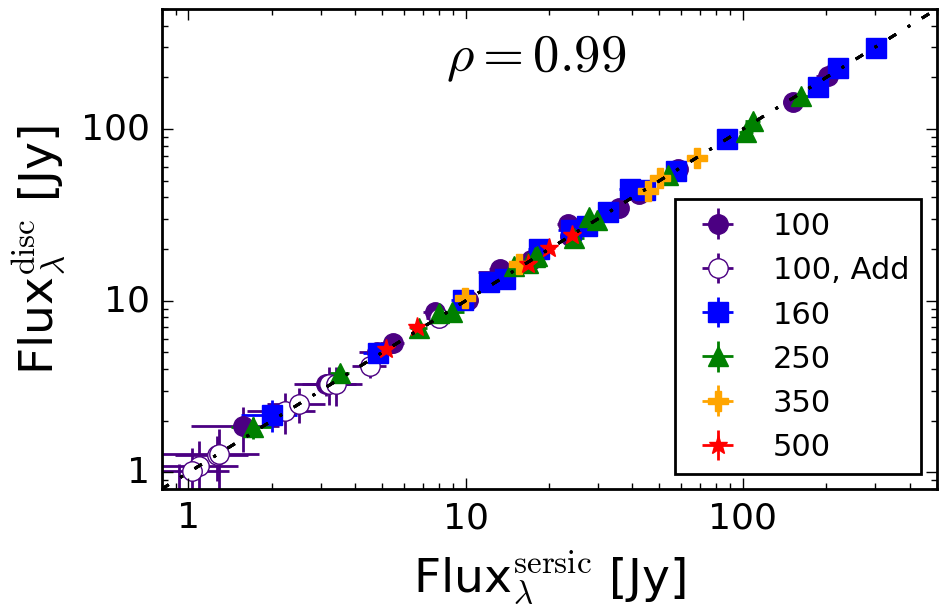}
\includegraphics[height=3.8cm]{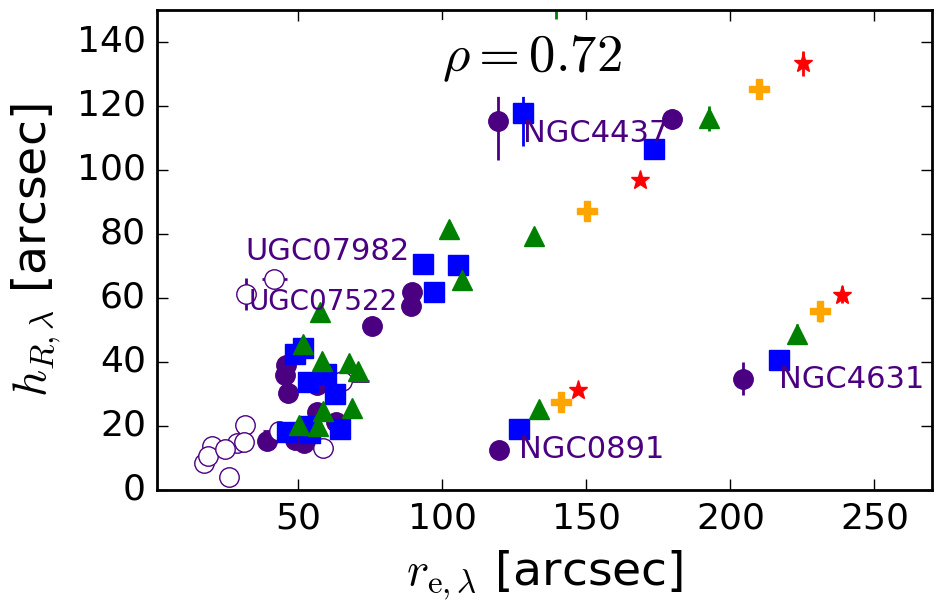}
\includegraphics[height=3.8cm]{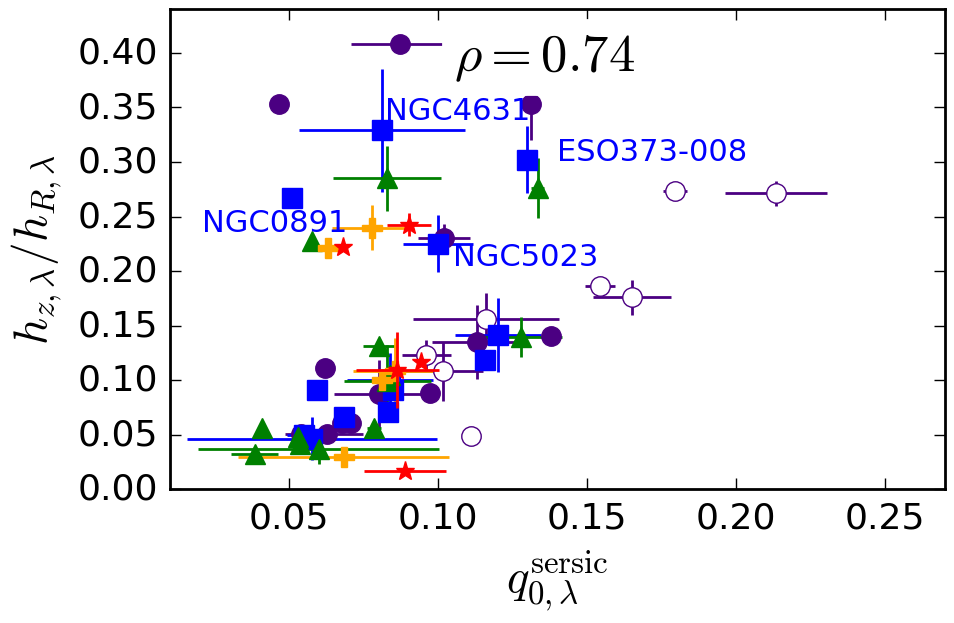}
\caption{{\it Left:} Comparison of the 3D disc and S\'ersic model fluxes within apertures with the semi-major and -minor axes from \citet{2018A&A...609A..37C}. {\it Middle:} Comparison of the 3D disc scale length and the effective radius of the S\'ersic model. {\it Right:} Comparison of the disc scale length-to-scale height ratio versus the intrinsic flattening for the S\'ersic model. The dot-dashed line shows a one-to-one relation.}
\end{figure*}

In Fig.~\ref{fig:dust_disk_compar}, we validate our fitting results for the 3D disc models as compared to the S\'ersic models. In the {\it left} panel we compare the modelled fluxes within the apertures provided in \citet{2018A&A...609A..37C}. As one can see, all data points follow a one-to-one relation. In the {\it middle} plot of Fig.~\ref{fig:dust_disk_compar}, we display how the disc scale length and the effective radius of the S\'ersic model (obtained in this study) agree with each other. Despite five outliers, a tight correlation between these parameters is apparent ($\rho=0.72$). NGC\,0891, NGC\,4437 and NGC\,4631 are well-resolved nearby galaxies with a complex structure for which a broken exponential profile with two radial scale lengths is more accurate than a simple S\'ersic function. UGC\,07522 and UGC\,07982 are, on the contrary, from the Additional sample and have rather large disc scale lengths with respect to their effective radii. Finally, in the {\it right} panel of Fig.~\ref{fig:dust_disk_compar}, we show the correlation between the intrinsic flattening for the 3D disc and S\'ersic models. Again, NGC\,0891 and NGC\,4631 deviate from the main trend ($\rho=0.74$), probably due to the presence of a thick dust disc and an unresolved superthin dust disc in NGC\,0891 \citep{2016A&A...586A...8B}. In general, we can conclude that both models agree fairly well with few exceptions.

\begin{figure*}
\label{fig:stellar_disk_compar}
\centering
\includegraphics[height=3.8cm]{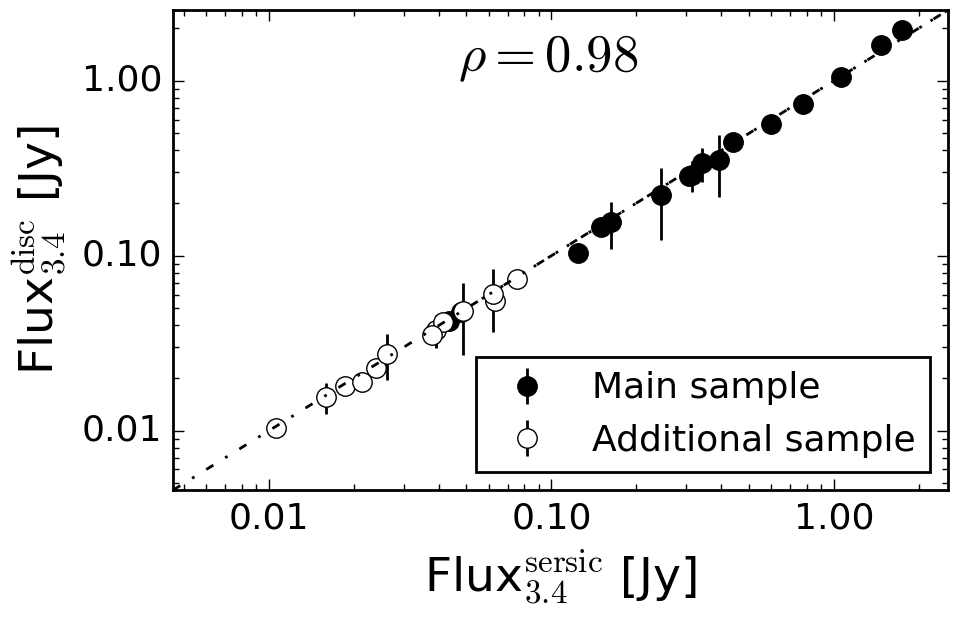}
\includegraphics[height=3.8cm]{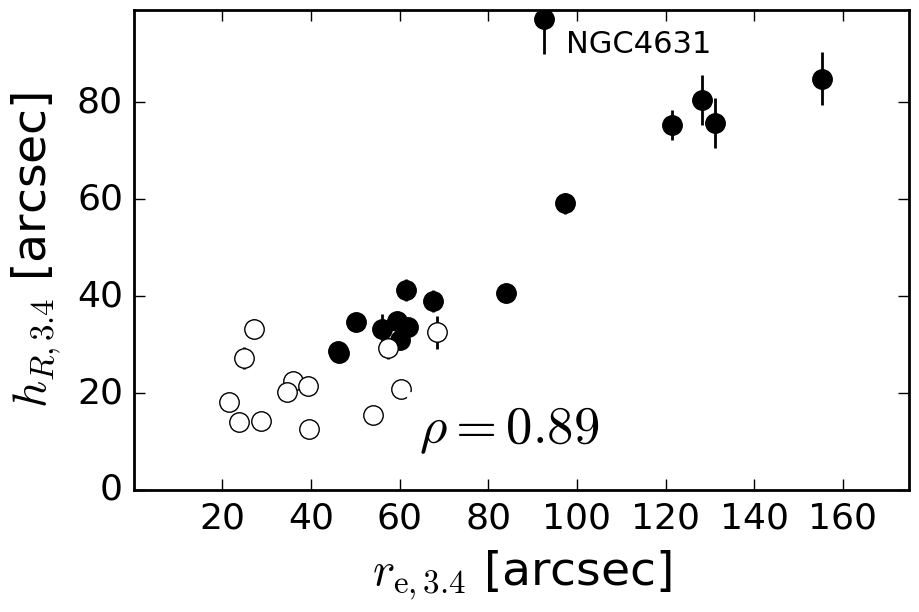}
\includegraphics[height=3.8cm]{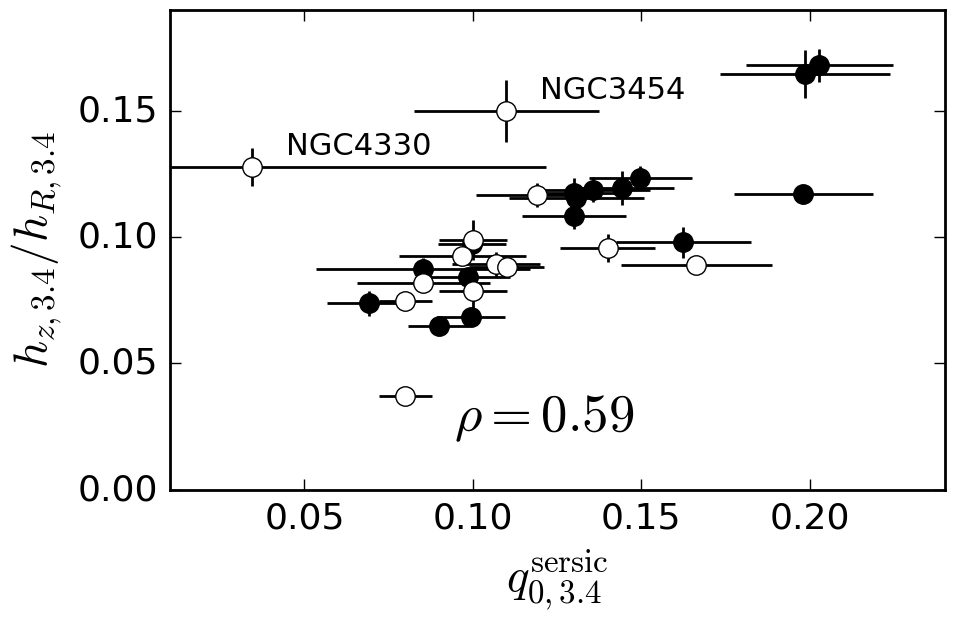}
\caption{Comparison of the 3D disc model obtained in this study with the S\'ersic model from \citetalias{2019A&A...622A.132M} for 3.4~$\mu$m data: the model fluxes ({\it left}), the disc scale length versus the S\'ersic effective radius ({\it middle}), and the disc relative thickness versus the intrinsic flattening for the S\'ersic model ({\it right}). The dot-dashed line shows a one-to-one relation.}
\end{figure*}

Fig.~\ref{fig:stellar_disk_compar} presents the comparison of our 3D disc models and S\'ersic models from \citetalias{2019A&A...622A.132M} for the WISE 3.4~$\mu$m data. As can be seen, the fluxes and geometrical parameters of the models are in good agreement, except for several outliers. Both NGC\,3454 and NGC\,4330 have puffed-up stellar discs with boxy isophotes which may explain why these galaxies do not follow the general trend between the intrinsic disc and S\'ersic flattenings. Also, 10 galaxies from our sample are in common with a sample of 175 edge-on galaxies from \citet{2010MNRAS.401..559M} who decomposed them into a bulge and disc using the Two Micron All Sky Survey (2MASS, \citealt{2006AJ....131.1163S}) $JHK_{s}$-band images. Our comparison with their study shows an excellent agreement for the disc scale length and scale height, with $\rho=0.99$ and $\rho=0.98$, respectively.

\subsection{Cumulative intensity profiles}
\label{sec:profiles}

In this subsection we describe cumulative radial and vertical intensity profiles with the 3D disc model shown in Fig.~\ref{fig:profiles} and in the online material. 

\begin{figure}
\label{fig:distr_rbr}
\centering
\includegraphics[width=8cm]{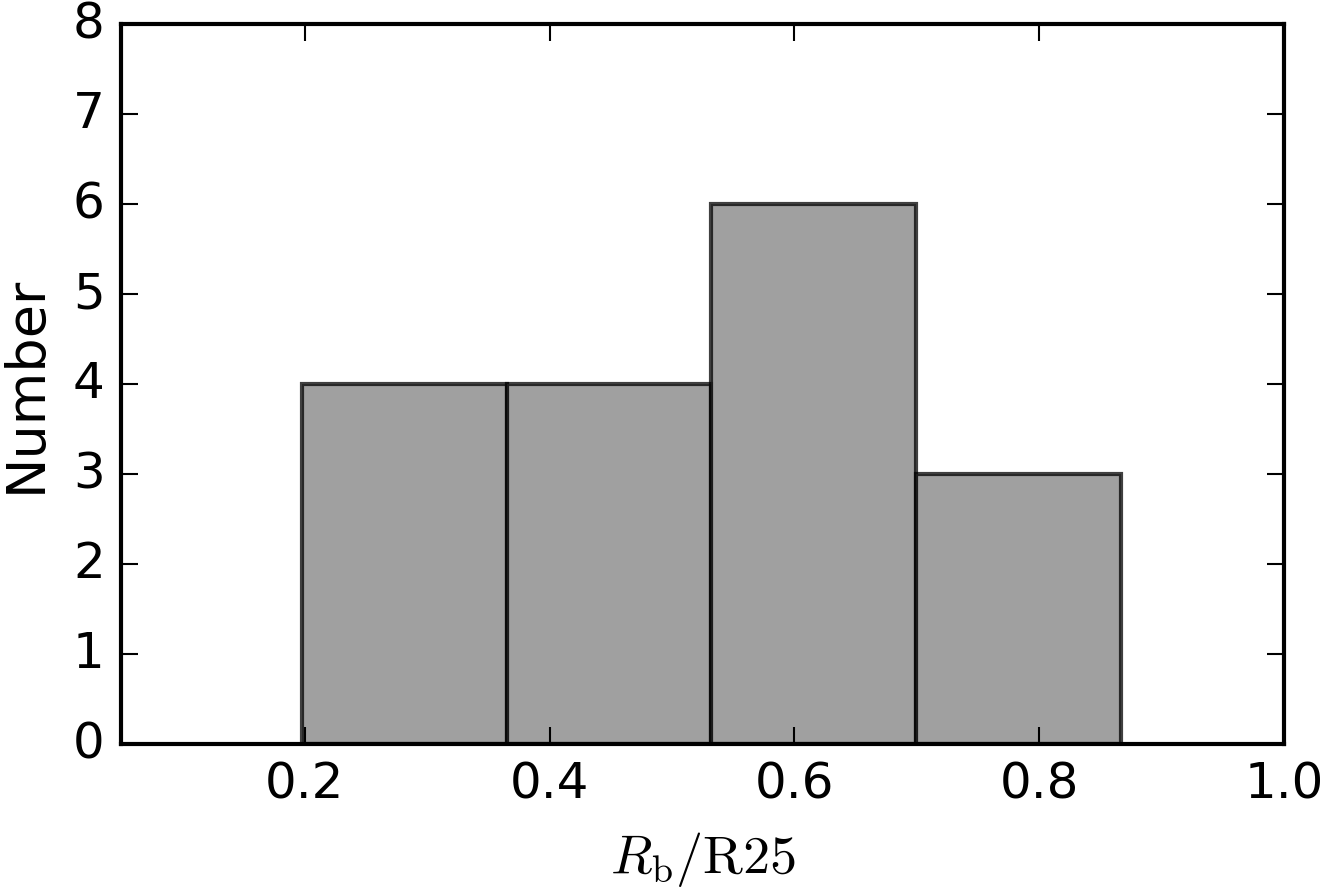}
\caption{Distribution by the break radius normalised by the optical radius R25=D25/2 (listed in Table~\ref{tab:sample}) for 17 galaxies in the Main and Additional samples which were fitted with a broken exponential disc model.}
\end{figure}

Seventeen galaxies out of 29 demonstrate a noticeable break in their radial profiles which we included in our broken exponential disc models for these galaxies. \citetalias{2019A&A...622A.132M} studied the distribution of such characteristic (break) radii for a large sample of DustPedia galaxies and found a bimodal distribution with one peak centred at $\sim$(0.3--0.4)$\,$\,R25 and another at $\sim0.6$\,R25, where R25 is half of the optical diameter D25. They attributed the inner (shorter) break radius to the region where the dust emission deficit is observed and the outer (longer) break radius to the break on the corresponding downbending (Type II, see e.g., \citealt{2008AJ....135...20E}) stellar profile. The number of galaxies in our sample is too low to study the statistics of break radii: they uniformly span from $\sim0.2$ to $\sim0.8\, \mathrm{R25}$ (see Fig.~\ref{fig:distr_rbr}). However, as shown in \citetalias{2019A&A...622A.132M}, the shortest break radii denote the region where the dust emission deficit should be observed (depression in the dust emission profile with a very large inner scale length $h_\mathrm{R,in}$ ---  e.g., for IC\,2531, NGC\,4437, NGC\,5023). This FIR emission deficit reflects the depression in the dust-mass density profile which, in its turn, generally follows the distribution of total (H{\sc i} and H$_2$) gas (see fig.~A2 in \citealt{2017A&A...605A..18C} where they compare the radial distribution of dust, stars, and gas in 18 face-on spiral galaxies). FIR surface brightness profiles without a depression in the centre are mostly found in galaxies with enhanced star formation in the central region which can be related to bar instability, formation of a pseudobulge, or AGN activity (see sect. 5.4.4 in \citetalias{2019A&A...622A.132M}). By comparing the profiles of the same galaxy in different {\it Herschel} bands, we can notice that this enhanced emission peak is usually very bright at 100~$\mu$m (where the contribution of the emission from stochastically heated warm dust is already small but not negligible) and gets lower and even disappears at longer wavelengths --- see, for example, the radial profiles of NGC\,3628, NGC\,4013, and NGC\,4631 in different FIR/submm wavebands. This dimming of the central source with increased wavelength stems from the fact that the contribution of the warm dust emission to the total emission decreases with wavelength and, thus, only the global cold dust disc is essentially traced in the SPIRE wavebands.

The longer break radii beyond $\sim0.5\,$R25  (e.g in NGC\,4217, NGC\,4631, NGC\,5907) characterise a smooth transition between the inner profile with a larger scale length and the outer region with a smaller scale length (all broken exponential disc models in our sample have downbending profiles). As strongly supported by \citetalias{2019A&A...622A.132M} (see their fig.~13), the dust and stellar radial profiles correlate with each other very well and are both observed in the radial range $\sim(0.4-0.8)\,$\,R25. For our sample, we also compared the radial profiles at 100~$\mu$m and 3.4~$\mu$m and found a good consistency between them, except for the cases where a central depression of dust emission is observed.

Among 16 galaxies in the Main sample, 10 galaxies have additional components apart from the dust disc (bright clumps, bright central point sources, bulges, and bars). For example, NGC\,0891 has a bright Gaussian source at the centre that is probably related to the bulge or a bar. NGC\,3628 is modelled with a bright point source at the centre and a S\'ersic bar. The NGC\,4244 model has five Gaussians to fit the brightest clumps of dust emission. NGC\,4631 hosts a bright S\'ersic bar, and NGC\,4437 has 9 point-like sources which describe bright emission clumps along the galaxy mid-plane. All these features are seen in our cumulative profiles. Galaxies in the Additional sample exhibit smoother profiles due to the worse spatial resolution and thus are only fitted with a single disc component.

The vertical profiles for 6 galaxies (ESO\,209-009, NGC\,0891, NGC\,3628, NGC\,4013, NGC\,4437, and NGC\,4631) show the presence of another (thicker) dust emission component in all or almost all wavebands. This component clearly dominates over the model at larger vertical distances above the galaxy mid-plane, usually starting at $z=1-2$~kpc. For these galaxies, the average $\chi^2=19.8\pm10.1$ compared to $\chi^2=1.04\pm0.45$ for the whole sample without these outliers. This proves that another dust component is required to better describe the observed profiles of these six galaxies. Four galaxies from the Main sample (NGC\,4244, ESO\,373-008, NGC\,5529, and NGC\,5746) show bumps in their cumulative vertical profiles above the model profile which can point to the existence of another vertically extended component. Six galaxies from the Additional sample (IC\,2233, NGC\,3592, UGC\,07321, UGC\,07387, UGC\,07522, and UGC\,09242) also show an excess of FIR emission compared to our models at large vertical distances. We note that this extraplanar FIR emission, which typically starts to dominate at heights 1--2~kpc above the mid-plane, cannot be explained by the scattered light of the PSF since we accounted for this effect while modelling. Also, variations in the galaxy inclination angle as described in Sect.~\ref{sec:method} cannot explain the observed excess of dust emission (the red spread along the model profile in Fig.~\ref{fig:profiles} shows a one-$\sigma$ spread of the model profiles for varied inclination angles, as described in Sect.~\ref{sec:3Ddisc}). This indicates that the observed extraplanar emission is real and produced by the dust at high distances above the mid-plane.

\subsection{Dependence of the structural parameters on wavelength}
\label{sec:pars_wave}

\begin{figure*}
\label{fig:pars_lambda}
\centering
\includegraphics[height=7cm]{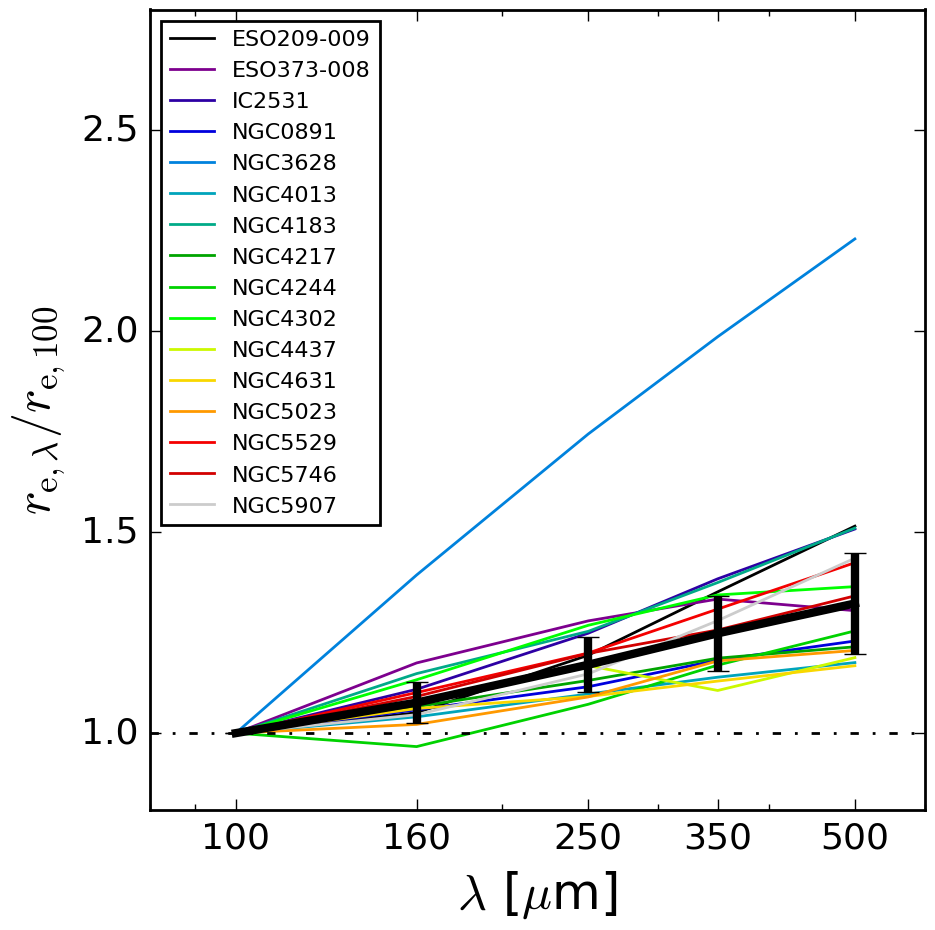}
\includegraphics[height=7cm]{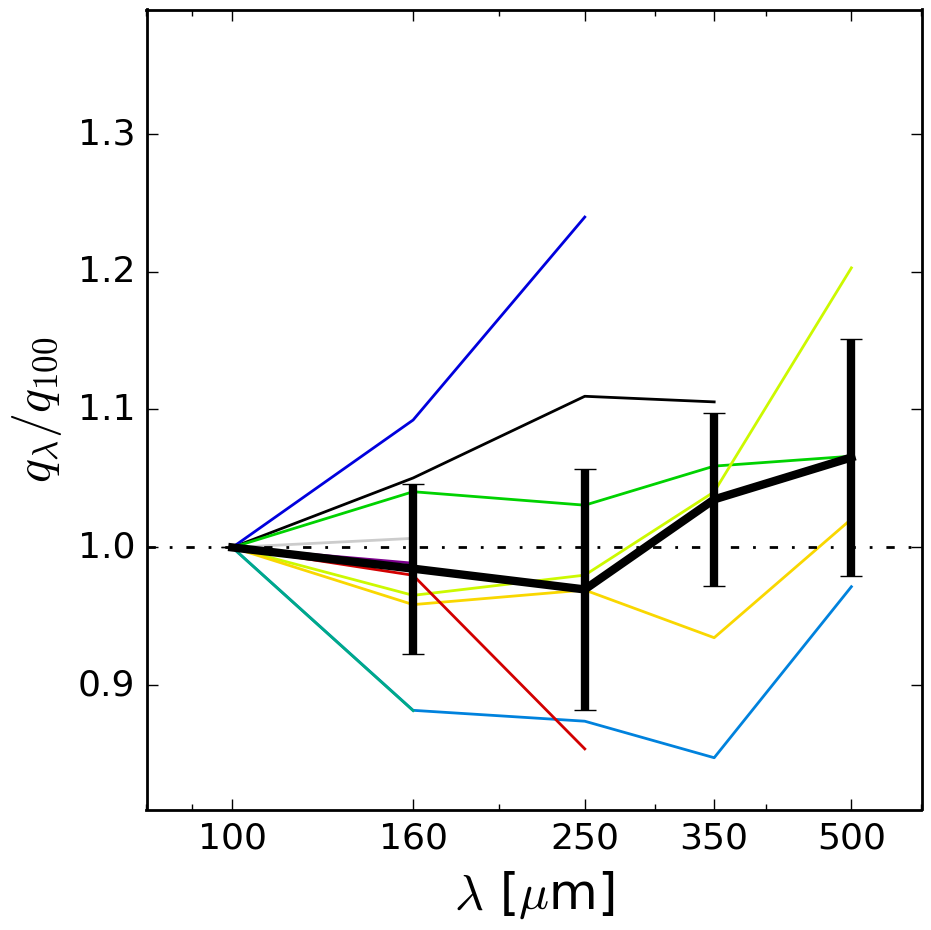}
\includegraphics[height=7cm]{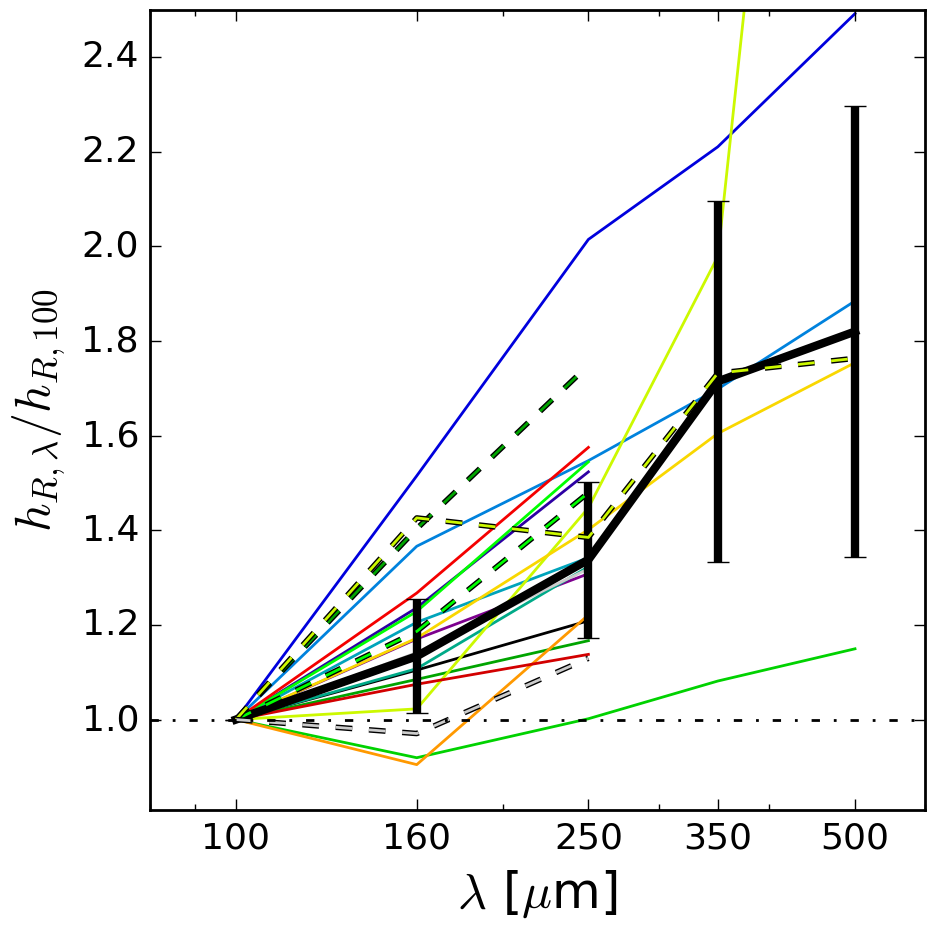}
\includegraphics[height=7cm]{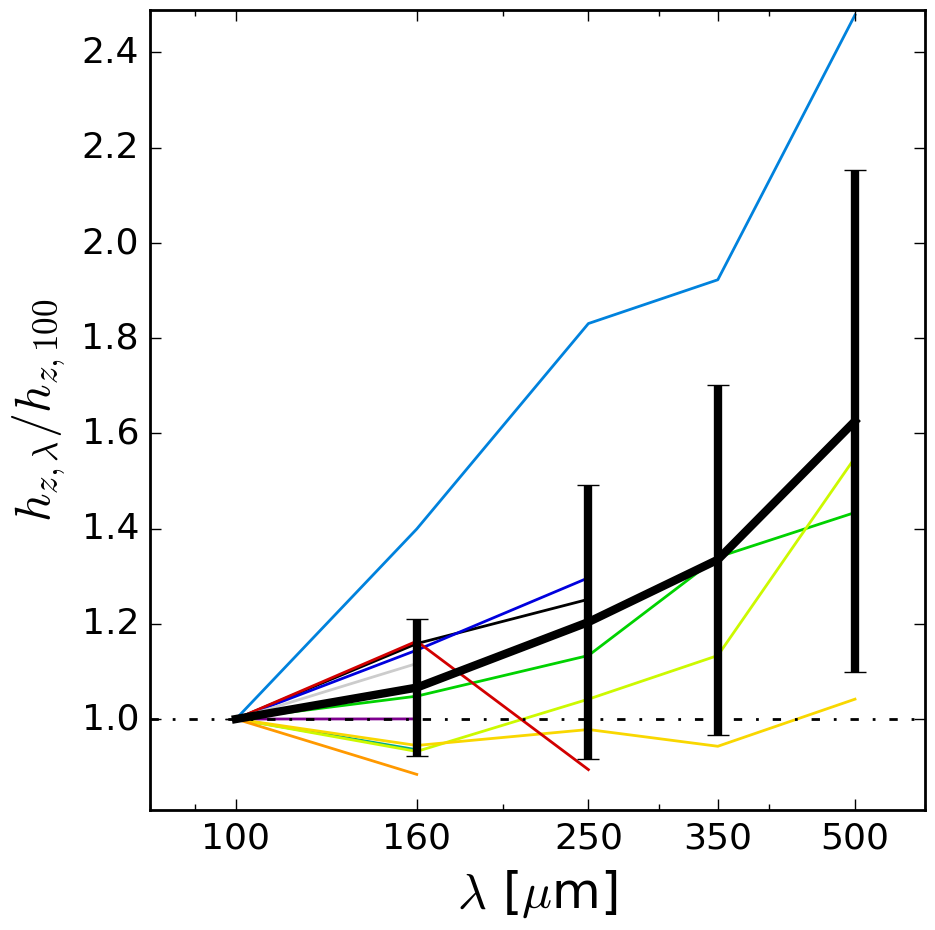}
\caption{Dependence of the S\'ersic and 3D disc model parameters on wavelength for each galaxy (coloured lines) and an average among all these galaxies for each band (black thick line, with error bars representing standard deviations). The dashed coloured lines correspond to the outer scale length $h_\mathrm{out}$ for four galaxies (NGC\,4217, NGC\,4302, NGC\,4437, and NGC\,5907) where no deficit in the emission profile is observed and $h_\mathrm{in}$ is comparable to $h_\mathrm{out}$.}
\end{figure*}

In Fig.~\ref{fig:pars_lambda}, we showcase how the structural parameters change with wavelength for both the S\'ersic and 3D disc models (see also Table~\ref{tab:Pars_wave} where we provide a quantitative description of the gradients). As one can see, the effective radius, on average, increases up to 30\% from PACS\,100 to SPIRE\,500 which is lower (probably because of the edge-on orientation of the galaxies in this study) than what was found in \citetalias{2019A&A...622A.132M} for the whole DustPedia sample: $r_\mathrm{e,160}/r_\mathrm{e,100}=1.10\pm0.27$, $r_\mathrm{e,250}/r_\mathrm{e,100}=1.23\pm0.29$, $r_\mathrm{e,350}/r_\mathrm{e,100}=1.38\pm0.34$, $r_\mathrm{e,500}/r_\mathrm{e,100}=1.57\pm0.41$. NGC\,3628 is an obvious outlier and shows an increase of the effective radius by a factor of 2.2 which is probably related to the peculiar structure of its dust (and stellar) disc perturbed by the interaction with a neighbour. 

A similar positive gradient is found for the disc scale length in Fig.~\ref{fig:pars_lambda}, {\it bottom left} panel. NGC\,891, NGC\,3628, and NGC\,4437, some of the best spatially resolved galaxies in our sample, exhibit a sharp increase of the disc scale length ($h_{R,500}/h_{R,100}\gtrsim1.8-2$) with wavelength. The dashed lines depict the change of the scale length for the outer part of the disc ($h_\mathrm{out}$) after the break. As one can see, the outer disc scale length also increases in the same way as the inner scale length. The agreement with the literature on the radial scale length is also good. For 18 face-on spiral galaxies, \citet{2017A&A...605A..18C} found the following ratios of the scale lengths at different wavelengths (we computed them using their table~7): $h_{R,160}/h_{R,100}=1.10\pm0.01$, $h_{R,250}/h_{R,100}=1.30\pm0.03$, $h_{R,350}/h_{R,100}=1.45\pm0.05$, and $h_{R,500}/h_{R,100}=1.60\pm0.01$. \citet{2016MNRAS.462..331S} found $h_{R,500}/h_{R,250}=1.07$ (see their table~1) for a combined radial profile of 45 large spiral galaxies versus 1.08 in this study.

As discussed in \citet{1998A&A...335..807A}, \citet{2017A&A...605A..18C}, and \citetalias{2019A&A...622A.132M}, the observed steady increase of the effective radius and the scale length with wavelength in the FIR domain is related to the dust heating (i.e., the gradient of the cold-dust temperature) by the diffuse ISRF which is gradually decreasing with radius. Also, the dust-mass surface density distribution profiles are flatter (i.e. have a larger scale length) than the stellar-mass density distribution (see table~7 in \citealt{2017A&A...605A..18C}). This effect was also demonstrated by \citet{2008A&A...490..461B} using a RT model of NGC\,0891 (see his fig.~10).

As shown in the {\it upper right} panel of Fig.~\ref{fig:pars_lambda}, the apparent flattening of the emission profiles does not change significantly with wavelength (typically, within 10\%, i.e. within the uncertainty of this parameter), although the statistics here are rather poor, especially at larger wavelengths. This suggests that the scale height of the dust emission profile should increase with wavelength, similar to what we observe for the radial extension. This is confirmed by the dependence of the disc scale height on wavelength from our 3D disc modelling (see the {\it bottom right} panel of Fig.~\ref{fig:pars_lambda}). Although the number of galaxies with the resolved vertical emission profile at larger FIR wavelengths is low, most galaxies in PACS\,160 and SPIRE\,250 clearly demonstrate a gradual increase of the dust disc thickness with wavelength. For the three largest galaxies, the scale height grows by more than 40\% from 100 to 500~$\mu$m. Again, NGC\,3628 with the very disturbed stellar and dust discs is an obvious outlier and demonstrates a substantial increase of the dust disc scale height with wavelength by a factor of 2.5. \citet{2016A&A...586A...8B} studied the vertical profile of NGC\,0891 in the NIR, MIR, and FIR and found the presence of thin and thick dust discs in this galaxy. They revealed that the scale height of the thin disc demonstrates a dramatic gradient through the FIR: it changes by a factor of two from 100 to 250~$\mu$m. The increase of the scale height for our single disc model of NGC\,0891 is less prominent (30\%) but is comparable to the change of the scale height of their thick disc.

The observed gradient of the scale height of the sample galaxies through the FIR could indicate that the dust temperature noticeably drops within its scale height, which naturally leads to an increase of the relative FIR emission with height at longer wavelengths. Another explanation could be the influence of the emission from point-like sources (dust clumps, ends of the spiral arms positioned along the line of sight, rings) which emit at relatively warm temperatures and thus become less pronounced at longer wavelengths (plus the poorer resolution smears them out). This can increase the scale height at larger wavelengths and create the observed gradient. \citet{2016A&A...586A...8B} also noted for NGC\,0891 that contamination from an unresolved super thin disc, which represents the collective emission from dust heated by nearby hot stars in star-forming regions, may also produce this effect.

\begin{table}
  \centering
  \caption{Distributions of the fit parameters (mean value and standard deviation) depending on wavelength.} 
  \label{tab:Pars_wave}
  \begin{tabular}{lcccc}
    \hline \hline\\[-2ex]
    Band & $r_{\mathrm{e},\lambda}/r_{\mathrm{e},100}$ & $q_{\lambda}/q_{100}$ & $h_{R,\lambda}/h_{R,100}$ & $h_{z,\lambda}/h_{z,100}$ \\
    \hline\\[-1.5ex]   
PACS\,160 & $1.08 \pm 0.05$& $0.98 \pm 0.06$& $1.13 \pm 0.12$& $1.07 \pm 0.14$ \\ 
SPIRE\,250 & $1.17 \pm 0.07$& $0.97 \pm 0.09$& $1.34 \pm 0.16$& $1.20 \pm 0.29$ \\ 
SPIRE\,350 & $1.25 \pm 0.09$& $1.03 \pm 0.06$& $1.71 \pm 0.38$& $1.33 \pm 0.37$ \\ 
SPIRE\,500 & $1.32 \pm 0.12$& $1.07 \pm 0.09$& $1.82 \pm 0.48$& $1.63 \pm 0.53$ \\ 
    \hline\\
  \end{tabular}
\end{table}

\section{Scaling relations}
\label{sec:relations}

One of the aims of this paper is exploring how the retrieved parameters of the dust disc correlate with each other, the stellar parameters, and the general galaxy parameters. Below we consider the results only for the PACS\,100 band in which we obtained models for all galaxies from the Main and Additional samples. However, we ensured that the scaling relations presented below are essentially valid in the other {\it Herschel} wavebands.

General properties of galaxies (such as colour, mass, luminosity, size, rotational velocity, velocity dispersion, etc.) are not distributed randomly, but are mutually related and form specific scaling relations (see e.g., \citealt{2010MNRAS.401..559M,2014MNRAS.441.1066M,2015MNRAS.451.2376M} with some caveats on their use and physical meaning). These relations provide important constraints on the proposed scenarios of galaxy formation and evolution: any theory and numerical and cosmological hydrodynamical simulations should be able to reproduce the observed scaling relations (see e.g., \citealt{2016MNRAS.462.1057C}, \citealt{2020MNRAS.494.2823T}, \citealt{2022MNRAS.512.2728C}). Below we present galaxy scaling relations which characterise the connection between the global dust and stellar structure in spiral galaxies. 

\subsection{Dust versus stellar structural parameters}
\label{sec:cor_pars}

\begin{figure*}
\label{fig:dust_vs_stellar_pars}
\centering
\includegraphics[height=6cm]{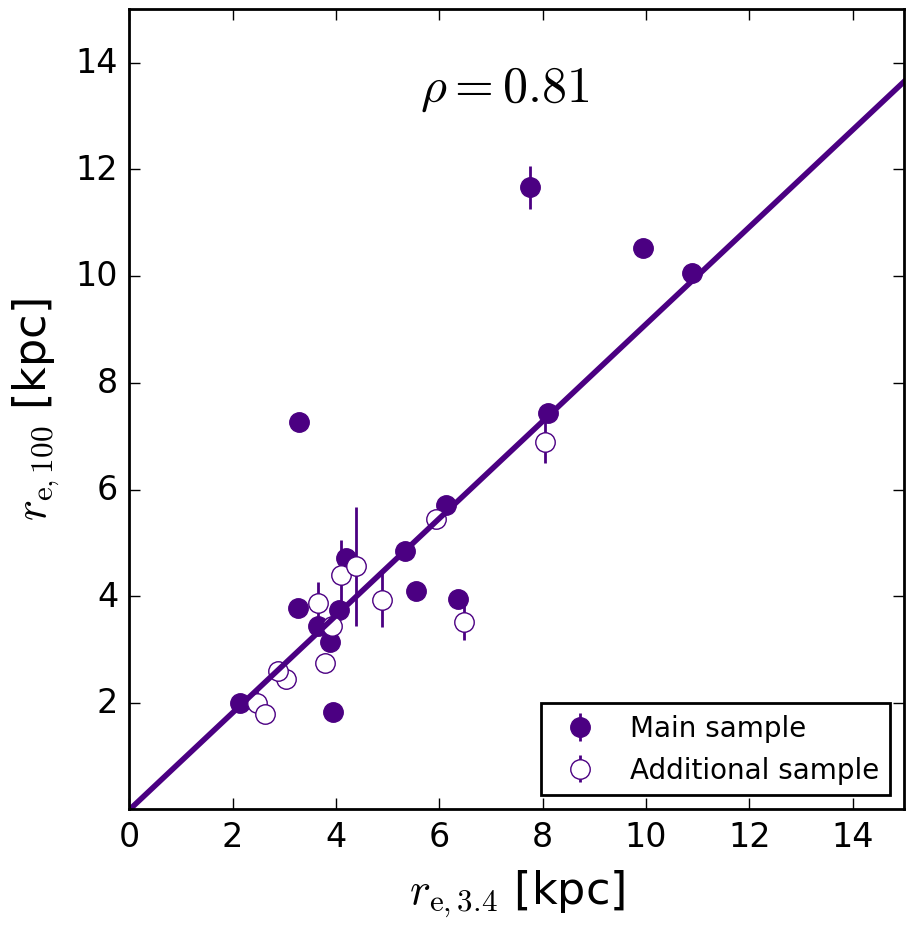}
\includegraphics[height=6cm]{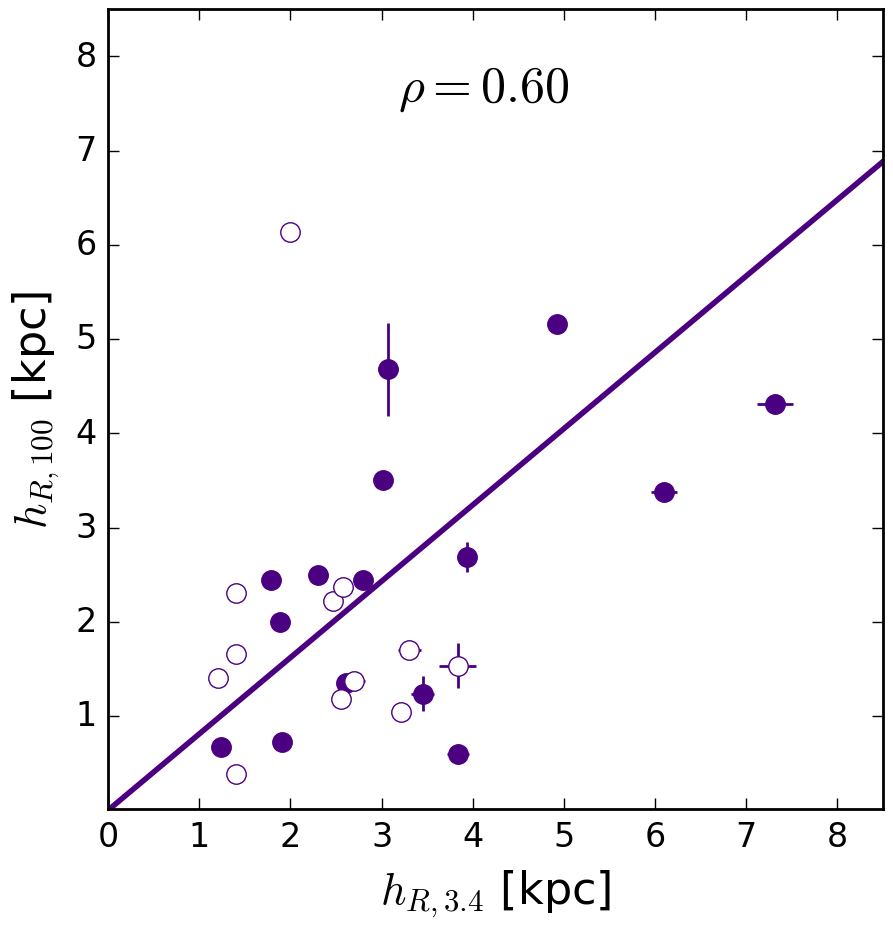}
\includegraphics[height=6cm]{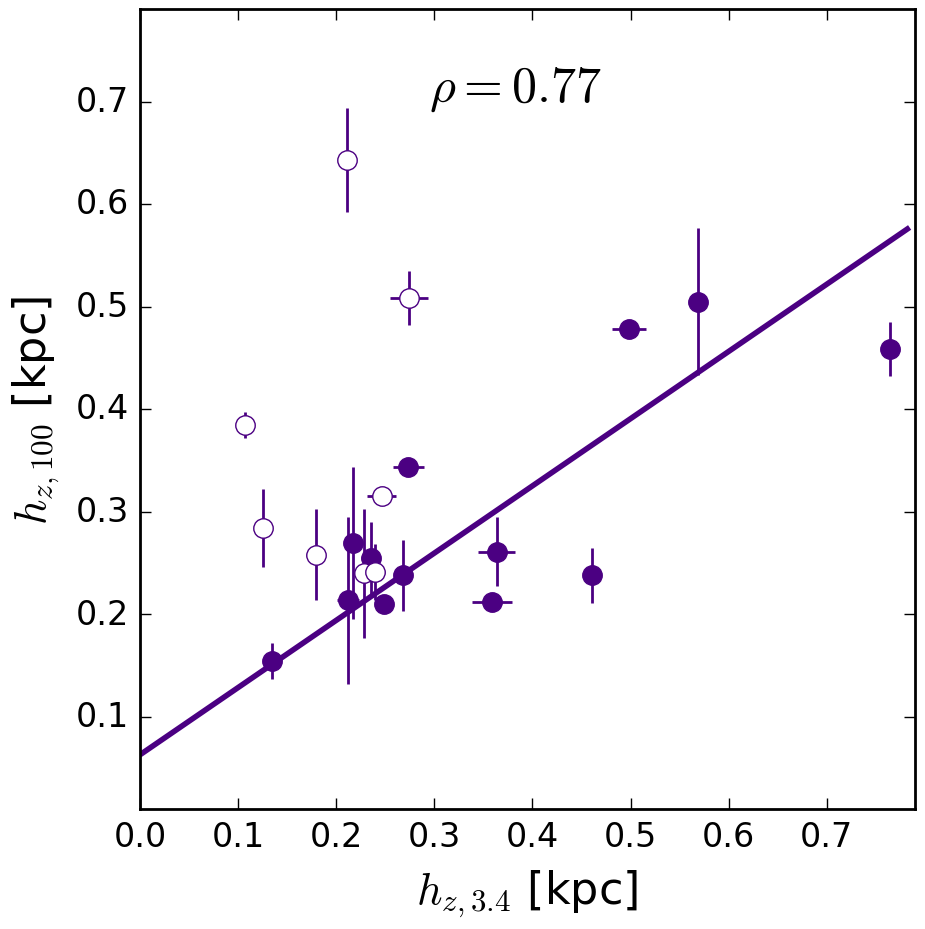}
\caption{Dependencies between the structural parameters at 3.4 and 100~$\mu$m: for the effective radii ({\it left} panel), for the radial scale length ({\it middel} panel), and for the disc scale heights ({\it right} panel). The thick lines depict the corresponding linear regression lines.}
\end{figure*}

In Fig.~\ref{fig:dust_vs_stellar_pars} we show how the dust and stellar radial and vertical scales correlate with each other. We can see that the effective radii of our S\'ersic models at 100 and 3.4~$\mu$m agree very well ($\rho=0.81$) which has been noted in \citetalias{2019A&A...622A.132M} for DustPedia galaxies with arbitrary inclination angles. For our edge-on galaxies, we obtain $\langle r_\mathrm{e,100}/r_\mathrm{e,3.4} \rangle=0.91\pm0.21$ versus $0.90\pm0.27$ from \citetalias{2019A&A...622A.132M}.

Similarly, the disc scale lengths at 3.4 and 100~$\mu$m are correlating with $\rho=0.60$: more extended stellar discs host more extended dust discs. This has been noted earlier by \citet{2017A&A...605A..18C} who found $\langle h_{R,100}/h_{R,3.6} \rangle=0.83\pm0.05$ compared to our 
$\langle h_{R,100}/h_{R,3.4} \rangle=0.81\pm0.31$.

The main ambition of this study is to explore the vertical structure of the dust disc. In Fig.~\ref{fig:dust_vs_stellar_pars}, {\it right} plot, one can see how the scale heights at 3.4 and 100~$\mu$m compare with each other. The relation between the vertical scales appears to be even tighter ($\rho=0.77$) than for the disc scale lengths: the thicker the stellar disc, the thicker the dust disc, on average. We can see that the scale height at 100~$\mu$m is similar to the scale height at 3.4~$\mu$m: $\langle h_{z,100}/h_{z,3.4} \rangle = 0.90 \pm 0.24$. However, there are some obvious outliers, mostly low-mass galaxies from the Additional sample, which do not follow this relation. The reason for this will be seen in Sect.~\ref{sec:mass_size}.

\begin{figure}
\label{fig:dust_vs_stellar_flatness}
\centering
\includegraphics[width=8cm]{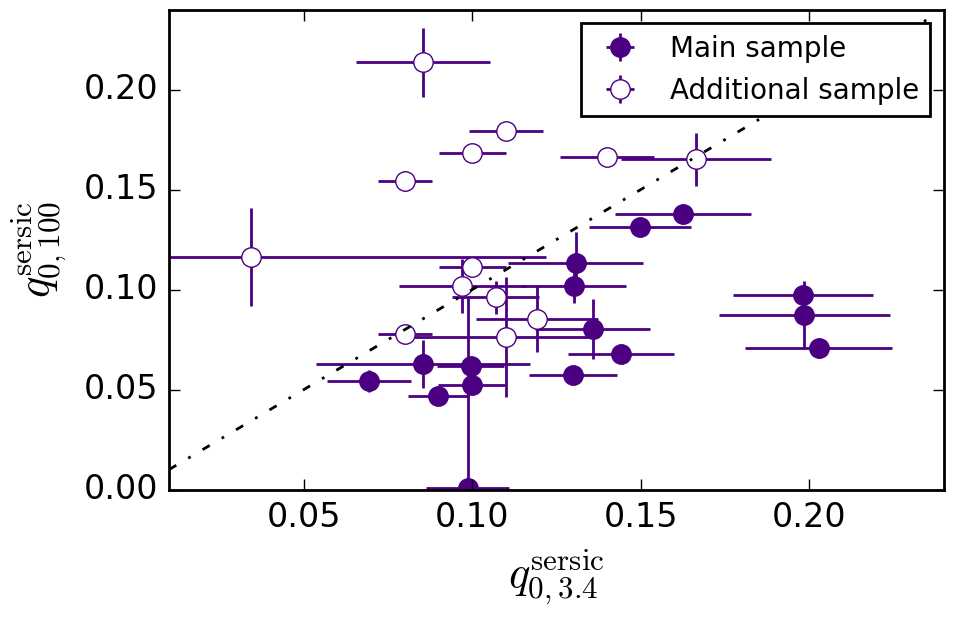}
\includegraphics[width=8cm]{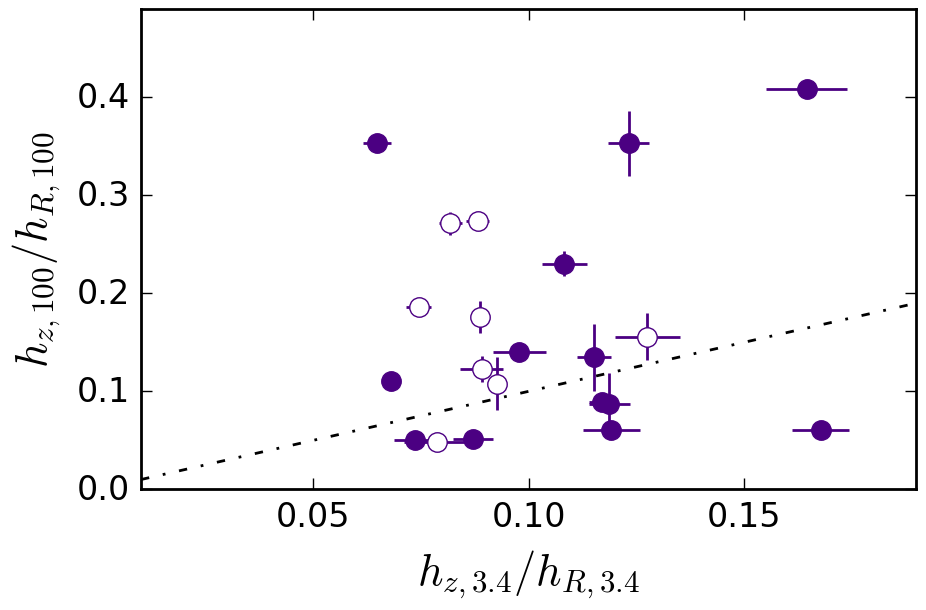}
\caption{Correlations between the intrinsic flattening at 3.4 and 100~$\mu$m for the S\'ersic model ({\it upper}) and 3D disc model ({\it bottom}). The dot-dashed line shows a one-to-one relation.}
\end{figure}

Interestingly, the intrinsic flattening of the dust disc does not correlate with the intrinsic flattening of the stellar disc, based on both the S\'ersic and 3D disc models (see Fig.~\ref{fig:dust_vs_stellar_flatness}). This indicates that the relative thickness of the dust disc can be rather different from the relative thickness of the stellar disc (see also \citealt{2015MNRAS.451.2376M} who considered the dependence of the disc relative thickness on Hubble stage and bulge-to-total luminosity ratio). Also, note that for some dust discs the relative thickness $h_{z,100}/ h_{R,100}$ is exceptionally large and reaches 0.3--0.4 versus $h_{z,3.4}/ h_{R,3.4}=0.10-0.15$ for the stellar disc.

\begin{figure}
\label{fig:length_vs_height}
\centering
\includegraphics[height=8cm]{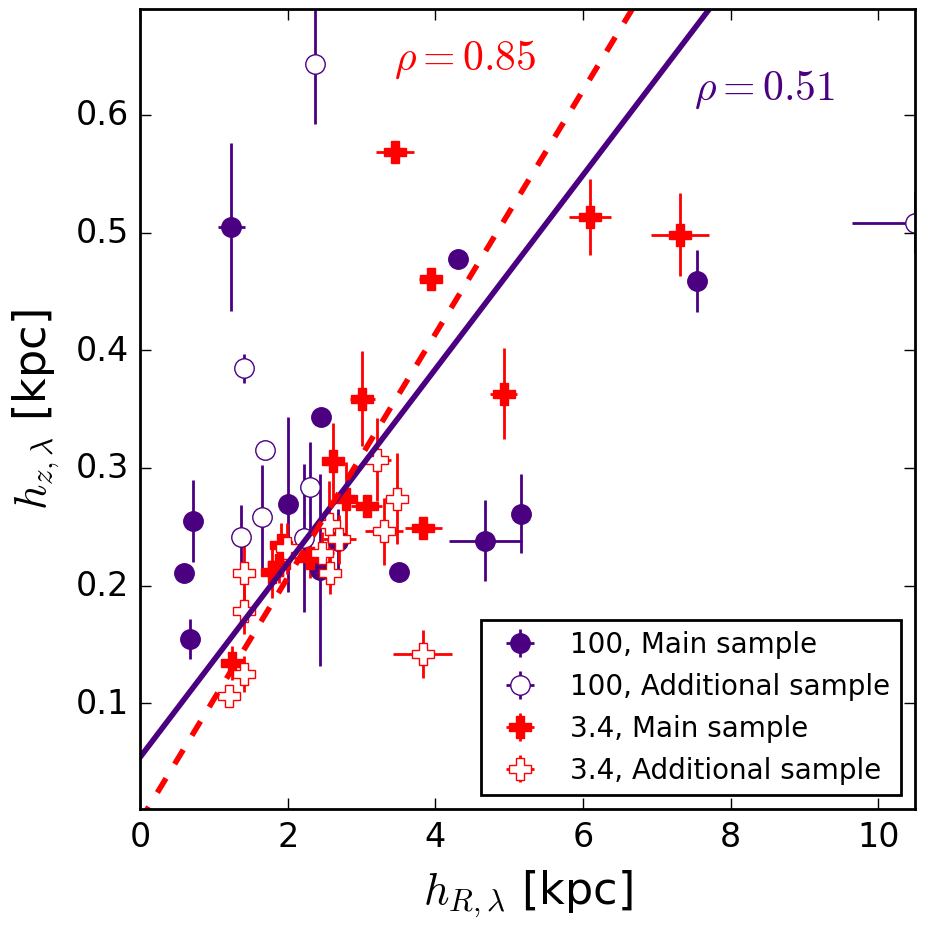}
\caption{Relation between the disc scale length and scale height at both 3.4 and 100~$\mu$m. The thick solid line depicts the linear regression line for the 100~$\mu$m data, whereas the red dashed line corresponds to the linear regression fit for 165 edge-on galaxies in the $K_s$ band from \citet{2010MNRAS.401..559M}.}
\end{figure}

Finally, in Fig.~\ref{fig:length_vs_height} we can see the dependence of the disc scale height on the radial scale length for both the 3.4~$\mu$m and 100~$\mu$m models. This is a typical scaling relation between the geometrical scales of spiral galaxies: the larger the galaxy in the radial direction, the thicker, on average, it should be. This scaling effect is well-seen for the stellar discs in \citet{2010MNRAS.401..559M} (see their fig.~10). Interestingly, both the dust and stellar discs from our sample follow similar relations linking the scale length and scale height ($\rho=0.51$, with slope $k=0.098$ and $\rho=0.85$, with slope $k=0.103$, respectively).

\subsection{Dust mass --- Size relation}
\label{sec:mass_size}

\begin{figure*}
\label{fig:dust_mass_vs_par}
\centering
\includegraphics[height=5.3cm]{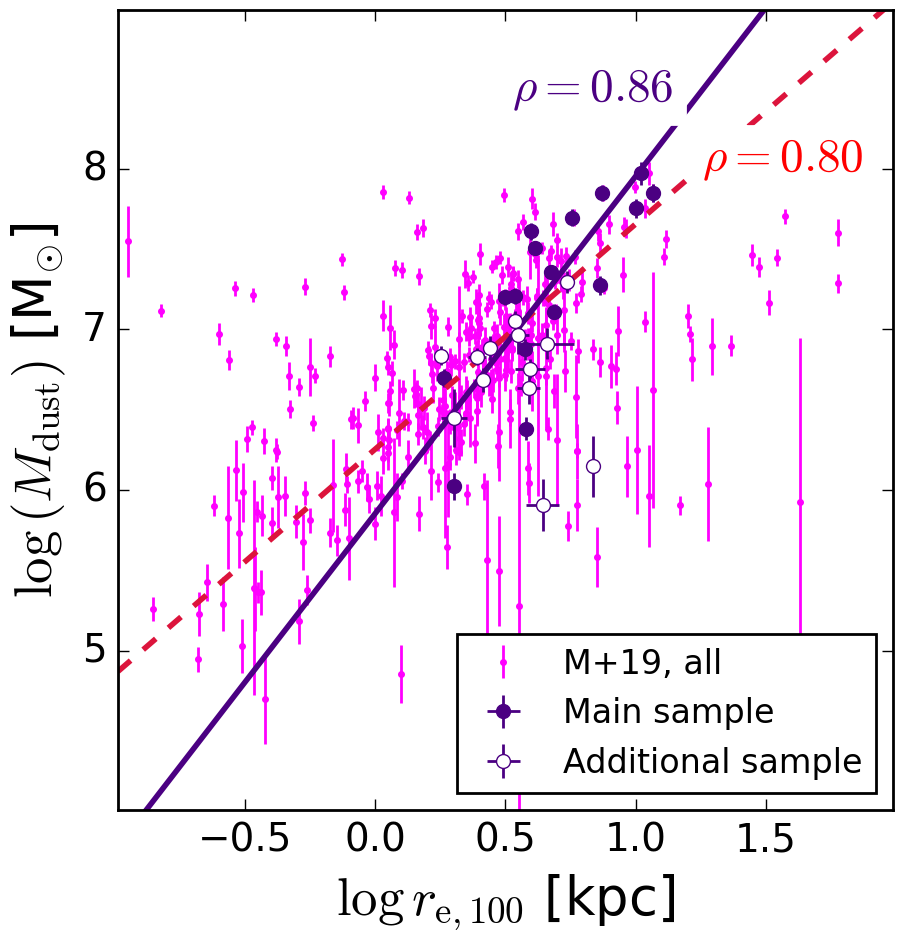}
\includegraphics[height=5.9cm]{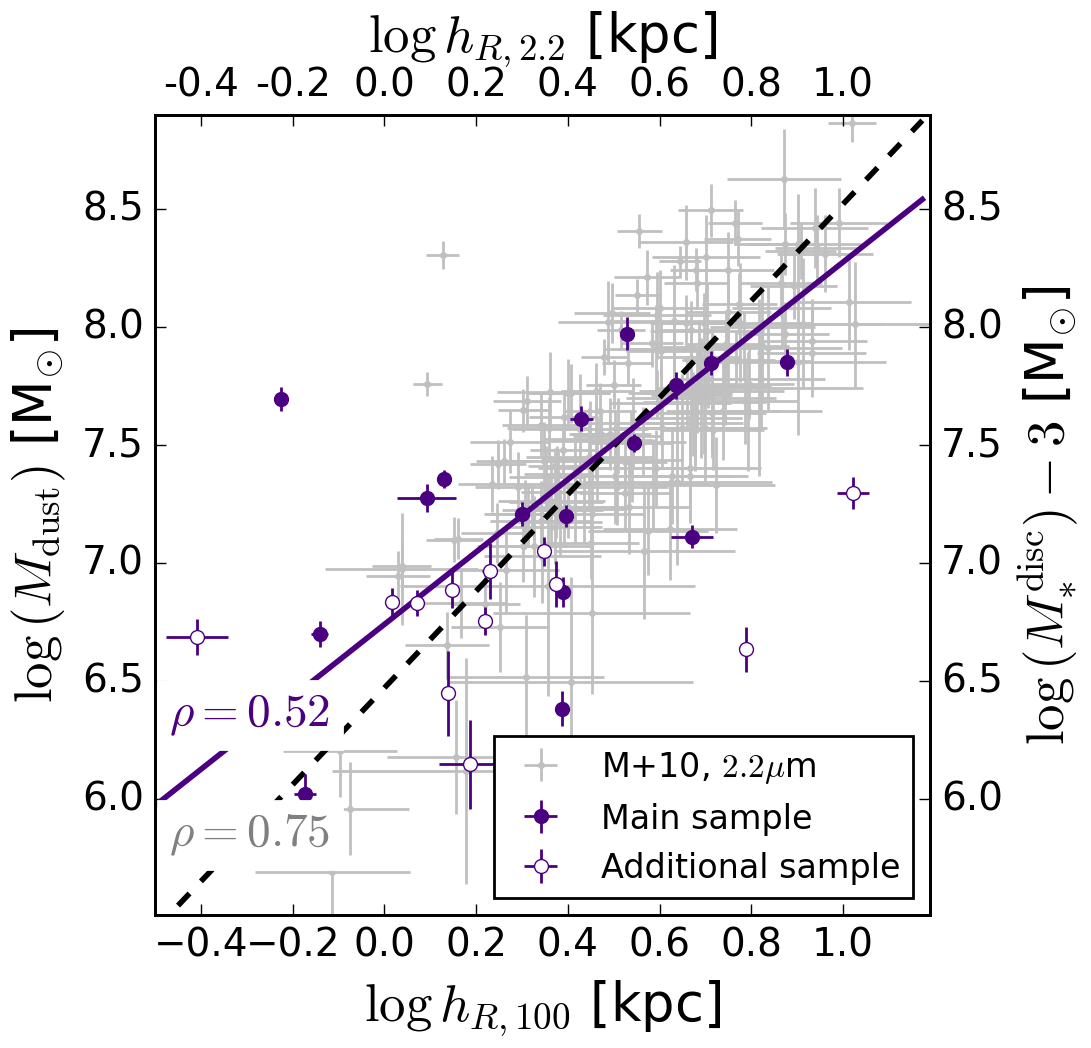}
\includegraphics[height=5.9cm]{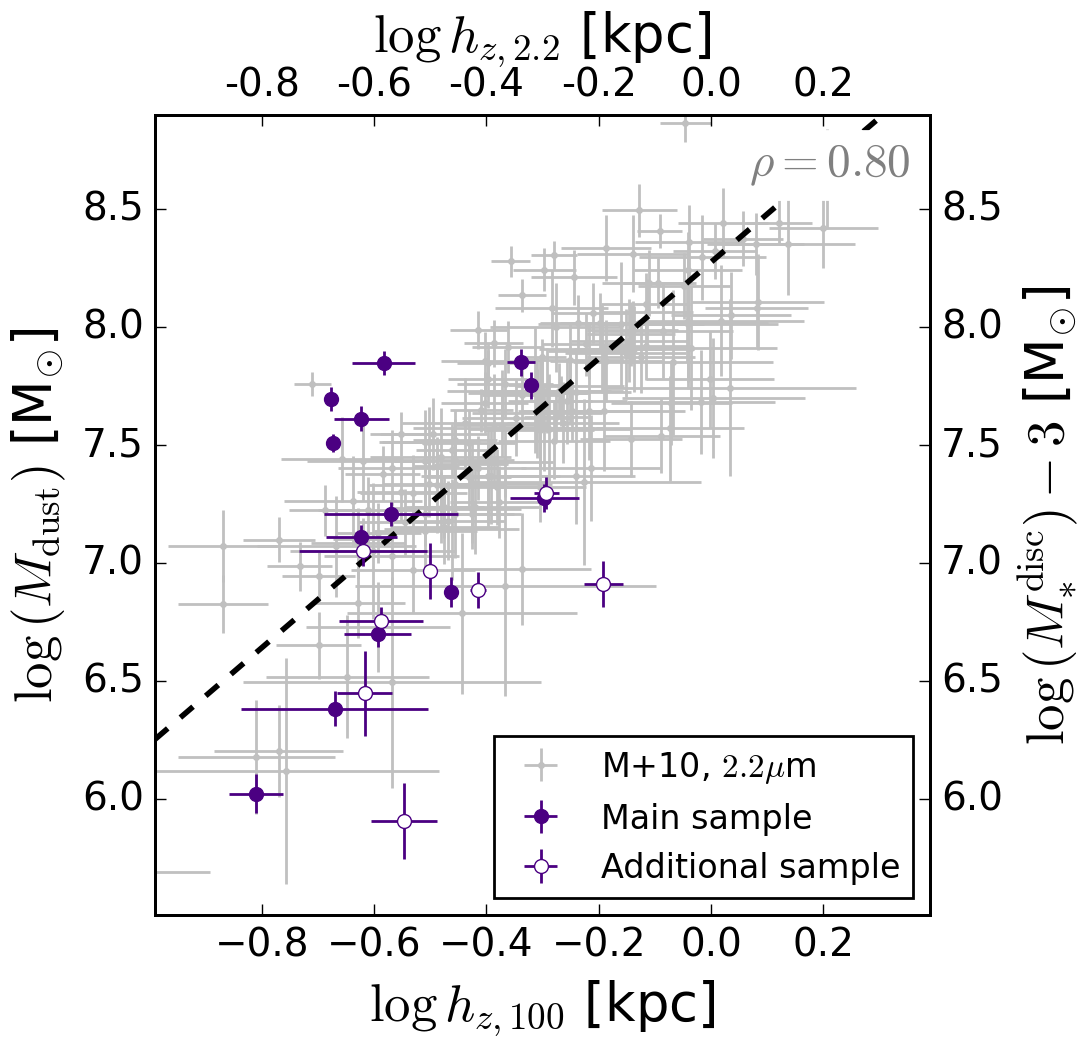}
\caption{Dependences between the total dust mass from \citet{2019A&A...624A..80N} on the S\'ersic effective radius ({\it left}), disc scale length ({\it middle}), and disc scale height ({\it right}) --- all for 100~$\mu$m. In the {\it left} panel we also show galaxies from the whole DustPedia sample in the same PACS\,100 waveband. In the {\it middle} and {\it right} panels, we also display the scaling relations for the stellar discs from \citet{2010MNRAS.401..559M} in the 2MASS $K_s$ (2.2~$\mu$m) band. The stellar masses are shifted downward by 3 dex in both plots. The dashed lines are linear regression lines for the large samples of galaxies, whereas the solid lines represent linear regression lines for our samples.}
\end{figure*}

It is now interesting to explore how the total mass of the dust disc depends on its size. In Fig.~\ref{fig:dust_mass_vs_par} we consider the dependence of the dust mass on the effective radius for the S\'ersic model ({\it left} panel) and the scale length ({\it middle}) and scale height ({\it right}) for the 3D disc model. These can be considered as scaling relations since we can expect that a more massive dust disc should have a larger radius given that the mean surface density does not change dramatically from galaxy to galaxy (indeed, the average dust surface density profile in fig.~5 from \citealt{2016MNRAS.462..331S} does not show a significant spread within the optical radius for 45 large spiral galaxies). Therefore, the correlation between the effective radius and the dust mass naturally occurs for both the edge-on galaxies ($\rho=0.86$) and the whole DustPedia sample ($\rho=0.80$). Similarly, we can see a correlation between the disc scale length and the dust mass, albeit with a larger scatter ($\rho=0.52$). The noisier correlation compared to the effective radius correlation can be explained by the more complex structure of the broken-exponential model as compared to the simple single S\'ersic law with only one scaling parameter $r_\mathrm{e}$. The stellar discs follow a similar relation ($\rho=0.75$).

Interestingly, there is no significant correlation ($\rho=0.31$) between the scale height and the total mass of the dust disc (see Fig.~\ref{fig:dust_mass_vs_par}, {\it right} plot): the most important structural parameter related to the total dust mass is the radial extent of the dust disc, not its vertical size. This is in contrast to what is observed for the stellar discs for which the scale height increases with the mass (as shown by gray symbols based on the results taken from \citealt{2010MNRAS.401..559M}), with a similar scatter as for the stellar mass--scale length relation. This is one of the main differences between the scaling relations for the dust and stellar discs in our study.

\subsection{Flattening of the dust disc}
\label{sec:mass_size}

\begin{figure*}
\label{fig:dust_mass_vs_flatness}
\centering
\includegraphics[height=5.5cm]{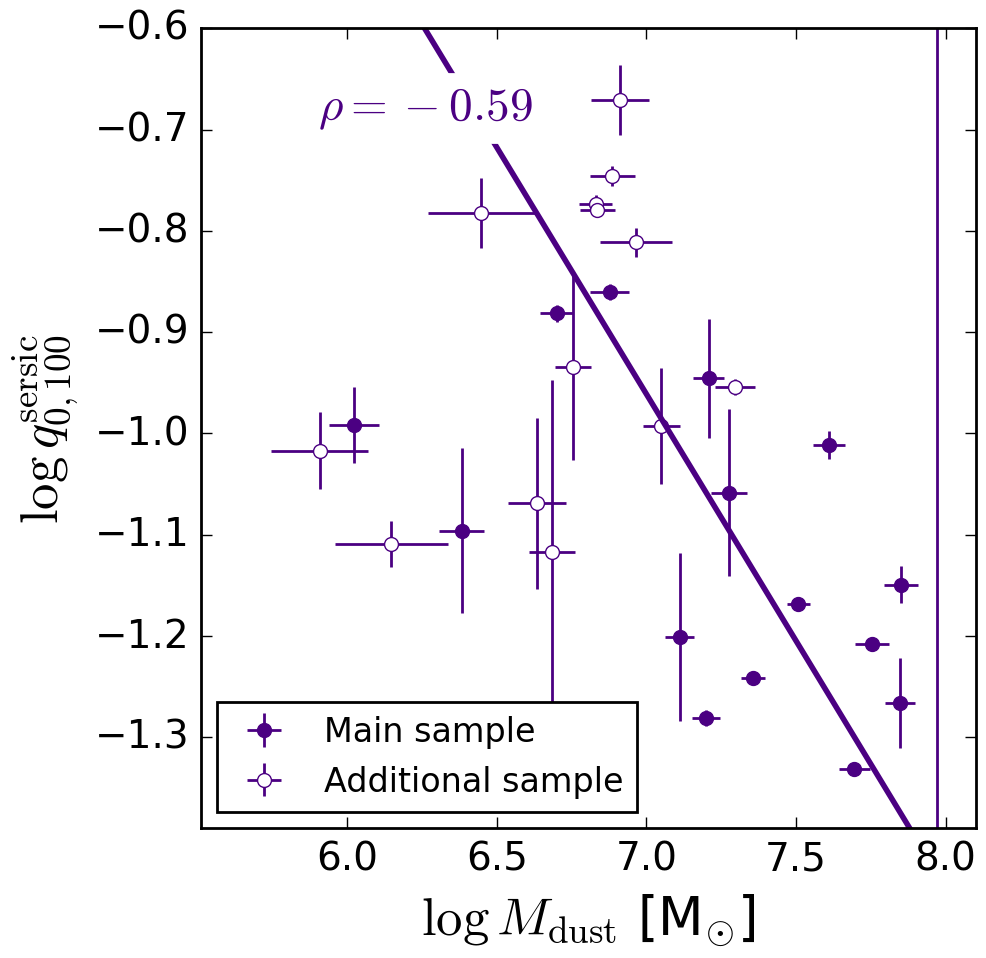}
\includegraphics[height=5.5cm]{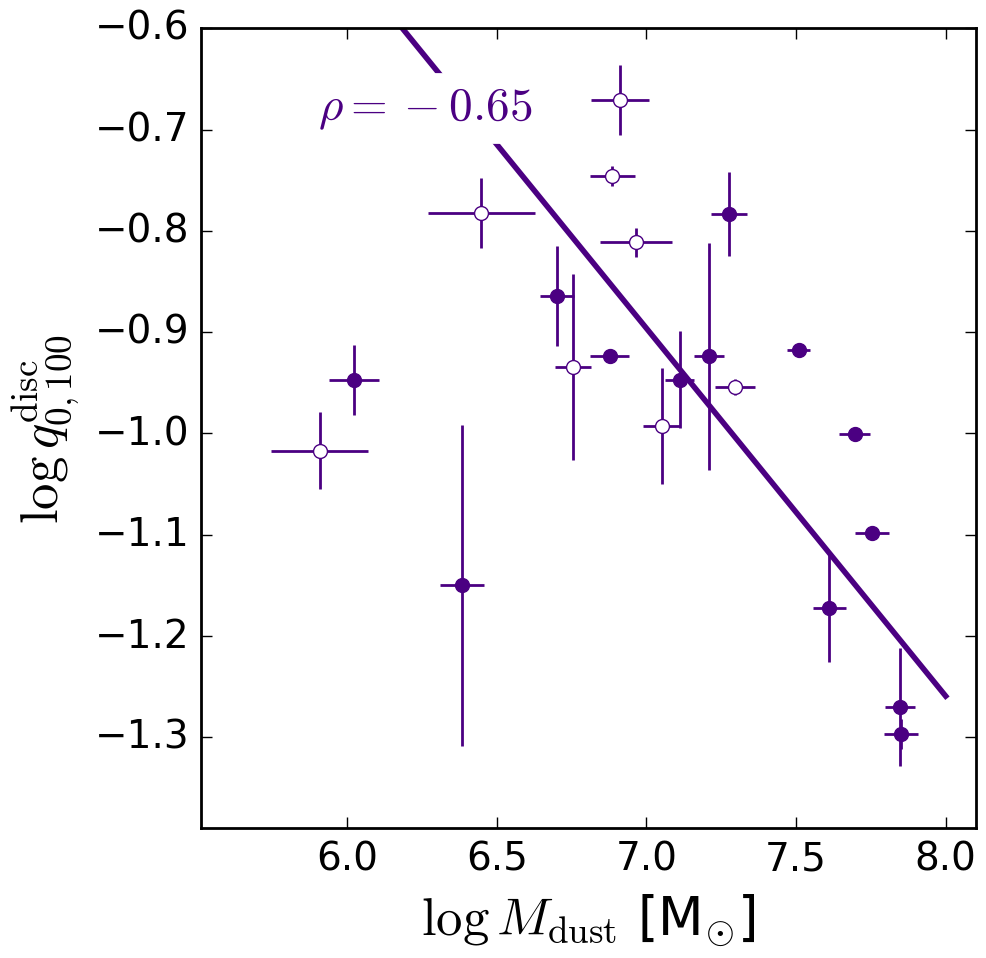}
\includegraphics[height=5.5cm]{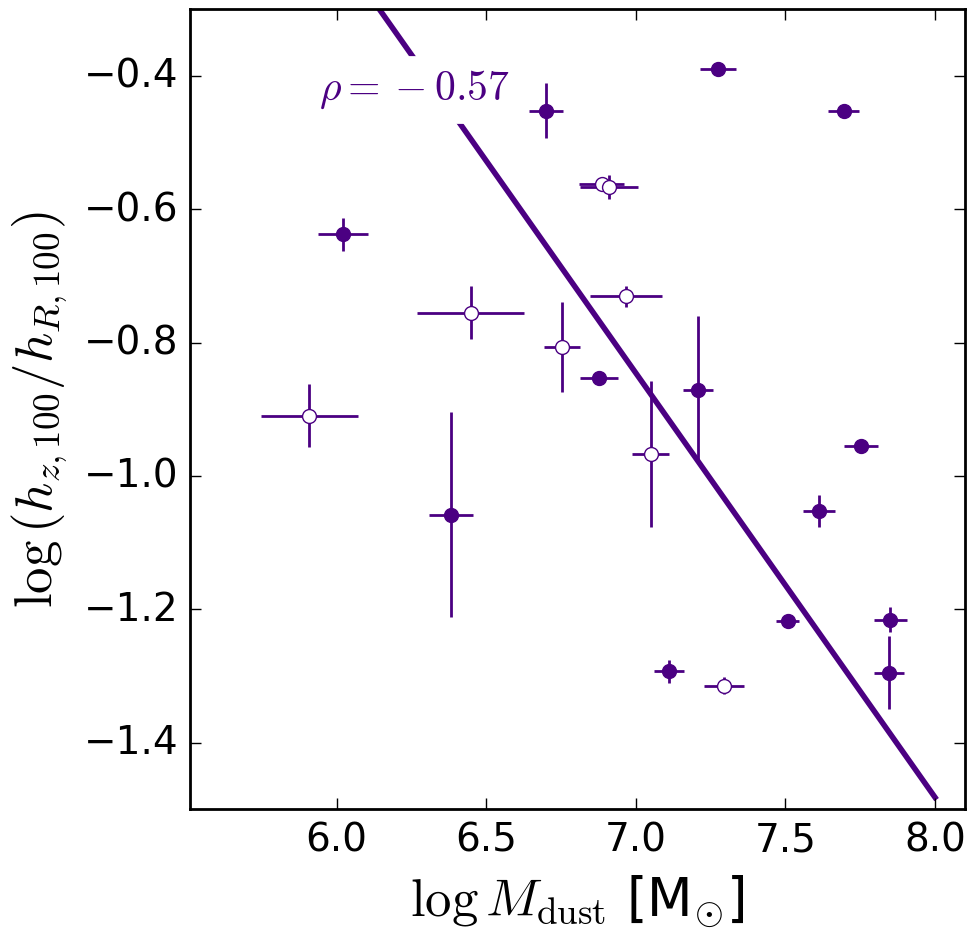}
\caption{Dependences between the total dust mass from \citet{2019A&A...624A..80N} on the intrinsic flattening for the S\'ersic  ({\it left}) and 3D disc ({\it middle}) models and the disc scale height-to-length ratio ({\it right}). The solid lines depict linear regression lines.}
\end{figure*}

\begin{figure*}
\label{fig:vrot_vs_flatness}
\centering
\includegraphics[height=5.5cm]{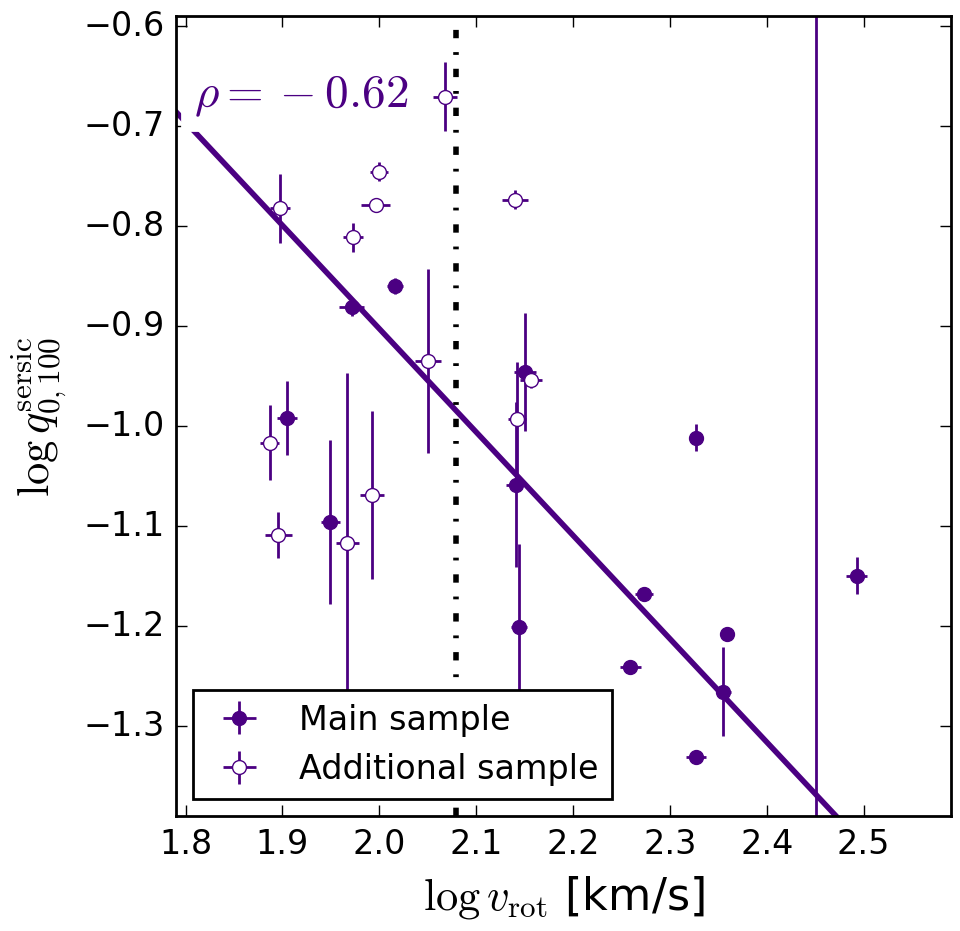}
\includegraphics[height=5.5cm]{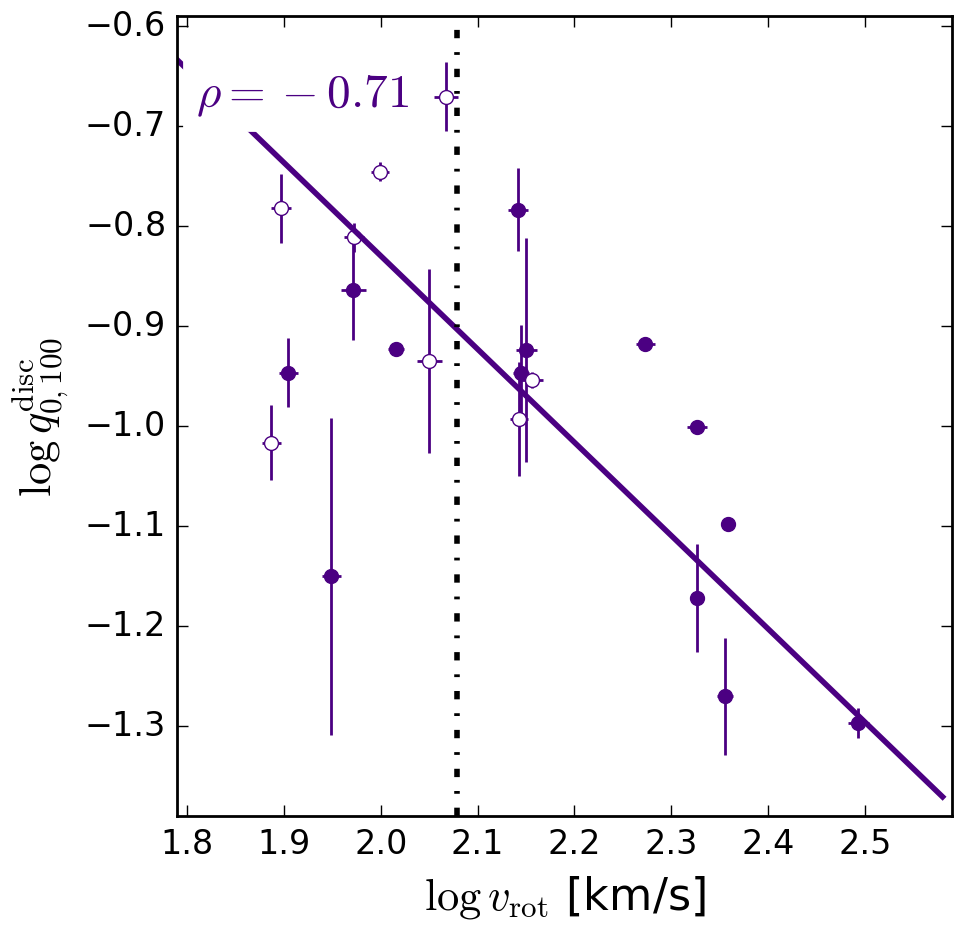}
\includegraphics[height=5.5cm]{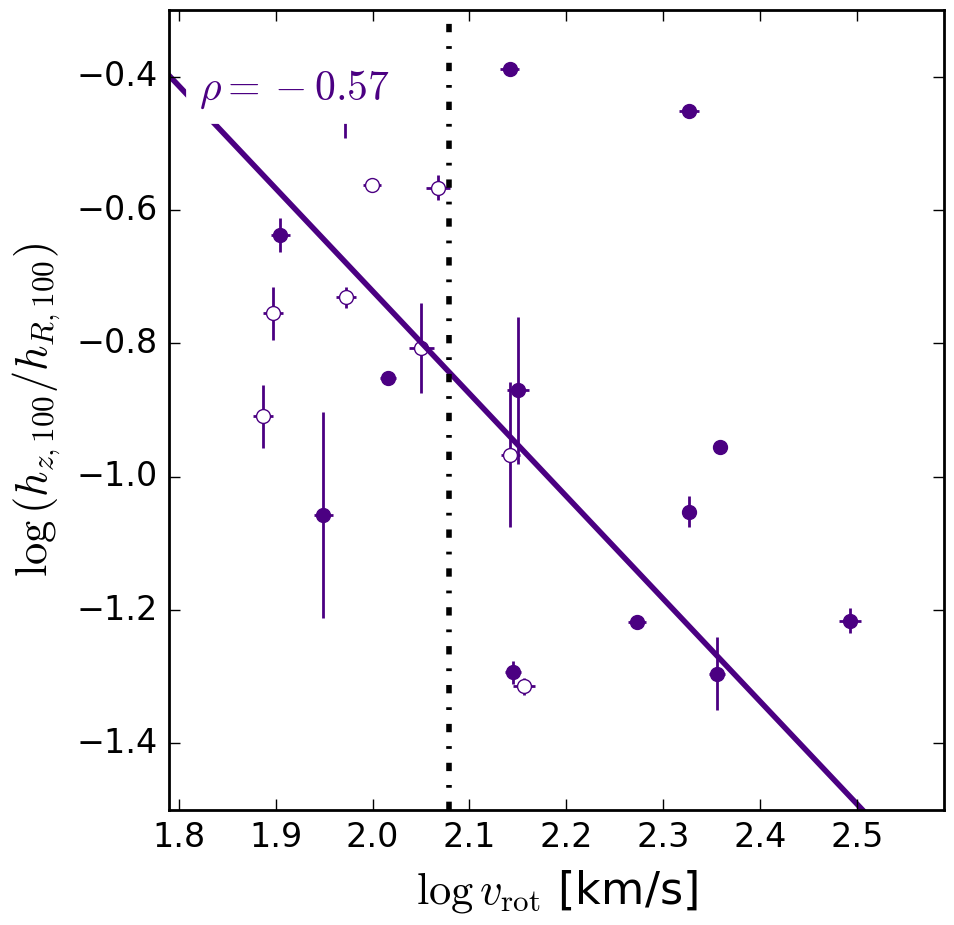}
\caption{Dependences between the rotation velocity from HyperLeda and the intrinsic flattening for the S\'ersic  ({\it left}) and disc ({\it middle}) models and the disc scale height-to-length ratio ({\it right}). The solid lines depict linear regression lines. The dash-dotted lines show 120~km$\,\mathrm{s}^{-1}$.}
\end{figure*}

\begin{figure*}
\label{fig:dust_stellar_height_vs_dust_mass}
\centering
\includegraphics[width=7cm]{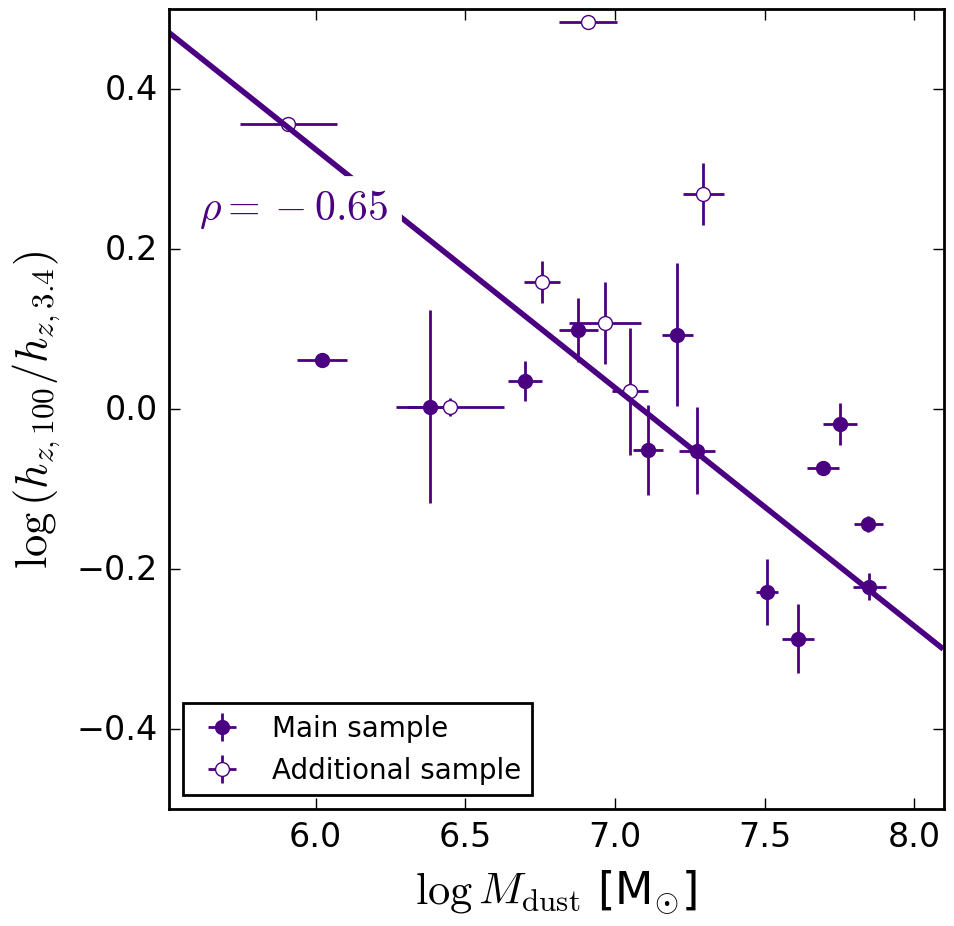}
\includegraphics[width=7cm]{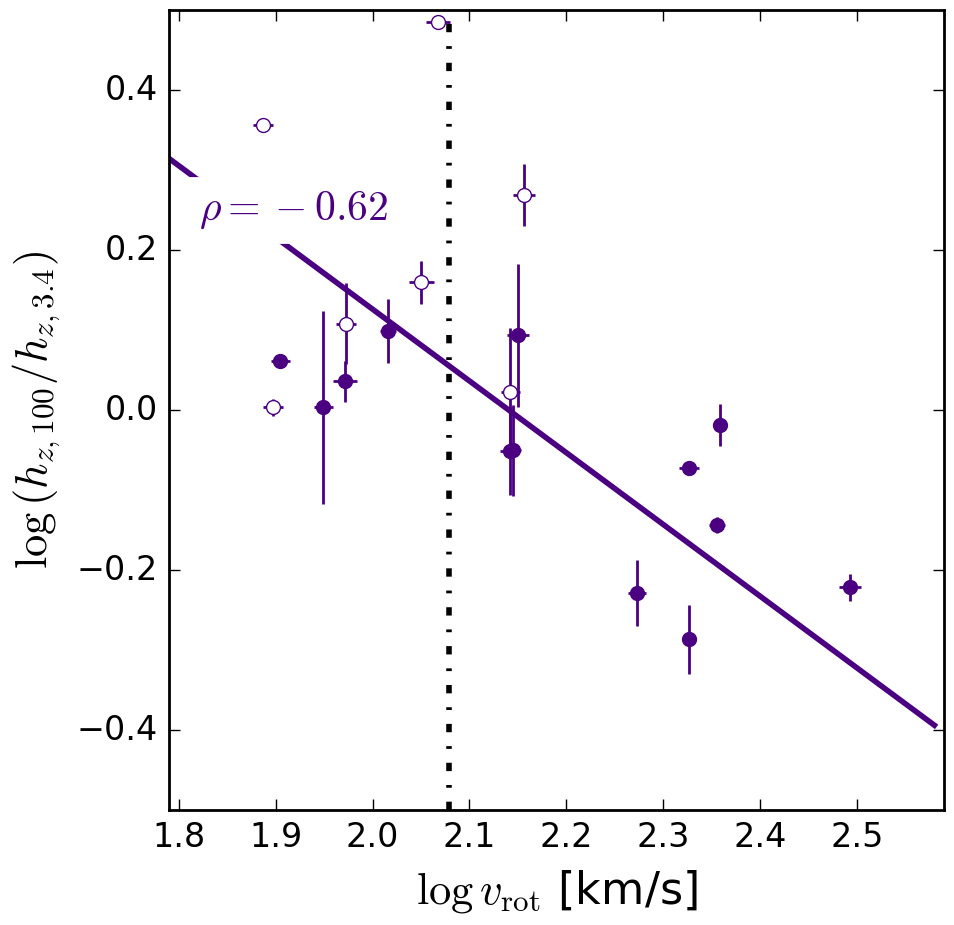}
\caption{Correlations between the ratio of the scale heights at 100 and 3.4~$\mu$m and the total dust mass ({\it upper} panel) and the rotation velocity ({\it bottom} panel). The solid lines depict linear regression lines. The dash-dotted line shows 120~km$\,\mathrm{s}^{-1}$.}
\end{figure*}

Finally, let us consider how the flattening of the dust disc relates to its mass. In Fig.~\ref{fig:dust_mass_vs_flatness} we show the intrinsic flattening for the S\'ersic ({\it left} panel) and 3D disc ({\it middle} panel) models as well as the ratio of the disc scale height to scale length ({\it right} panel). In all these scatter plots we can see a certain trend which indicates that more massive dust discs tend to be flatter ($\rho=-0.59$ for the S\'ersic flattening, -0.65 for the disc flattening, -0.57 for the scale height-to-scale length disc ratio). Furthermore, the relative thickness of the disc and the maximum rotation velocity of the galaxy also negatively correlate with each other (see Fig.~\ref{fig:vrot_vs_flatness}) --- $\rho=-0.62$, -0.71, and -0.57, for the S\'ersic flattening, 3D disc flattening, and the scale height-to-scale length disc ratio, respectively. 
This suggests that the flatness of the general dust disc is also related to the total mass of the galaxy ($v_\mathrm{rot}^2$ is proportional to the mass): flatter dust discs tend to be found in more massive spiral galaxies compared to relatively thick diffuse dust discs in less massive galaxies.
Finally, we find that the ratio between the scale height of the dust disc and that of the stellar disc also negatively correlates with the dust mass ($\rho=-0.65$) and maximum rotation velocity ($\rho=-0.62$, see Fig.~\ref{fig:dust_stellar_height_vs_dust_mass}): less massive dust discs have a larger scale height with respect to the host stellar discs, and for slow-rotating spiral galaxies the scale height of the dust disc can be similar or even larger than that of the stellar disc (this has been shown in Fig.~\ref{fig:dust_vs_stellar_pars}). In other words, in low-mass galaxies the dust is relatively more distributed throughout the height of the hosting stellar disc, whereas in high-mass galaxies, the dusty ISM is concentrated in the plane of the thin disc and appears as a prominent dust lane if the galaxy is oriented edge-on with respect to the observer.

This observational fact was first revealed by \citet{2004ApJ...608..189D}. They concluded that in slowly rotating galaxies with $v_\mathrm{rot}<120$~km$\,\mathrm{s}^{-1}$ (i.e. in low-mass spiral galaxies), the dust distribution has a much larger scale height than in more massive, more rapidly rotating spiral galaxies (see also \citealt{2019AJ....158..103H}). In the optical, such low-mass galaxies, being observed edge-on, have ill-defined or no mid-plane dust lanes at all. They attributed this phenomenon to the vertical stability of the gas and stellar disc: in galaxies with $v_\mathrm{rot}>120$~km$\,\mathrm{s}^{-1}$, the high stellar mass surface density creates instabilities in the cold ISM, which vertically collapses into a thin stellar disc with the formation of a thin dust disc. For galaxies with a lower circular velocity, the dust settles in a clumpy structure, which is well-seen in NGC\,4244 and NGC\,4437 (the brightest clumps were fitted on top of the S\'ersic and disc components) from our Main sample. For the other two galaxies with low rotation velocities (ESO\,373-008 and NGC\,5023), the resolution is not sufficient to see the details, and we only trace the global, smoothed structure of the dust disc.

\citet{2019AJ....158..103H} examined the frequency of the occurrence of dust lanes in edge-on galaxies and found that bulge morphology correlates with the presence of dust lanes: dust lanes are present more often in galaxies with large, round bulges, whereas the presence of boxy bulges anti-correlates with the identification of dust lanes in such galaxies. They also concluded that dust lanes are mostly observed in galaxies where no signs of interaction with a neighbour are found. Our sample is too small to make any conclusions on this matter in general, but several examples contradict the mentioned findings of \citet{2019AJ....158..103H}. For instance, NGC\,3628 is a well-known galaxy with a boxy/peanut-shaped (B/PS) bulge and with apparent signs of a merger and a well-defined dust disc (see e.g., \citealt{2020MNRAS.494.1751M}). IC\,2531, NGC\,0891, NGC\,4013 NGC\,5529, and NGC\,5746 also demonstrate the presence of a B/PS bulge, but, at the same time, exhibit highly contrasted dust lanes.

\section{Conclusions}
\label{sec:conclusions}

In this paper, we have extended the study by \citetalias{2019A&A...622A.132M} with a main focus on the vertical structure of the dust discs using FIR/submm {\it Herschel} observations. We considered 16 large (Main sample) and 13 less spatially resolved (Additional sample) edge-on spiral galaxies in each of the five 100--500$\mu$m Herschel bands where possible. We carefully examined the background in the images to ensure that it is flat and does not affect the results of our modelling. We used both S\'ersic and 3D (broken) exponential disc profiles to fit the galaxy images. For galaxies with bright sources within the plane, they were simultaneously fitted along with the target galaxy to minimise the influence of the light from such sources on the galaxy fitting parameters. This analysis enabled us to directly study not only the radial distribution of dust, but also the distribution that extends out perpendicular to the galaxy mid-plane. The cumulative galaxy profiles with the superimposed models are presented in Fig.~\ref{fig:profiles} and Table~A1 (both published in their entirety as online material). We also used WISE 3.4~$\mu$m imaging of the galaxies to retrieve the structural parameters of their old stellar distribution. We can summarise our results as follows.

\begin{enumerate}
\item The obtained S\'ersic and 3D (broken) exponential disc models describe the observed emission profiles equally well, albeit the advantage of using a 3D disc over a S\'ersic model is in retrieving the scale height, an independent parameter of the 3D disc geometry.
\item We find that the effective radius of the dust emission increases with wavelength, whereas the apparent flattening of the S\'ersic model remains almost constant (Fig.~\ref{fig:pars_lambda}, two upper panels). We also point out an increase of both the disc scale length and scale height with wavelength (Fig.~\ref{fig:pars_lambda}, two bottom panels). The first effect is related to the dust heating by the diffuse ISRF (coupled with the fact that dust discs are more widespread than their stellar counterparts), the second one --- with a dramatic drop in dust temperature within the disc scale height and/or the presence of unresolved clumps or other unresolved global structural components (ring, superthin disc, spiral arms) in the FIR images.
\item For 6 (and possibly additional four) of the 16 large galaxies in our sample, we clearly detect a second vertically-extended component which can be interpreted as a thick dust disc or a halo (see Fig.~\ref{fig:profiles}). These galaxies are NGC\,0891, NGC\,3628, NGC\,4013, ESO\,209-009, NGC\,4437, NGC\,4631 and NGC\,4244, ESO\,373-008, NGC\,5529, NGC\,5746 with a possible extraplanar emission. In the Additional sample, we identify six galaxies (IC\,2233, NGC\,3592, UGC\,07321, UGC\,07387, UGC\,07522, and UGC\,09242) demonstrating an excess of the FIR emission.
\item Similar to the stellar discs, the dust discs grow proportionally in the vertical and radial directions (Fig.~\ref{fig:length_vs_height}).
\item We found moderate positive correlations between the disc scale lengths in the NIR and FIR and between the scale heights in the same domains (Fig.~\ref{fig:dust_vs_stellar_pars}), confirming that the dust mass distribution correlates with the total stellar distribution \citep{1998A&A...335..807A,2009ApJ...701.1965M,2014A&A...565A...4H,2017A&A...605A..18C}, probably through the population of AGB stars. 
\item Based on the dust emission in the FIR, we found an anti-correlation between the intrinsic flattening (relative thickness) of the dust disc and the dust mass (Fig.~\ref{fig:dust_mass_vs_flatness}). Also, we report that the dust disc flatness and the ratio between the dust disc scale height and the stellar disc scale height anti-correlate with the maximum galaxy rotation velocity (Figs.~\ref{fig:vrot_vs_flatness} and ~\ref{fig:dust_stellar_height_vs_dust_mass}, respectively): spiral galaxies with higher dust (and total) masses possess flatter dust discs, whereas low-mass spiral galaxies host puffed-up diffuse dust discs distributed throughout the stellar disc. This is a confirmation of the findings by \citet{2004ApJ...608..189D} and \citet{2019AJ....158..103H}, but is seen for the first time in dust emission.
\end{enumerate}

In our next articles in this series dedicated to the study of the dust distribution in edge-on galaxies, we will carry out a detailed analysis of the structure for the most resolved galaxies in our sample which, as we showed in this paper, demonstrate an extraplanar dust emission above a single exponential disc. Also, it is particularly important to study how contemporary dust RT models, which are built upon the observed dust attenuation in edge-on galaxies in the UV, optical, and NIR, compare with the results of this study. For example, the well-known dust-energy balance problem, which states that RT models underestimate the observed FIR emission by a factor of two to four \citep[see e.g.,][]{2000A&A...362..138P,2015MNRAS.451.1728D,2018A&A...616A.120M}, has not been entirely solved.

\section*{Acknowledgements}
We acknowledge financial support from the Russian Science Foundation (grant no. 20-72-10052).

We thank the anonymous referee whose comments and
valuable suggestions improved the paper.

We acknowledge the usage of the HyperLeda database (http://leda.univ-lyon1.fr).
Funding for the Sloan Digital Sky Survey IV has been provided by the Alfred P. Sloan Foundation, the U.S. Department of Energy Office of Science, and the Participating Institutions. SDSS-IV acknowledges
support and resources from the Center for High-Performance Computing at
the University of Utah. The SDSS web site is www.sdss.org.

SDSS-IV is managed by the Astrophysical Research Consortium for the 
Participating Institutions of the SDSS Collaboration including the 
Brazilian Participation Group, the Carnegie Institution for Science, 
Carnegie Mellon University, the Chilean Participation Group, the French Participation Group, Harvard-Smithsonian Center for Astrophysics, 
Instituto de Astrof\'isica de Canarias, The Johns Hopkins University, Kavli Institute for the Physics and Mathematics of the Universe (IPMU) / 
University of Tokyo, the Korean Participation Group, Lawrence Berkeley National Laboratory, 
Leibniz Institut f\"ur Astrophysik Potsdam (AIP),  
Max-Planck-Institut f\"ur Astronomie (MPIA Heidelberg), 
Max-Planck-Institut f\"ur Astrophysik (MPA Garching), 
Max-Planck-Institut f\"ur Extraterrestrische Physik (MPE), 
National Astronomical Observatories of China, New Mexico State University, 
New York University, University of Notre Dame, 
Observat\'ario Nacional / MCTI, The Ohio State University, 
Pennsylvania State University, Shanghai Astronomical Observatory, 
United Kingdom Participation Group,
Universidad Nacional Aut\'onoma de M\'exico, University of Arizona, 
University of Colorado Boulder, University of Oxford, University of Portsmouth, 
University of Utah, University of Virginia, University of Washington, University of Wisconsin, 
Vanderbilt University, and Yale University.

The Digitized Sky Surveys were produced at the Space Telescope Science Institute under U.S. Government grant NAG W-2166. The images of these surveys are based on photographic data obtained using the Oschin Schmidt Telescope on Palomar Mountain and the UK Schmidt Telescope. The plates were processed into the present compressed digital form with the permission of these institutions.

The National Geographic Society - Palomar Observatory Sky Atlas (POSS-I) was made by the California Institute of Technology with grants from the National Geographic Society.

The Second Palomar Observatory Sky Survey (POSS-II) was made by the California Institute of Technology with funds from the National Science Foundation, the National Geographic Society, the Sloan Foundation, the Samuel Oschin Foundation, and the Eastman Kodak Corporation.

The Oschin Schmidt Telescope is operated by the California Institute of Technology and Palomar Observatory.

The UK Schmidt Telescope was operated by the Royal Observatory Edinburgh, with funding from the UK Science and Engineering Research Council (later the UK Particle Physics and Astronomy Research Council), until 1988 June, and thereafter by the Anglo-Australian Observatory. The blue plates of the southern Sky Atlas and its Equatorial Extension (together known as the SERC-J), as well as the Equatorial Red (ER), and the Second Epoch [red] Survey (SES) were all taken with the UK Schmidt.

Supplemental funding for sky-survey work at the STScI is provided by the European Southern Observatory.

This research makes use of data products from the Wide-field Infrared Survey Explorer, which is a joint project of the University of California, Los Angeles, and the Jet Propulsion Laboratory/California Institute of Technology, and NEOWISE, which is a project of the Jet Propulsion Laboratory/California Institute of Technology. WISE and NEOWISE are funded by the National Aeronautics and Space Administration.

\textit{Herschel} is an ESA space observatory with science instruments provided by European-led Principal Investigator consortia and with important participation from NASA. The \textit{Herschel} spacecraft was designed, built, tested, and launched under a contract to ESA managed by the \textit{Herschel}/\textit{Planck} Project team by an industrial consortium under the overall responsibility of the prime contractor Thales Alenia Space (Cannes), and including Astrium (Friedrichshafen) responsible for the payload module and for system testing at spacecraft level, Thales Alenia Space (Turin) responsible for the service module, and Astrium (Toulouse) responsible for the telescope, with an excess of a hundred subcontractors.

\section*{Data availability}
The data underlying this article are available in the article and in its online supplementary material.

\bibliographystyle{mnras}
\interlinepenalty=10000
\bibliography{dust}

\appendix

\section{Results of the fitting}
\label{sec:models}

\begin{table*}
	\label{tab:decomp_results_all}	
	\caption{The results of our fitting for galaxies from the Main and Additional samples. The results for the S\'ersic modeling at 3.4~$\mu$m are taken from \citetalias{2019A&A...622A.132M}. * means that this parameter can be unreliable due to the poor vertical resolution. $\infty$ denotes that the inner disc profile is a plateau with a very large inner disc scale length $h_\mathrm{in}$. The complete table is available online.}
	\begin{tabular}{cccccccccc}
		\hline
		Galaxy & band & $n$ & $r_\mathrm{e}$ & $q_0^\mathrm{sersic}$ & $h_\mathrm{in}$ & $h_\mathrm{out}$ & $R_\mathrm{b}$ & $h_z$ & $q_0^\mathrm{disc}$ \\
        & & & (kpc) &  & (kpc) & (kpc) & (kpc) & (kpc) &  \\ 
		\hline
ESO209-009&3.4&$1.38$&$3.66$&$0.13\pm0.02$&$1.89\pm0.07$&---&---&$0.22\pm0.02$&---\\
&100&$1.00$&$3.45\pm0.01$&$0.11\pm0.02$&$2.00\pm0.05$&---&---&$0.27\pm0.07$&$0.12\pm0.03$\\
&160&$1.00$&$3.63\pm0.02$&$0.12\pm0.01$&$2.21\pm0.03$&---&---&$0.31\pm0.08$&$0.12\pm0.03$\\
&250&$1.00\pm0.08$&$4.12\pm0.02$&$0.13\pm0.01$&$2.42\pm0.02$&---&---&$0.34\pm0.05$&$0.12\pm0.02$\\
&350&$1.00$&$4.67\pm0.03$&$0.13\pm0.01$&---&---&---&---&---\\
...&...&...&...&...&...&...&...&...&...\\
\hline		
	\end{tabular}
\end{table*}

\begin{figure*}
\centering
\includegraphics[height=4.2cm]{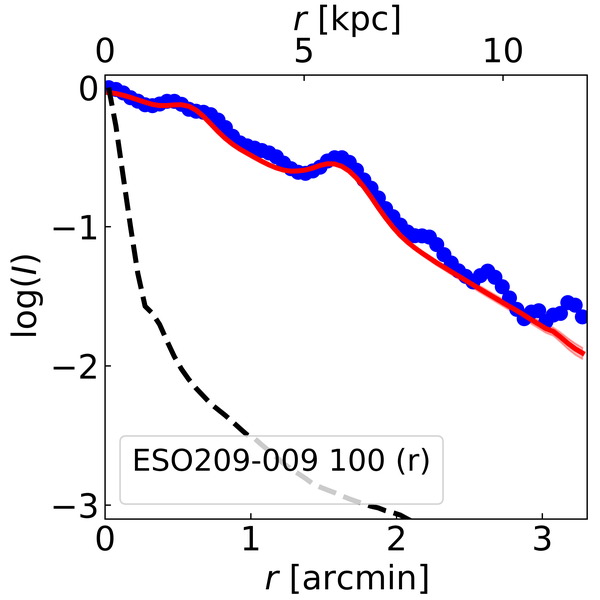}
\includegraphics[height=4.2cm]{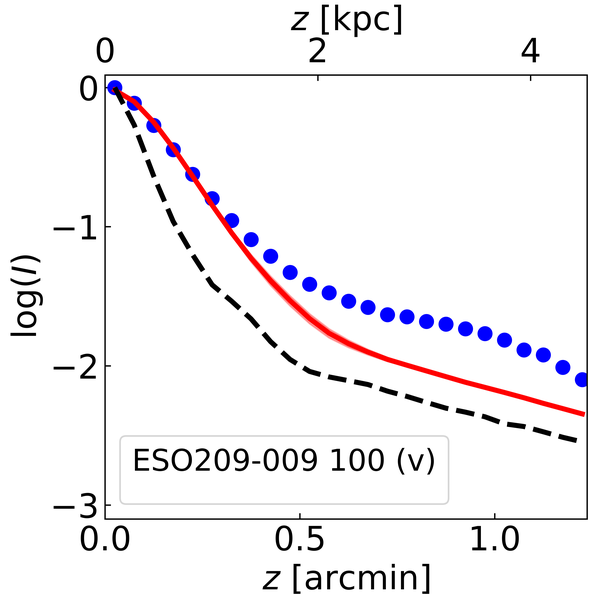}
\includegraphics[height=4.2cm]{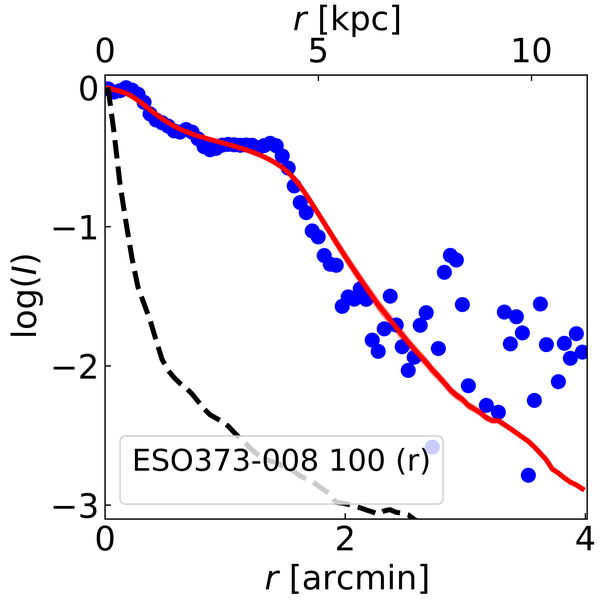}
\includegraphics[height=4.2cm]{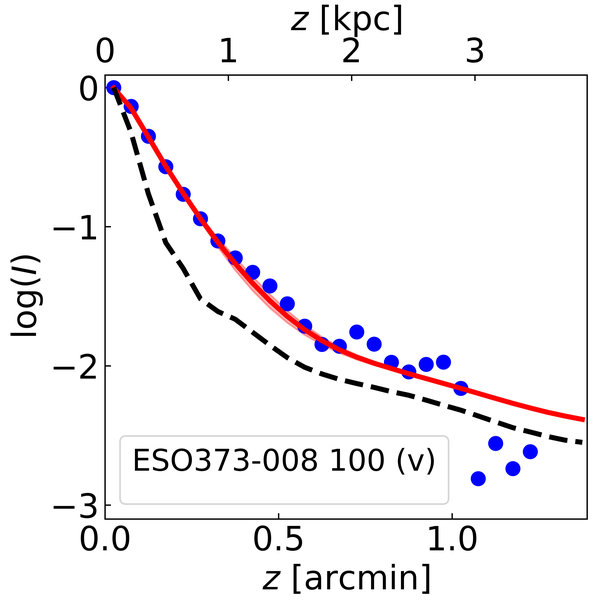}
\includegraphics[height=4.2cm]{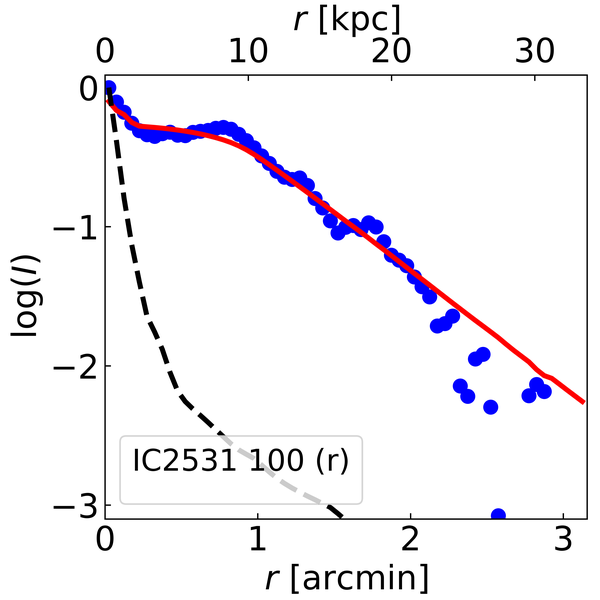}
\includegraphics[height=4.2cm]{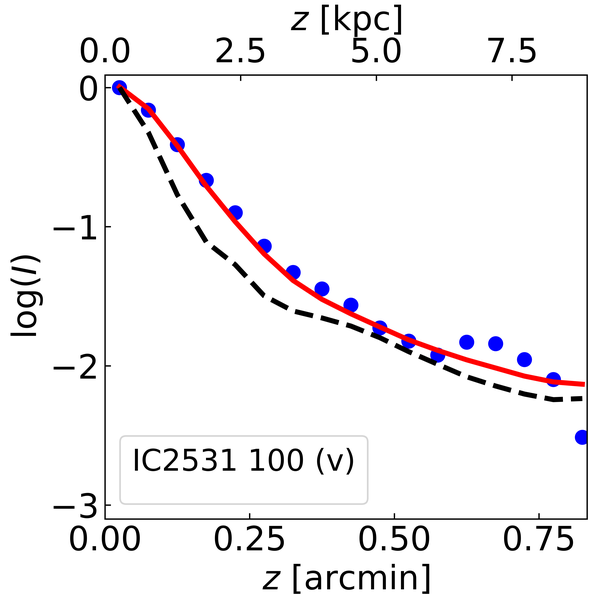}
\includegraphics[height=4.2cm]{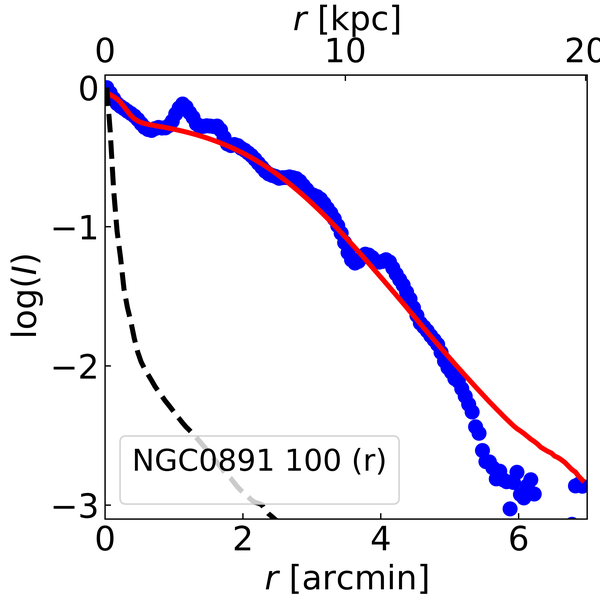}
\includegraphics[height=4.2cm]{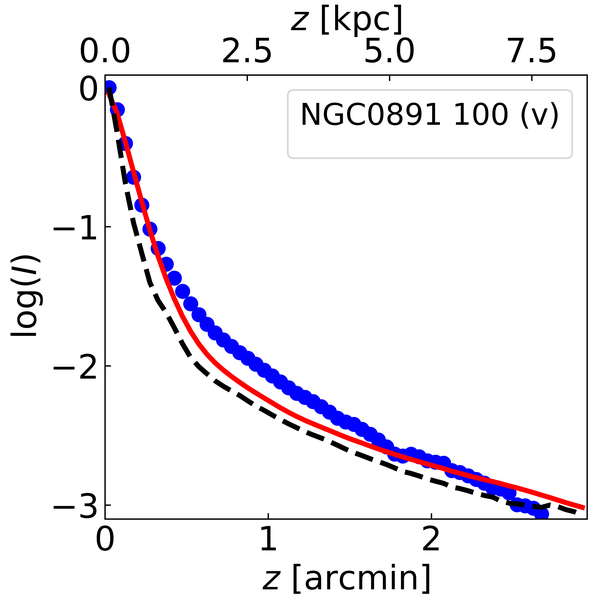}
\includegraphics[height=4.2cm]{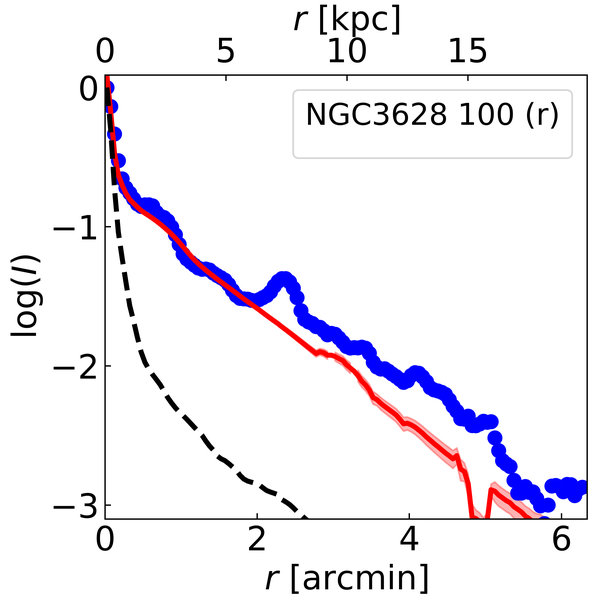}
\includegraphics[height=4.2cm]{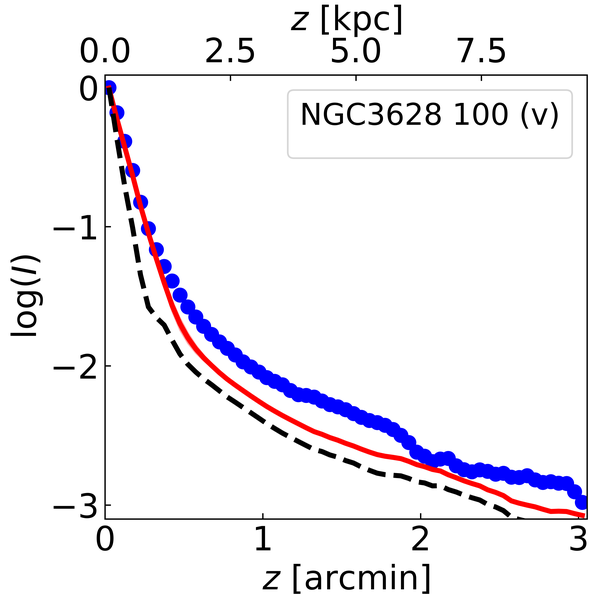}
\includegraphics[height=4.2cm]{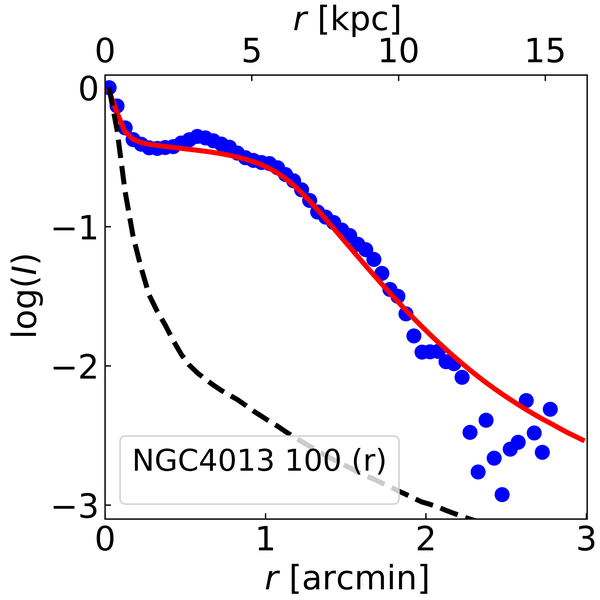}
\includegraphics[height=4.2cm]{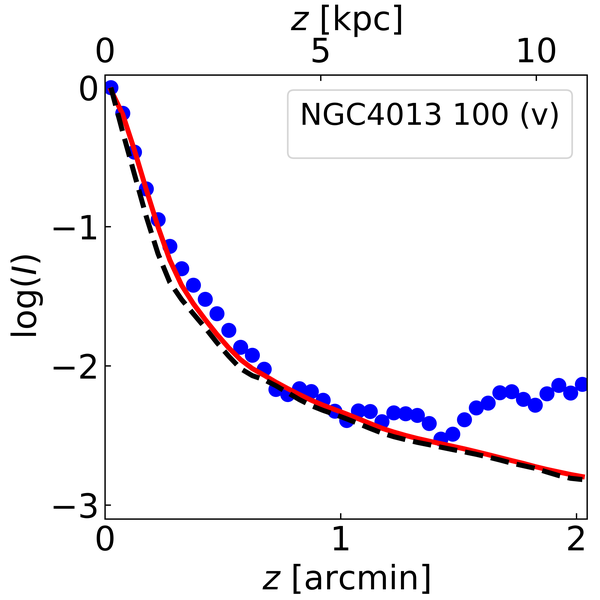}
\includegraphics[height=4.2cm]{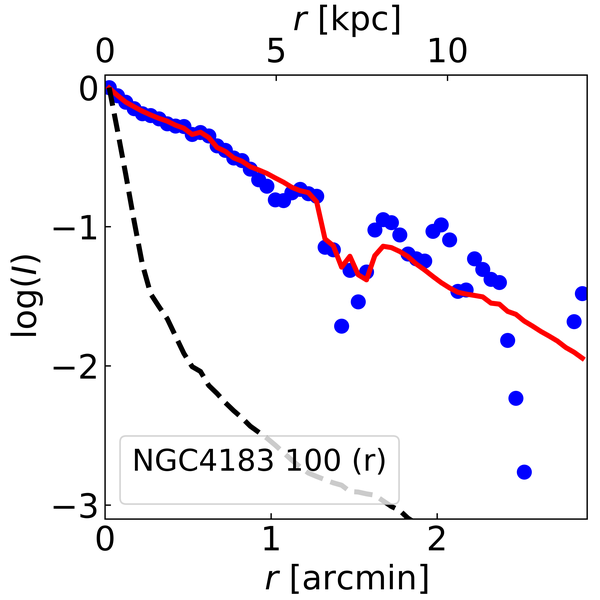}
\includegraphics[height=4.2cm]{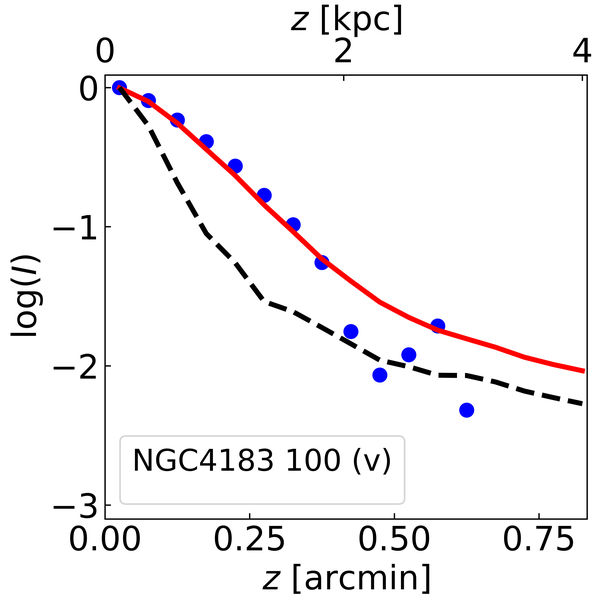}
\includegraphics[height=4.2cm]{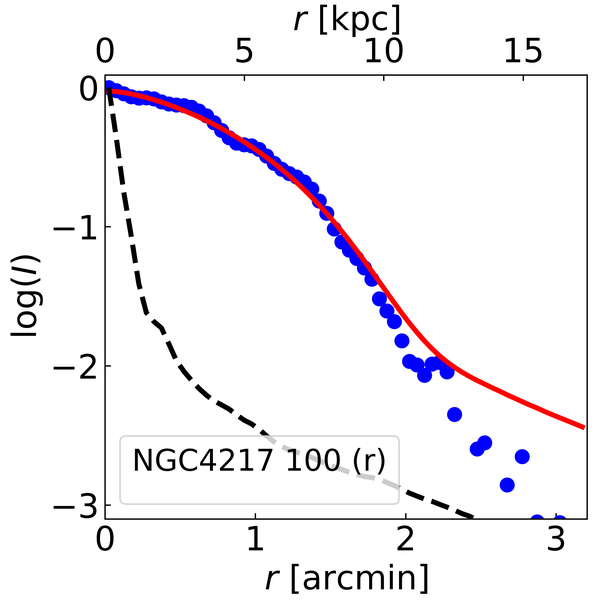}
\includegraphics[height=4.2cm]{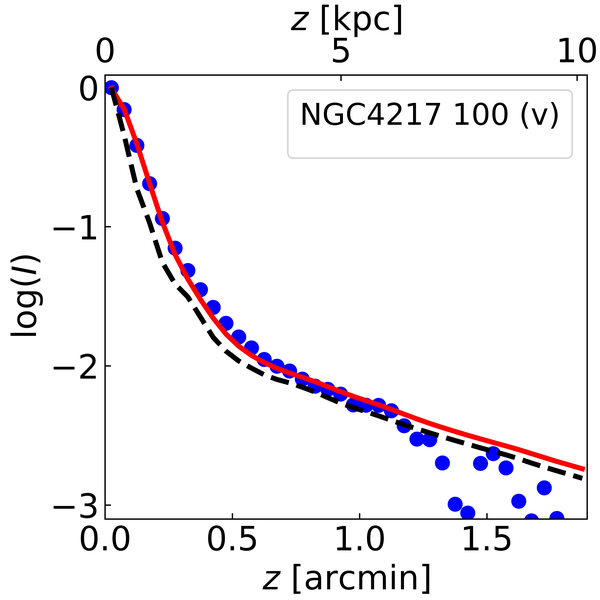}
\includegraphics[height=4.2cm]{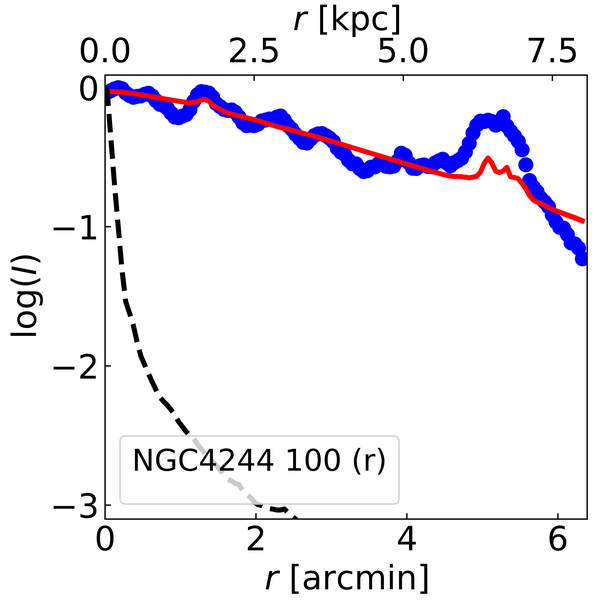}
\includegraphics[height=4.2cm]{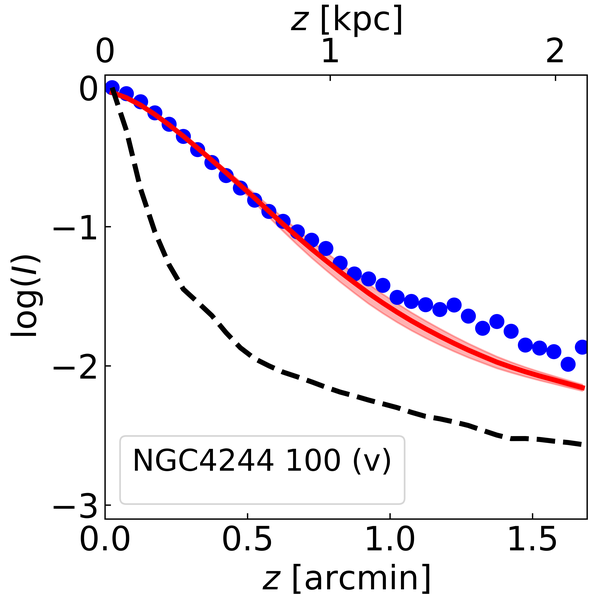}
\includegraphics[height=4.2cm]{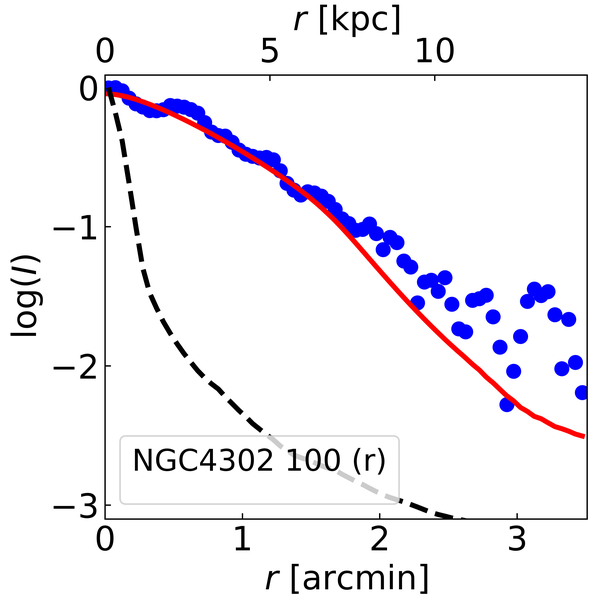}
\includegraphics[height=4.2cm]{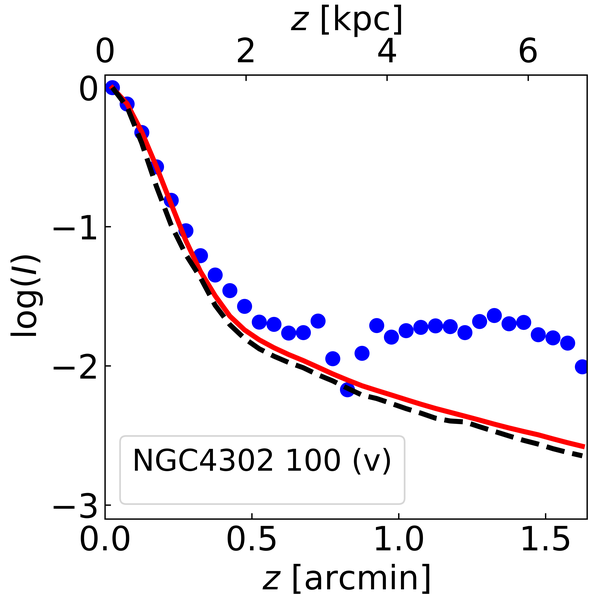}
\caption{Cumulative radial (r) and vertical (v) profiles of galaxies from the Main and Additional samples for the PACS\,100 waveband. In each plot, the blue dots represent the galaxy profile, the red solid line -- its 3D disc model, the reddish spread shows a 1$\sigma$ spread of the model profiles for varied inclination angles (if the spread is not seen, its width is less than the width of the red solid line), and the black dashed line -- the corresponding LSF profile. All profiles are normalised by the maximum of the galaxy profile. Profiles for the other {\it Herschel} wavebands and for the S\'ersic modelling are available online.}
\label{fig:profiles}
\end{figure*}

\addtocounter{figure}{-1}
\begin{figure*}
\centering
\includegraphics[height=4.2cm]{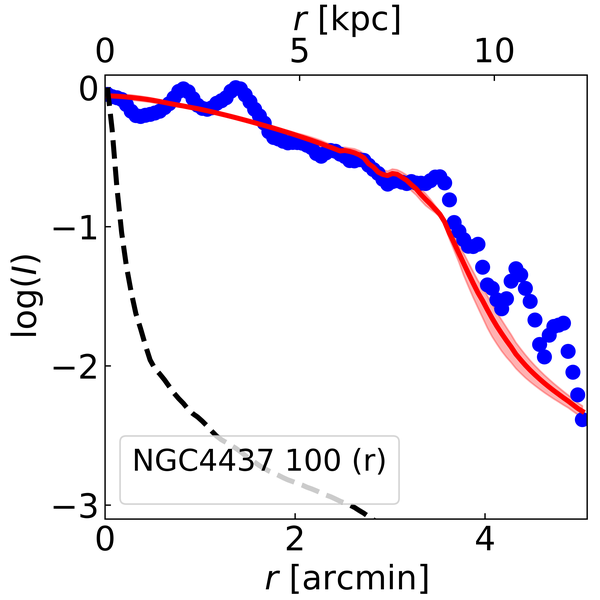}
\includegraphics[height=4.2cm]{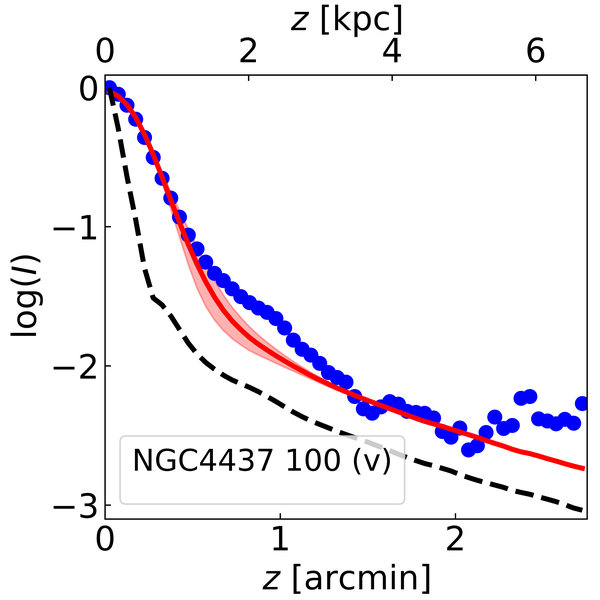}
\includegraphics[height=4.2cm]{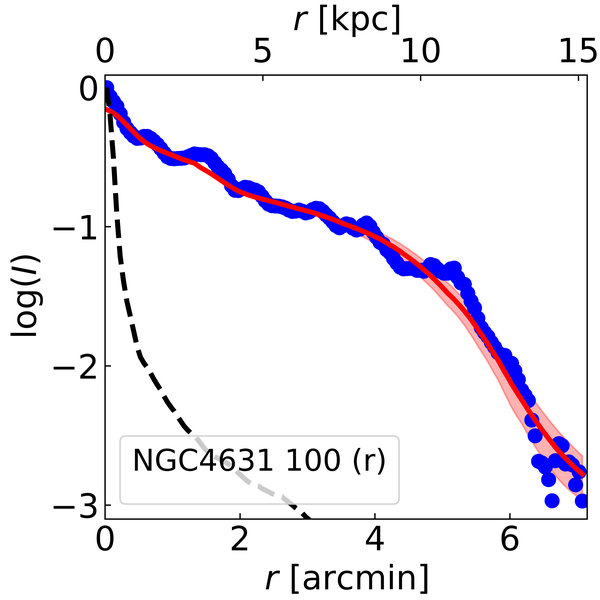}
\includegraphics[height=4.2cm]{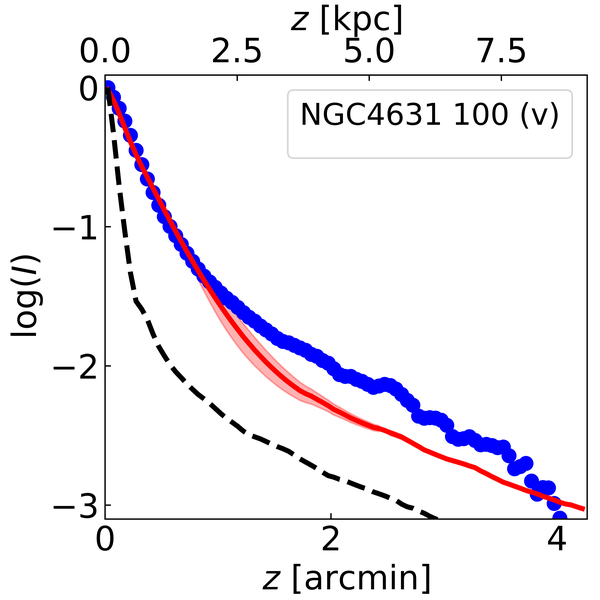}
\includegraphics[height=4.2cm]{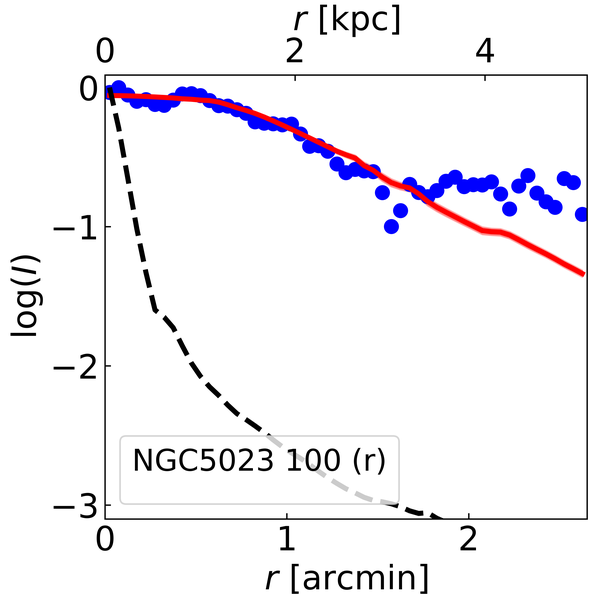}
\includegraphics[height=4.2cm]{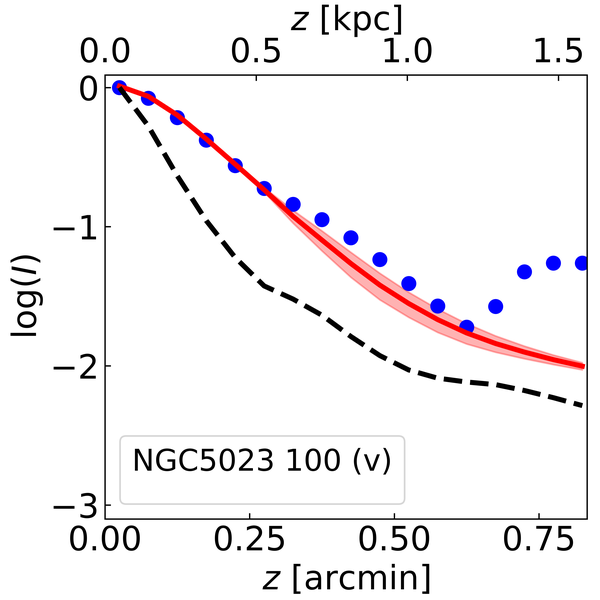}
\includegraphics[height=4.2cm]{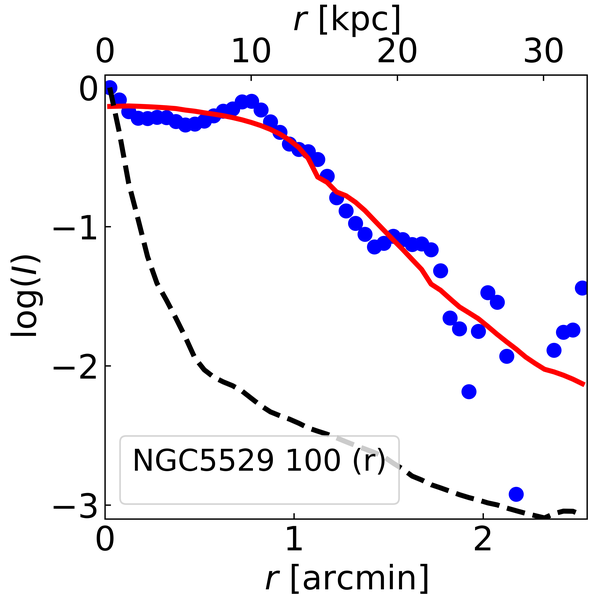}
\includegraphics[height=4.2cm]{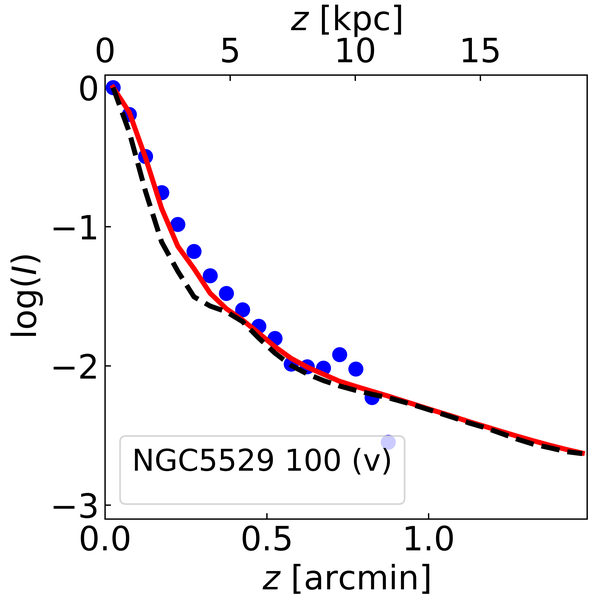}
\includegraphics[height=4.2cm]{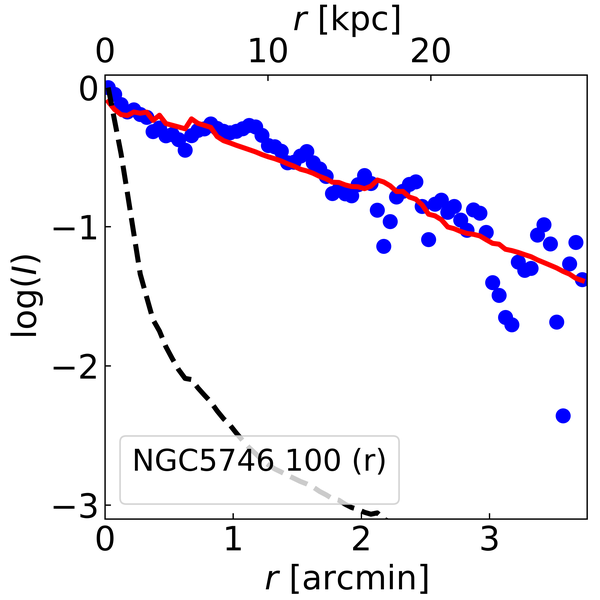}
\includegraphics[height=4.2cm]{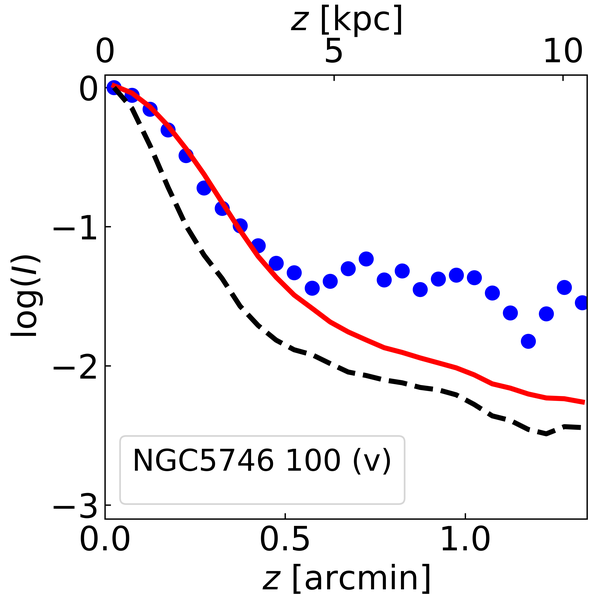}
\includegraphics[height=4.2cm]{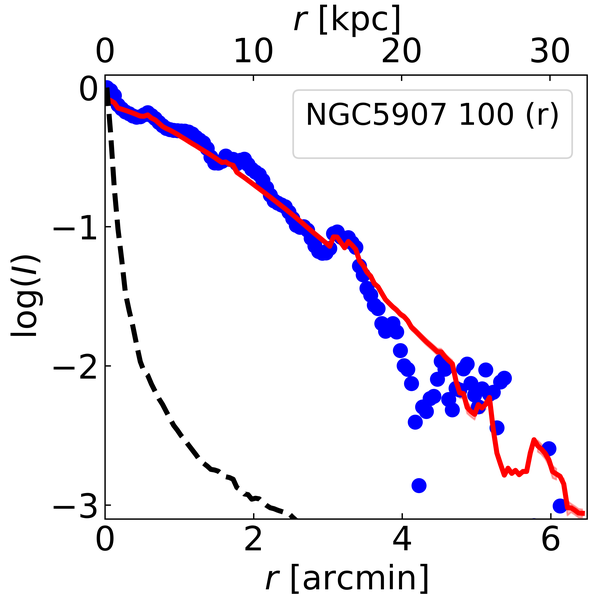}
\includegraphics[height=4.2cm]{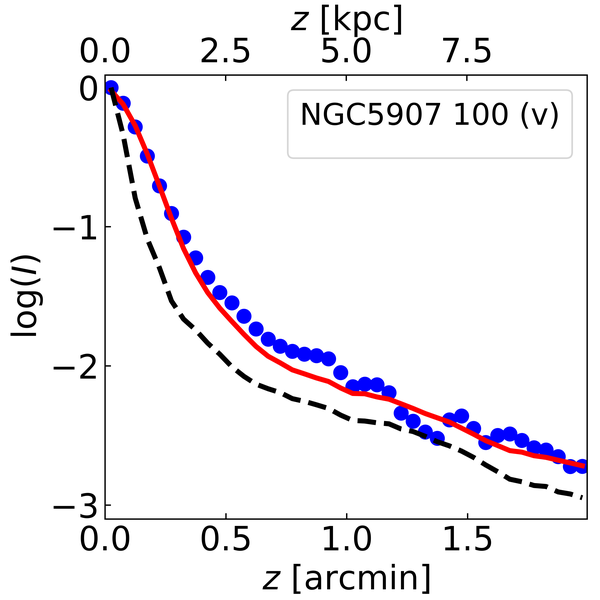}
\includegraphics[height=4.2cm]{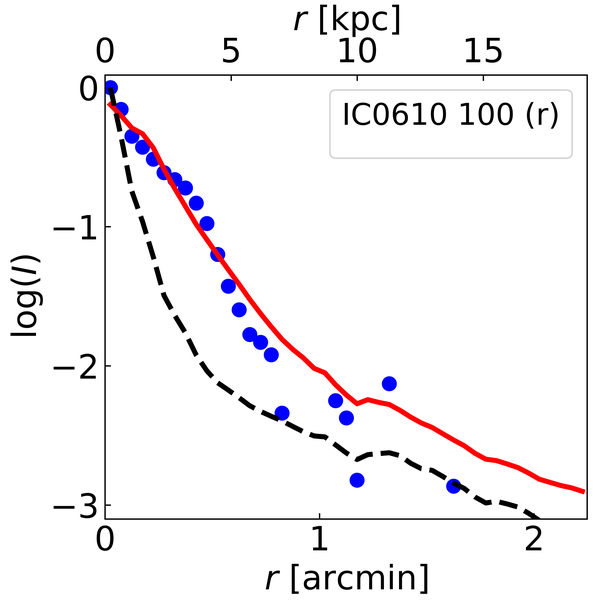}
\includegraphics[height=4.2cm]{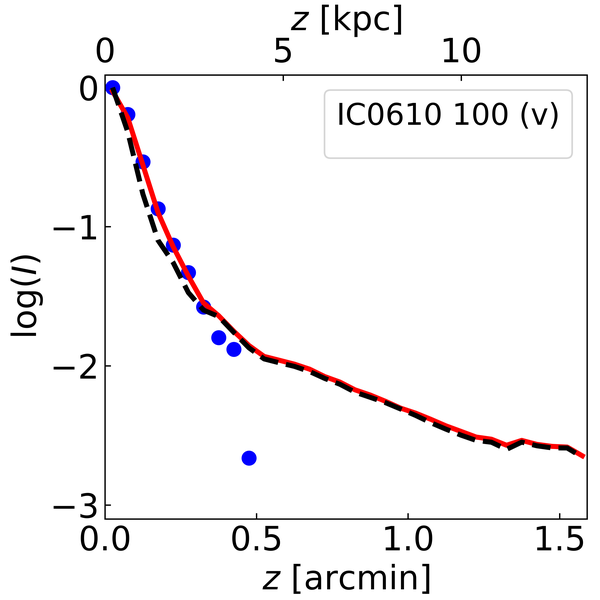}
\includegraphics[height=4.2cm]{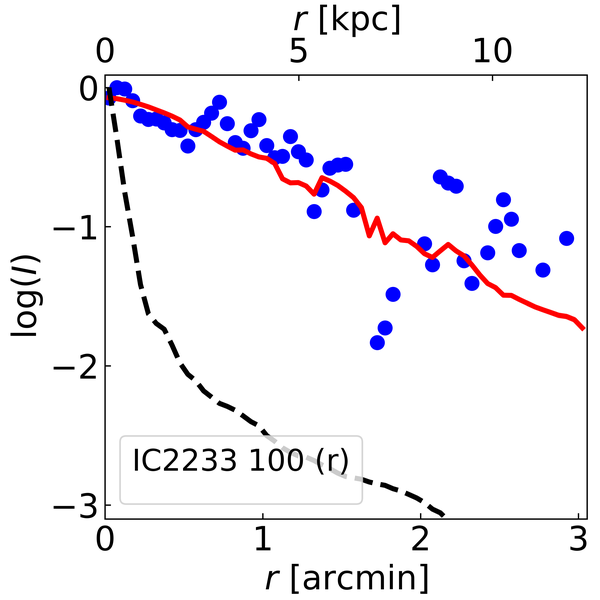}
\includegraphics[height=4.2cm]{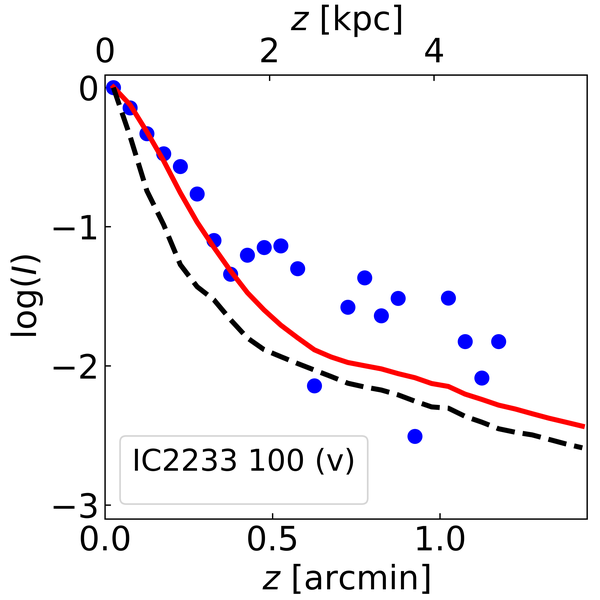}
\includegraphics[height=4.2cm]{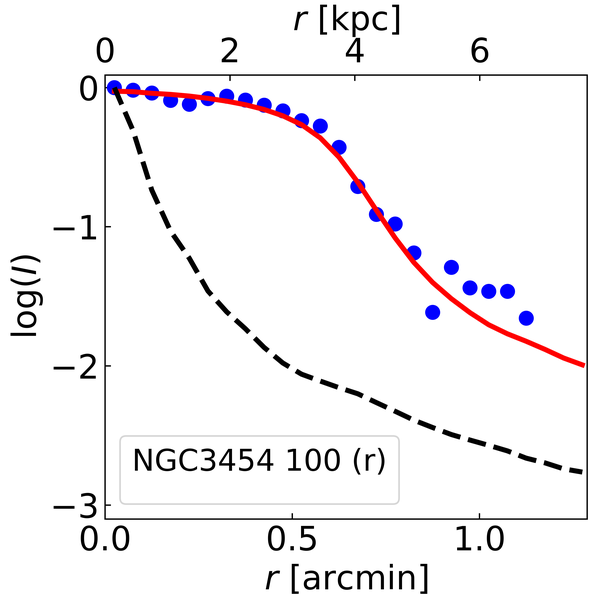}
\includegraphics[height=4.2cm]{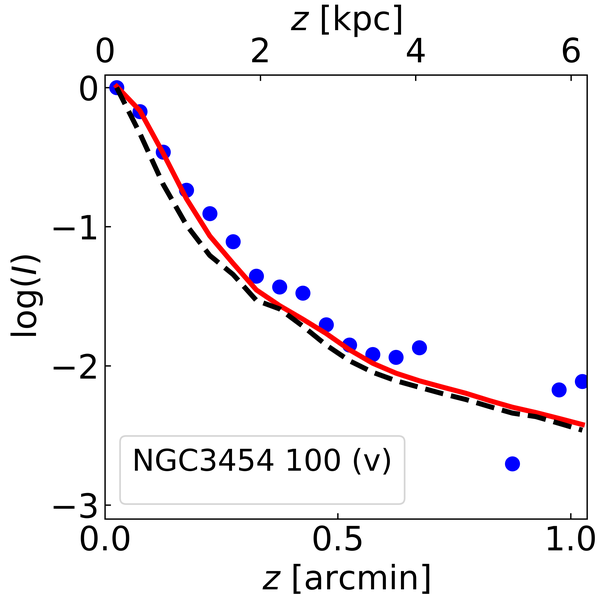}
\includegraphics[height=4.2cm]{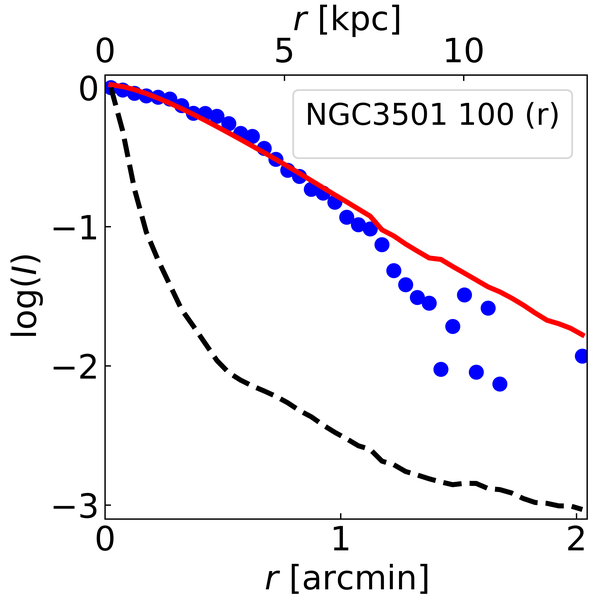}
\includegraphics[height=4.2cm]{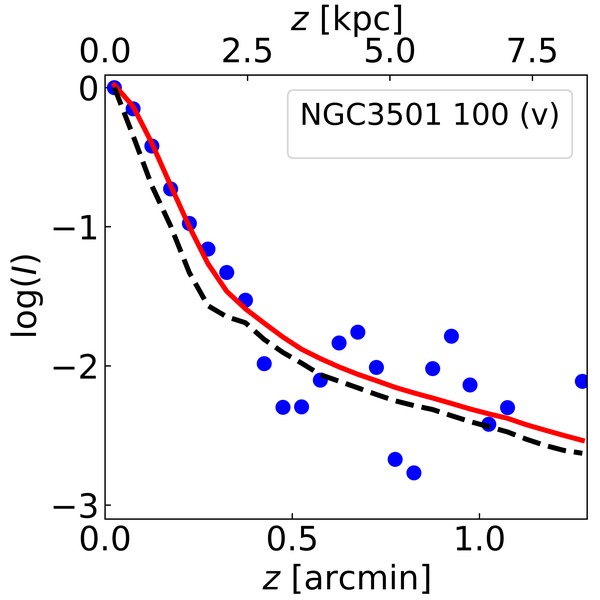}
\caption{(continued)}
\end{figure*}

\addtocounter{figure}{-1}
\begin{figure*}
\centering
\includegraphics[height=4.2cm]{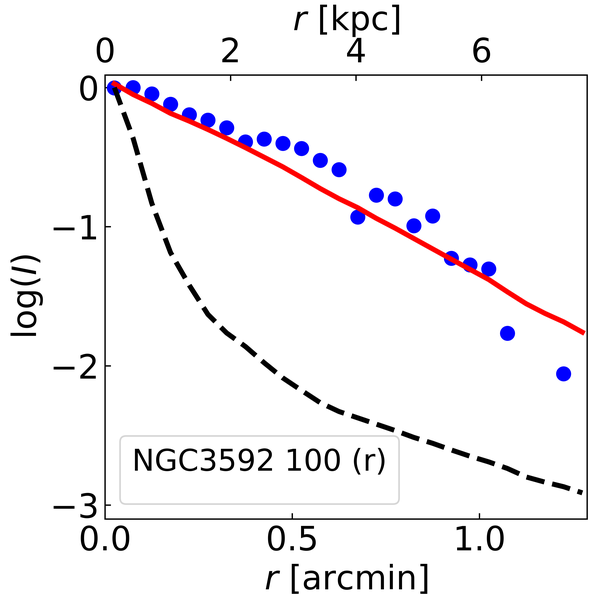}
\includegraphics[height=4.2cm]{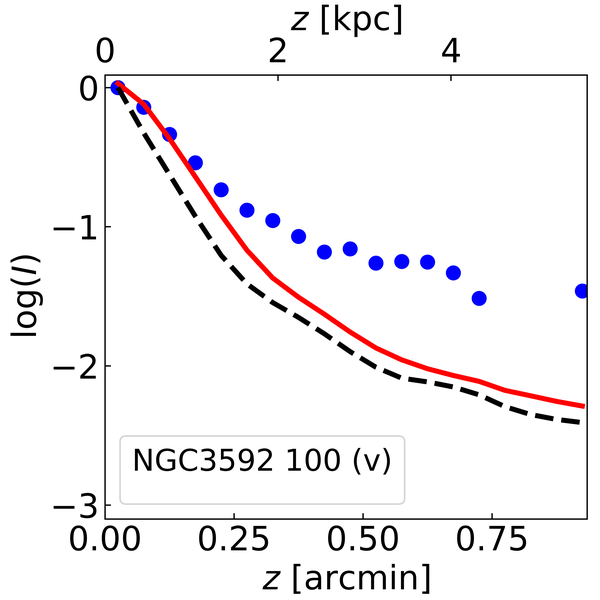}
\includegraphics[height=4.2cm]{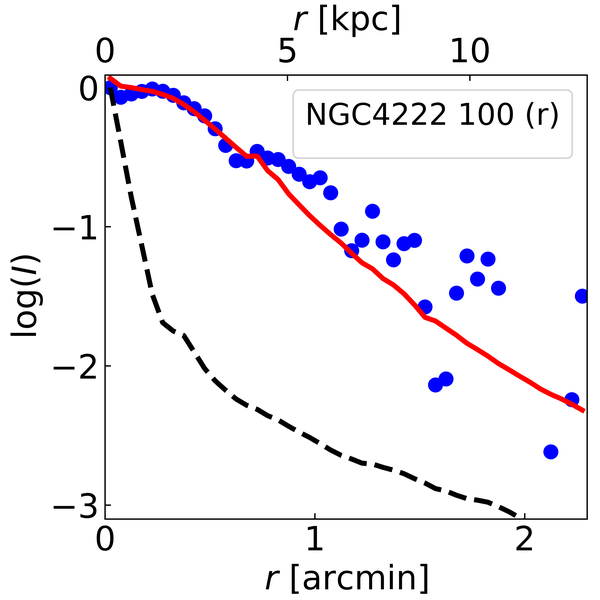}
\includegraphics[height=4.2cm]{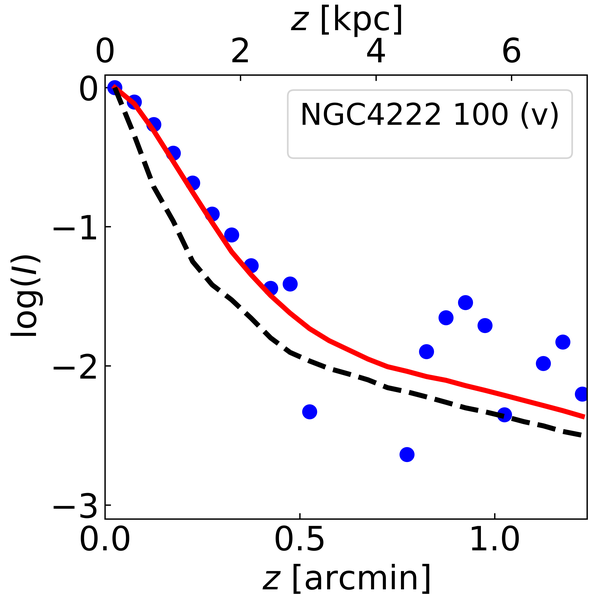}
\includegraphics[height=4.2cm]{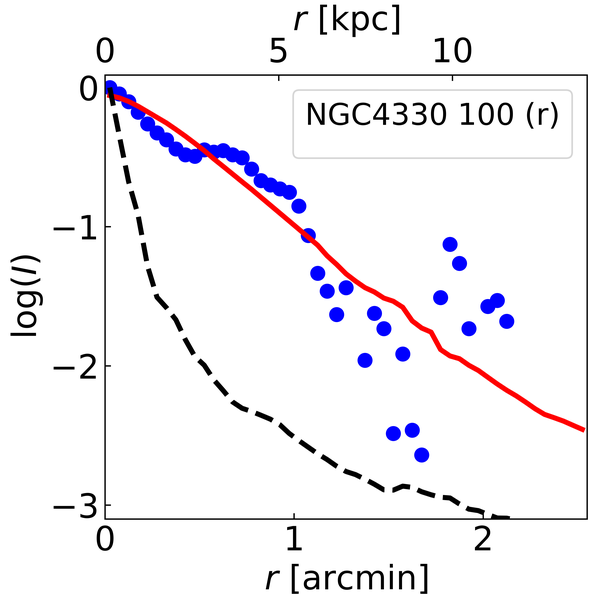}
\includegraphics[height=4.2cm]{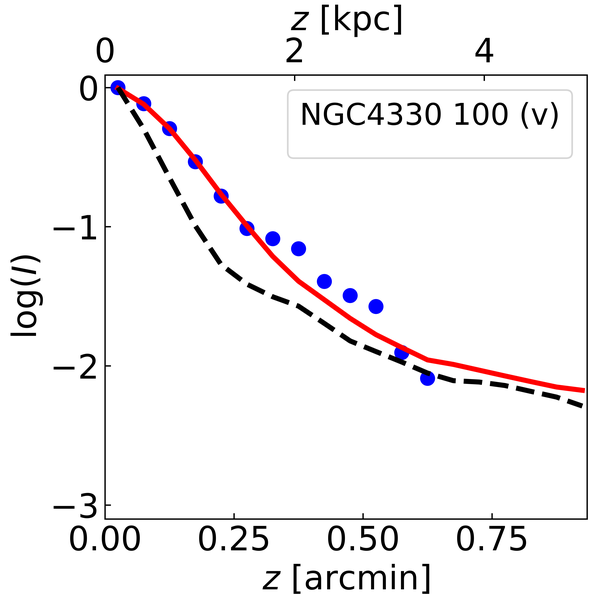}
\includegraphics[height=4.2cm]{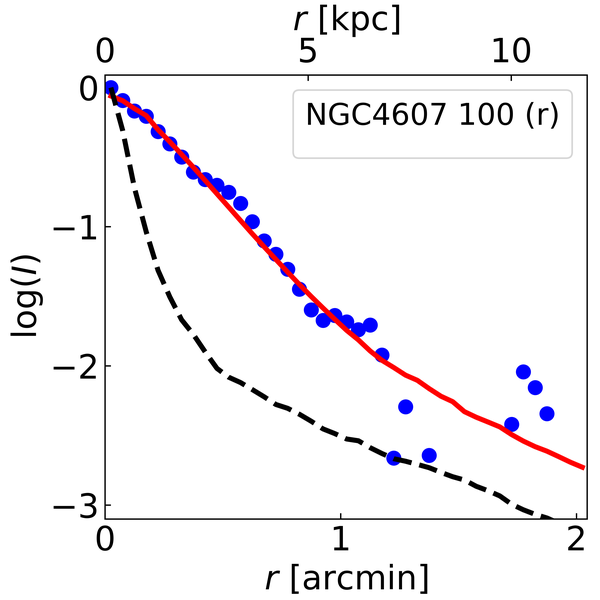}
\includegraphics[height=4.2cm]{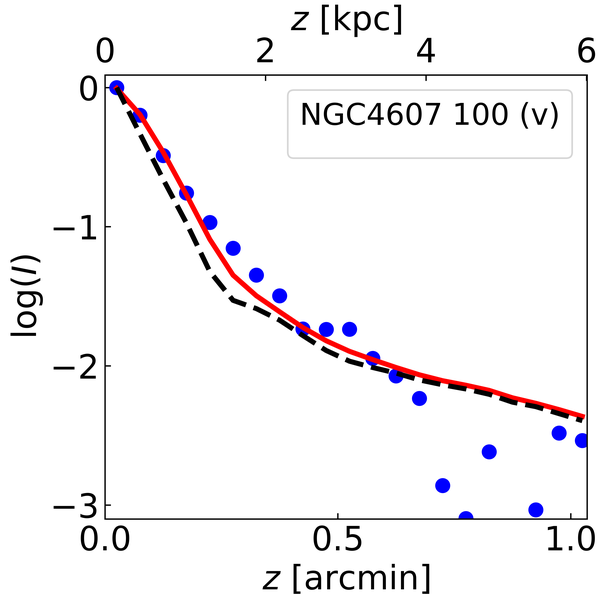}
\includegraphics[height=4.2cm]{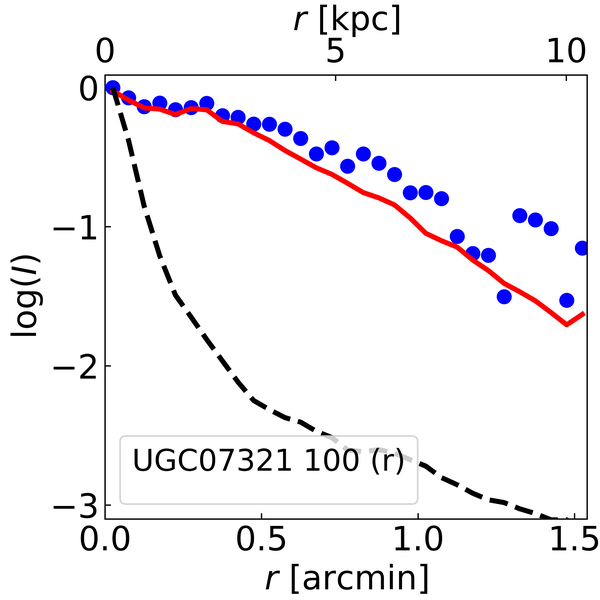}
\includegraphics[height=4.2cm]{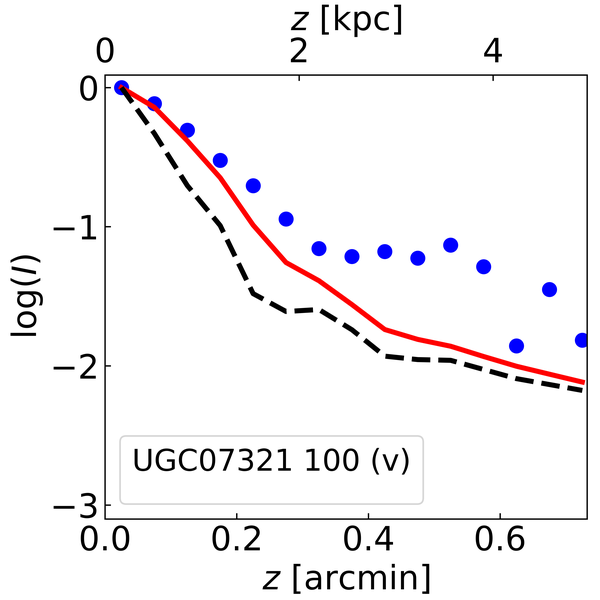}
\includegraphics[height=4.2cm]{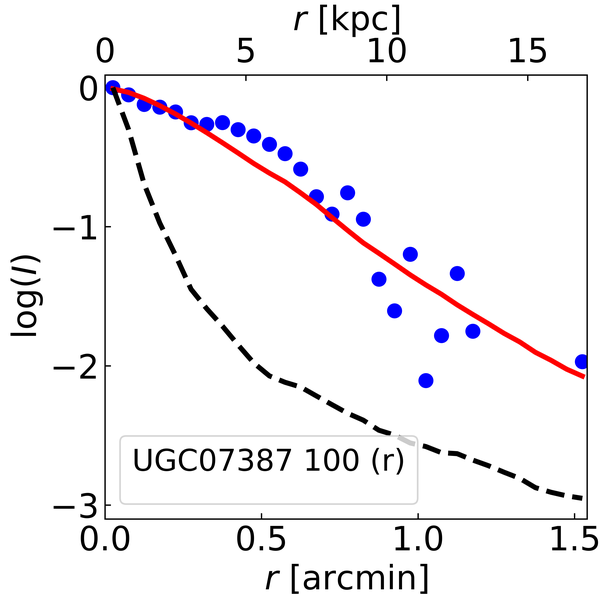}
\includegraphics[height=4.2cm]{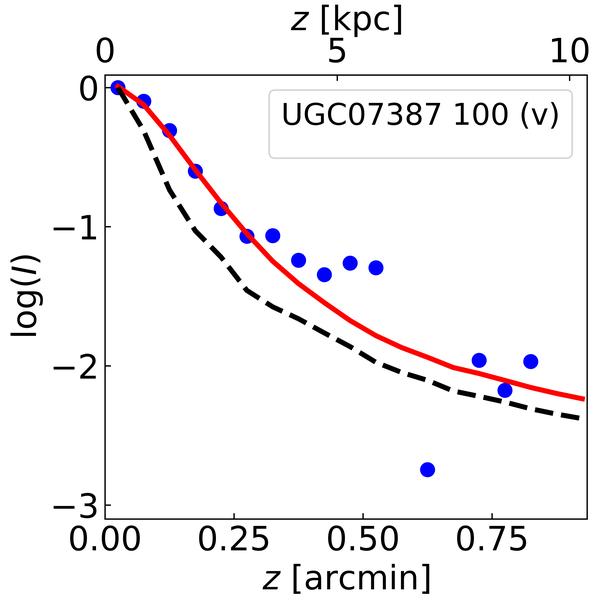}
\includegraphics[height=4.2cm]{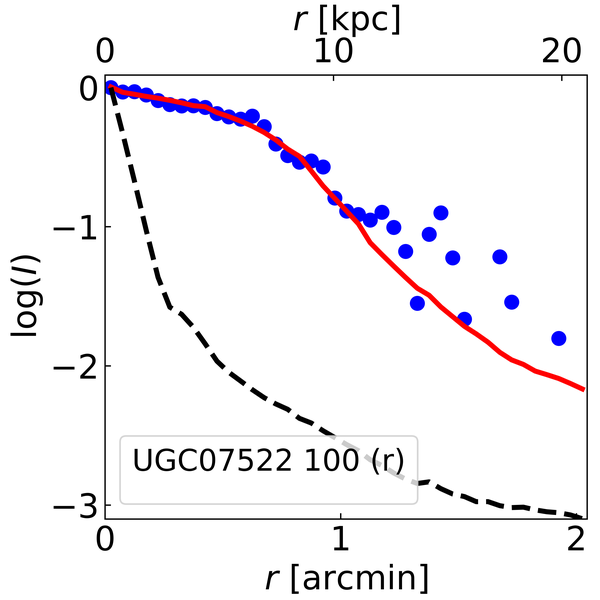}
\includegraphics[height=4.2cm]{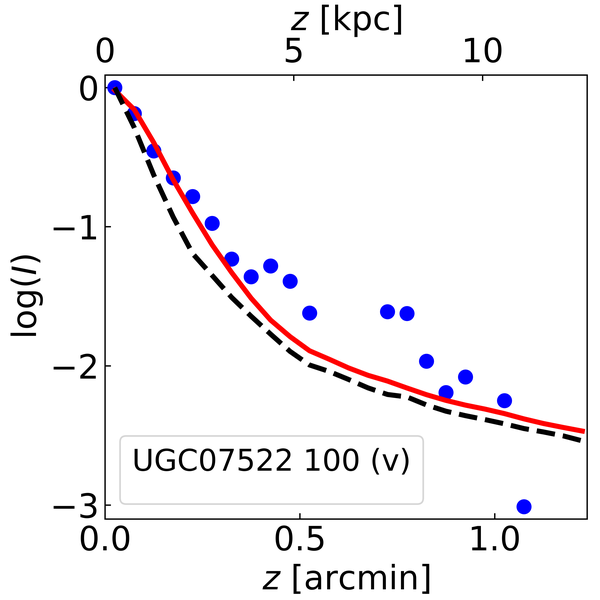}
\includegraphics[height=4.2cm]{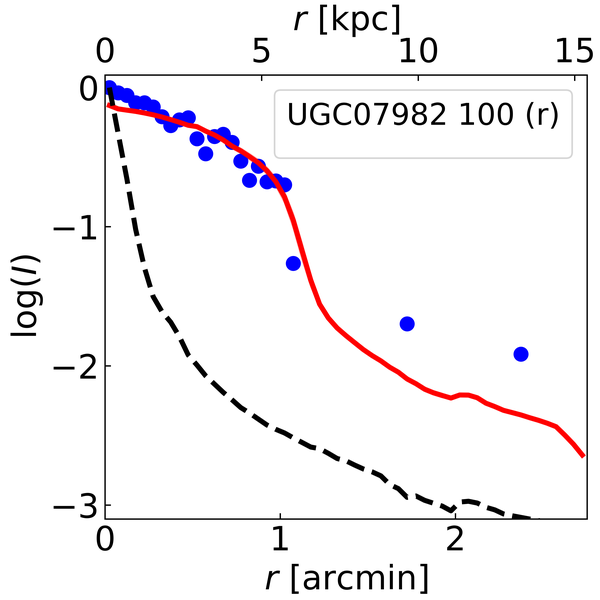}
\includegraphics[height=4.2cm]{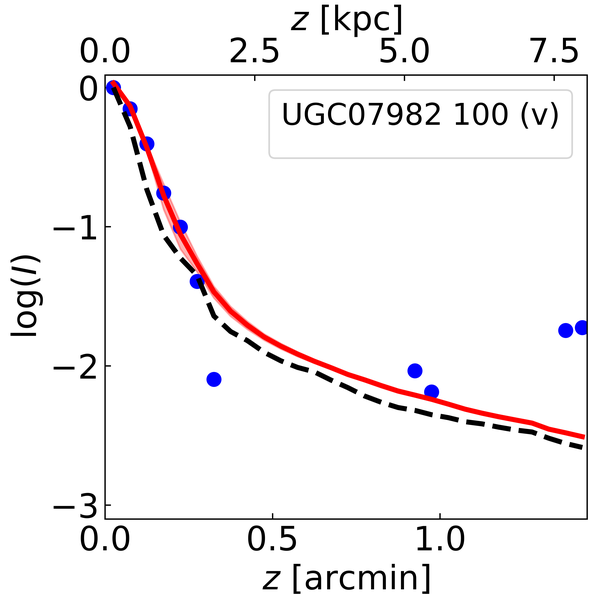}
\includegraphics[height=4.2cm]{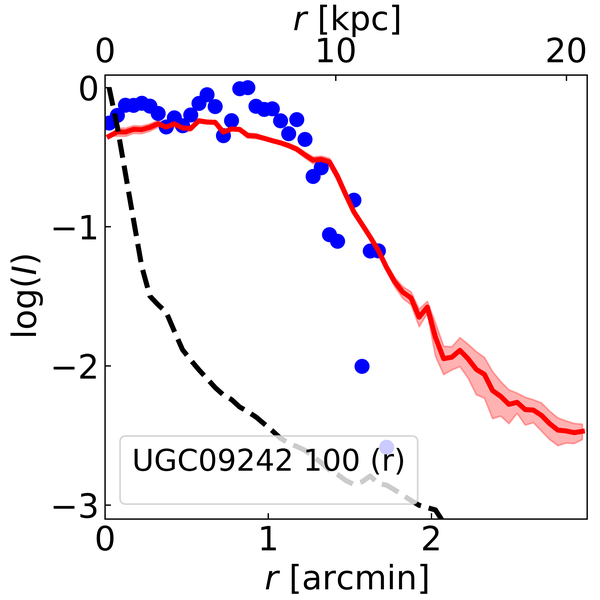}
\includegraphics[height=4.2cm]{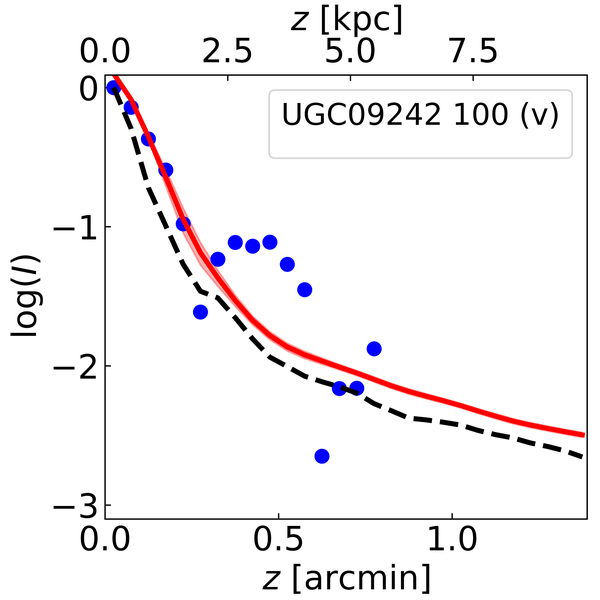}
\caption{(continued)}
\end{figure*}

\newpage

\bsp	
\label{lastpage}
\end{document}